\documentclass[bibyear]{aa}

\usepackage{graphicx}
\usepackage{txfonts}
\usepackage{natbib}
\bibpunct{(}{)}{;}{a}{}{,}
\usepackage{lscape}
\usepackage{scrextend}			
\usepackage{tikz}
\usepackage{caption,subcaption}
\usepackage{multirow}
\usepackage{gensymb}
\begin{document} 

\newcommand{\ma}[1]{\mathrm{#1}}
\newcommand{\uma}[1]{^{\mathrm{#1}}}
\newcommand{\dma}[1]{_{\mathrm{#1}}}
\newcommand{\be}{$\beta$}
\newcommand{\R}{$R_0$}
\newcommand{\mum}{\,$\mu$m}

\title{Hot exozodiacal dust : an exocometary origin?}

   \author{\'E. Sezestre\inst{1}
          \and
          J.-C. Augereau\inst{1}
          \and
          P. Th\'ebault\inst{2}
          }
          
  \offprints{\'E. Sezestre,\\ email: elie.sezestre@univ-grenoble-alpes.fr}

   \institute{Univ. Grenoble Alpes, CNRS, IPAG, 38000 Grenoble, France 
     \and
     LESIA, Observatoire de Paris, CNRS, Universit\'e Paris Diderot,
     Universit\'e Pierre et Marie Curie, 5 place Jules Janssen, 92190
     Meudon, France
   }

   \date{Received -; accepted -}

 
   \abstract
%
   {Near- and mid-infrared interferometric observations have revealed
     populations of hot and warm dust grains populating the inner
     regions of extrasolar planetary systems. These are known as
     exozodiacal dust clouds, or exozodis, reflecting the similarity
     with the Solar System's zodiacal cloud. Radiative transfer models
     have constrained the dust to be dominated by tiny
     submicron-sized, carbon-rich grains that are accumulated very
     close to the sublimation radius.  The origin of this dust is an
     unsolved issue.
   }
%
%
   {We aim to explore two exozodiacal dust production mechanisms,
     first re-investigating the Poynting-Robertson drag pile-up
     scenario, and then elaborating on the less explored, but
     promising exocometary dust delivery scenario.}
%
%
   {We developped a new versatile, numerical model that calculates the
     dust dynamics, with non orbit-averaged equations for the grains
     close to the star. The model includes dust sublimation and
     incorporates a radiative transfer code for direct comparison to
     the observations. We consider in this study four stellar types,
     three dust compositions, and we assume a parent belt at 50\,au.}
%
%
   {We find that, in the case of the Poynting-Robertson drag pile-up
     scenario, it is impossible to produce long-lived submicron-sized
     grains close to the star. The inward drifting grains fill in the
     region between the parent belt and the sublimation distance,
     producing an unrealistically strong mid-infrared excess compared to
     the near-infrared excess. The dust pile-up at the sublimation
     radius is by far insufficient to boost the near-IR flux of the
     exozodi to the point where it dominates over the mid-infrared
     excess. In the case of the exocometary dust delivery scenario, we
     find that a narrow ring can form close to the sublimation zone,
     populated with large grains several tens to several hundred of
     micrometers in radius. Although not perfect, this scenario provides a better
     match to the observations, especially if the grains are
     carbon-rich. We also find that the required number of active
     exocomets to sustain the observed dust level is reasonable.}
%
%
   {We conclude that the hot exozodiacal dust detected by
     near-infrared interferometry is unlikely to result from inwards
     grains migration by Poynting-Robertson drag from a distant parent
     belt, but could instead have an exocometary origin.}


   \maketitle


\section{Introduction}
\label{sec:Intro}

Hot exozodiacal dust ("exozodis") has been detected, by means of
interferometric observations in the near-infrared (near-IR, H- or
K-band), around about 25 main sequence stars \citep{Absil2013,
  Ertel2014, Ertel2016, Kral2017, Nunez2017}. These exozodis are very
bright, amounting to $\sim 1\%$ of the stellar flux in the K-band,
which is about 1000 times more than the solar system's own zodiacal
cloud in the same spectral range. For some of these systems, a "warm"
counterpart has also been detected in the mid-infrared \citep[mid-IR,
  8-20 $\mu$m, e.g.][]{Mennesson2013, Su2013, Mennesson2014,
  Ertel2018}, but this mid-IR exozodi to star flux ratio never exceeds
the flux ratios in the H- or K-band \citep[][]{Kirchschlager2017}.
Furthermore, for the handful of systems for which parametric modelling
based on radiative transfer codes has been performed \citep{Absil2006,
  difolco2007, absil2008, Akeson2009, Defrere2011, Lebreton2013,
  Kirchschlager2017}, the ratio between the fluxes in the near-IR and
mid-IR has constrained the dust to be dominated by tiny
submicron-sized grains that are accumulated very close to the
sublimation radius (hereafter $r\dma{s}$, typically a few stellar radii).

The presence of such large amounts of very small grains so close to
their star poses a challenge when it comes to explaining the exozodis'
origin.  Indeed, the canonical explanation invoked for "standard" cold
debris disks, i.e., the \textsl{in situ} steady production of small
grains by a collisional cascade starting from larger parent bodies
\citep[e.g.][]{Krivov2010}, cannot hold here because collisional
erosion is much too fast in these innermost regions to be sustained
over periods comparable to the system's age \citep{Bonsor2012,
  Kral2017}. Therefore, the long-term existence of a hot exozodi
requires both an external reservoir of material and an inward
transport mechanism, feeding with dust the region close to the
sublimation radius at a rate of about
$10^{-10}-10^{-9}~M_{\oplus}$/year \citep[e.g.][]{Absil2006,
  Kral2017}. A significant fraction (more than $\sim$20\%) of nearby
solar- and A-type stars does possess an extrasolar analog to the
Kuiper belt \citep{Montesinos2016,Sibthorpe2018,Thureau2014},
indicating that external reservoirs for exozodis are common. The
inward transport mechanism must then be sufficiently generic to affect
more than 10\% of the nearby stars, independent of their age and
spectral type \citep{Ertel2014, Nunez2017}. For instance, large-scale
dynamical instabilities in planetary systems, that could occur
randomly (e.g. the Late Heavy Bombardment in the Solar System), were
shown to indeed significantly increase the number of small bodies
scattered from an external Kuiper-like belt toward the star, but
because each event lasts less than a few million years, the
probability to observe hot exozodiacal dust produced during such an
event is less than 0.1\% \citep{Bonsor2013}. This mechanism cannot
explain the vast majority of the hot exozodis.

So far, two main categories of exozodi-origin scenarios have been
explored.  The first one assumes that the dust is collisionally
produced further out in the system (in an asteroidal or Kuiper-like
belt) and migrates inward, because of Poynting-Robertson drag
(hereafter PR-drag), until it reaches the sublimation distance
$r\dma{s}$. There, it starts to sublimate and shrink until radiation
pressure becomes significant and increases its orbital semi-major axis
and eccentricity, while keeping its periastron nearly the same. This
will slow-down the inward migration and thus potentially create a
pile-up of small grains close to $r\dma{s}$. This scenario follows the
pionieering work of \citet{Belton1966} predicting a density peak near
the sublimation distance in the Solar System, and those by
\citet{Mukai1974} and \citet{Mukai1979} attempting to explain the
observed flux bump at about $4\,R_{\odot}$ in the F-corona (the hot
component of the zodiacal dust cloud). However, the estimated
amplitude of this pile-up seems to be too weak to explain the observed
near-IR excesses in extrasolar systems 
\citep{Kobayashi2008,Kobayashi2009, Kobayashi2011, VanLieshout2014}.  
Another problem is
that this scenario does not seem to be able to produce grains that are as
small as those derived from radiative transfer modeling. However, it is worth noting
that these results were obtained using orbit-averaged equations of
motion that might become inaccurate close to $r\dma{s}$ because of the very
fast variations imposed by the sublimation.

A second way of delivering dust in the innermost regions of planetary
systems is by the sublimation of large asteroidal or cometary bodies,
originating in an external belt, and scattered inwards by a chain of
low-mass planets \citep{Bonsor2012, Bonsor2014, Raymond2014, Marboeuf2016}.
There are evidences for exocometary activity around other stars than
the Sun, through the observation of transient, Doppler-shifted gas
absorption lines \citep[e.g.][and references therein]{Beust2000,
  Kiefer2014a, Kiefer2014b}, and the analysis of Kepler transit light
curves attributed to trailing dust tails passing in front of the star
\citep[][with mass loss rates of $\sim 10^{-12}~M_{\oplus}$/year and
  $>10^{-10}~M_{\oplus}$/year, respectively]{Kiefer2017,
  Rappaport2018}. In the Solar System, comets are supposed to
contribute significantly to the zodiacal cloud
\citep[e.g.][]{Liou1995,Dermott1996}. \citet{Nesvorny2010} estimated
for example that $\sim 90\%$ of the zodiacal dust originates from
Jupiter family comets. The cometary hypothesis as a source of hot
exozodiacal dust has, however, never been tested quantitatively in
terms of the level of dustiness that can be obtained near the $r\dma{s}$
region.

This paper reinvestigates both these scenarios. For the PR-drag case
(Sec.~\ref{sec:PRdrag}), we use for the first time a sophisticated
numerical model that does not rely on orbit-averaged equations in the
crucial sublimation region (Sec.~\ref{sec:Model}).  We also explore
the potential role played by the Differential Doppler Effect (DDE)
evoked by Kimura et al. (2017, 10th meeting on Cosmic Dust,
Tokyo)\footnote{https://www.cps-jp.org/\textasciitilde dust}.  As for
the comet-delivery case, we perform the first quantitative exploration
of this scenario in the context of exozodis, following the fate of the
dust that is produced as the comet sublimates (Sec.~\ref{sec:Comets}).
For each scenario, we explore a wide range of possible grain
compositions and stellar types (Sec.~\ref{sec:Model}). Rather than
checking the validity of each scenario by assessing how well they can
reproduce the \emph{predictions} of radiative-model fits (grain
location and typical sizes), we chose to directly focus on the
observational constraints themselves, in particular the fluxes in the
near- and mid-IR.


\section{Numerical Model}
\label{sec:Model}

\begin{figure*}[tp!]
\centering
\hbox to \textwidth
{
\parbox{0.32\textwidth}{
\includegraphics[width=0.32\textwidth,trim=0.5cm .5cm .5cm .5cm,clip]{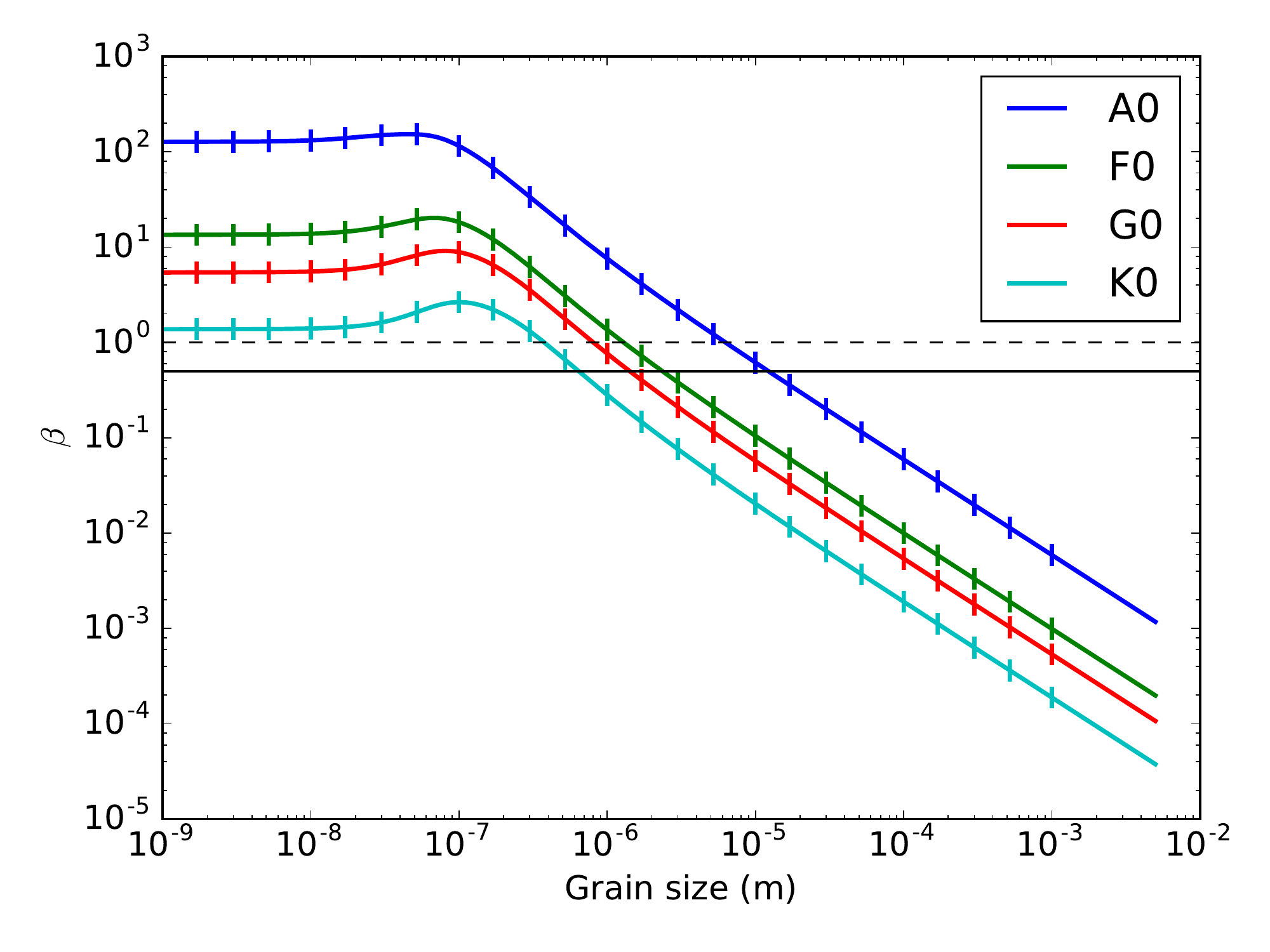}
\subcaption{Carbon}
\label{fig:betaCar}
}
\hfill
\parbox{0.32\textwidth}{
\includegraphics[width=0.32\textwidth,trim=0.5cm .5cm .5cm .5cm,clip]{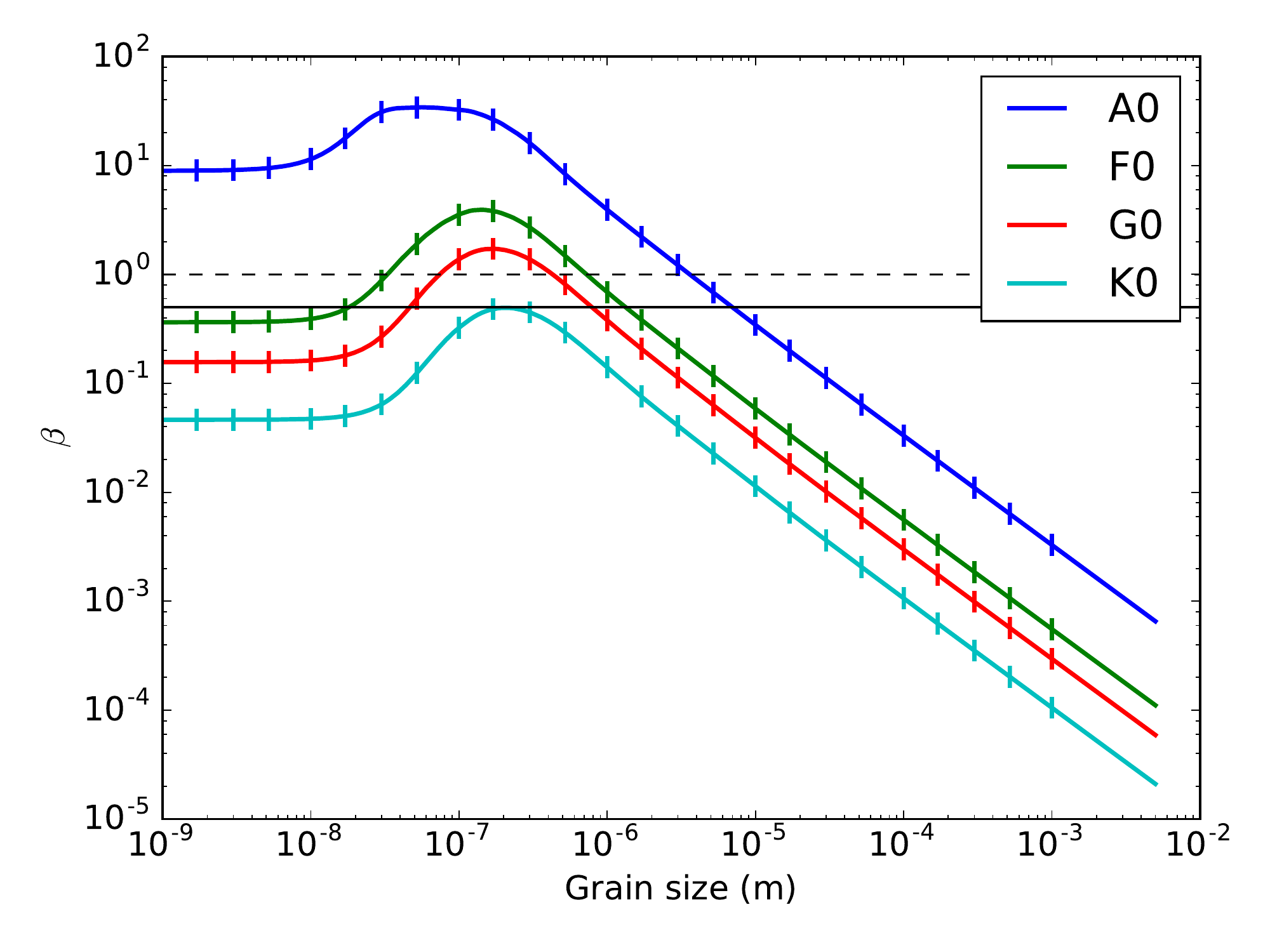}
\subcaption{Astrosilicate}
\label{fig:betaSi}
}
\hfill
\parbox{0.32\textwidth}{
\includegraphics[width=0.32\textwidth,trim=0.5cm .5cm .5cm .5cm,clip]{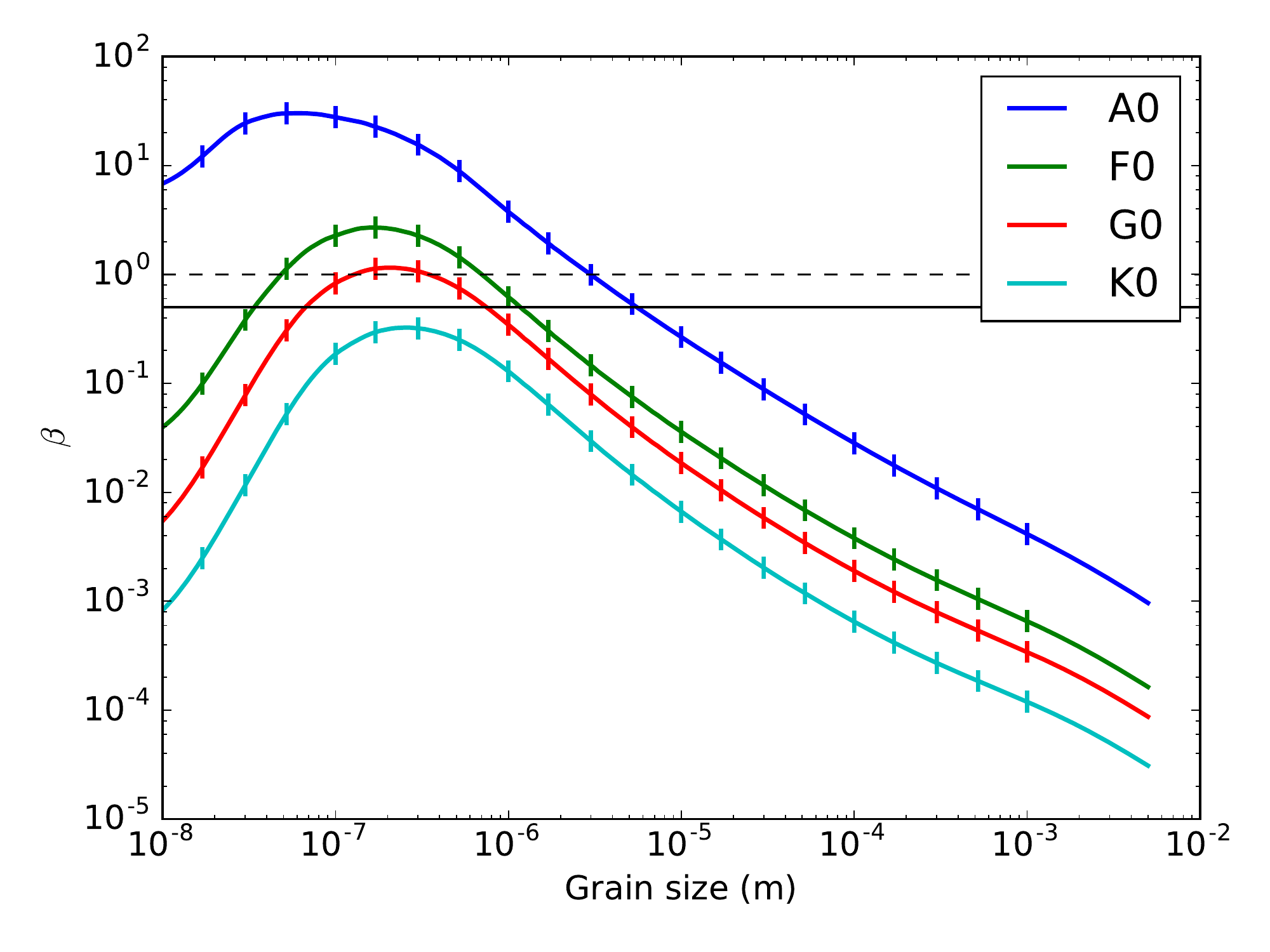}
\subcaption{Glassy silicate}
\label{fig:betaGla}
}
}
\caption{$\beta\dma{pr}$ for 3 different grain compositions around different spectral type stars
  (ticks correspond to the grain sizes used in our simulations).  
  The solid horizontal line is the limit $\beta = 0.5$: 
  grains over this value are blown-out by radiation pressure if produced from circular orbits.
  The dashed horizontal line is the limit $\beta = 1$: 
  grains above this value are always expelled, regardless the of the way they are produced.  
  }
\label{fig:BetaAll}
\end{figure*}

\subsection{General philosophy}
\label{subsec:GenePhilo}

We use in essence the same numerical code to investigate both the
PR-drag pile-up and the cometary delivery scenarios. Our model
performs a consistent treatment of a grain evolution, from its release
to its ejection, sublimation, or fall onto the star. We take into
account stellar gravity, stellar radiation/wind pressure, PR-drag and
sublimation. In a more advanced version, the stellar magnetic field
can be turned on, but this capability will not be used in this paper.

In this study, we chose to neglect collective effects such as mutual
collisions. This might appear as a step back when compared to the
studies of \cite{Kobayashi2009} and \cite{VanLieshout2014}, who did
take into account collisional effects (albeit in a very simplified
way) for the PR-drag pile-up scenario.  However, we believe that this
neglect of collisions does not radically bias our results.
\cite{VanLieshout2014} has indeed shown that, because of the
self-regulating interplay between collisions and PR-drag, collisional
effects will only play a significant role, potentially halting the
inward drift of grains, very close to the location of the parent body
belt releasing the dust grains.  As soon as the dust has migrated away
from the parent belt, its number density is always low enough for
mutual collisions not to have a major effect on its evolution
\citep[see Fig.\,2 of][]{VanLieshout2014}.  In this respect, the only
drawback of not taking into account collisions is that we cannot
derive the density and the mass of the dust-producing parent belt, but
this is not the main focus of our study, which concentrates on the
evolution of the dust once it has reached the inner regions of the
system.

For the dust evolution in these innermost regions, our code presents a
step forward as compared to previous studies because it does not rely
on orbit-averaged d$a$/d$t$ and d$e$/d$t$ estimates but integrates the
exact equations of motion up until the grain is removed.  This is a
crucial point in the critical region close to the sublimation radius,
where a grain radius can vary on timescales much smaller than the
local orbital period, thus inducing dynamical changes that cannot be
accounted for with averaged estimates. In addition, the orbit-averaged
d$e$/d$t$ estimates can lead to eccentricity values that can be
infinitely small, whereas in reality there is always a minimum
"residual" osculating eccentricity below which the particle's orbit
cannot go (see section \ref{subsec:GrainEvo}).

In addition to the dust evolution, the output of the code is a global
density map assuming the system is at steady state. This is used to
produce a synthetic spectrum of the exozodi that can be compared to
mesured spectra, and also flux levels mesured by interferometric
studies.  Since the mass of dust close to the star in the PR-drag
scenario is not constrained by the mass of the dust producing belt, we
try to reproduce the trend of the spectra, and use the mass of the
exozodi as a free parameter. More specifically, we scale the mass such
that the excess corresponds to observations at 2\mum ,
and we use the excess observed in mid-IR to discuss the
relevancy of the examined scenarios
\citep[around 1\% at 8--20\mum, e.g.][]{Kirchschlager2017}.
On the contrary, in the cometary
release scenario, the flux level can be estimated by the mass of the
releasing comet, giving constraints on its radius.

\subsection{Dynamical approach}

The code computes the dynamics of a set of compact dust grains with
initial sizes chosen to sample different dynamical behaviours. The
equation of motion is solved with a 4$\uma{th}$~order Runge-Kutta
integrator with an adaptive timestep.  The code is able to take into
account the stellar gravity ($\mathbf{F}\dma{grav}$), the radiation
pressure and the Poynting-Roberston drag ($\mathbf{F}\dma{PR}$), 
and the
Differential Doppler Effect
\citep[$\mathbf{F}\dma{DDE}$, e.g.][]{Burns1979}.  
Each of these effects can be individually
switched on or off at any time. 

The forces are expressed as follows :
\begin{eqnarray}
\mathbf{F}\dma{grav} & = & - \frac{G M_{\star} m }{r^2} \cdot \vec{e}_r 	\label{eqn:Fgrav} \\
\mathbf{F}\dma{PR} & = & \beta\dma{pr} \frac{G M_{\star} m }{r^2} \> \left[ \left( 1 - \frac{\dot{r}}{c} \right) \vec{e}_r - \frac{\mathbf{v}}{c} \right] 	\label{eqn:Fpr} \\
\mathbf{F}\dma{DDE} & = & -  \frac{\omega_{\star} R_{\star}^2 }{4}
\frac{\beta\dma{pr}}{\sqrt{1-\beta\dma{pr}} }
\sqrt{\frac{GM_{\star}}{r^5}} \cdot
\frac{\mathbf{v}}{c}	\label{eqn:Fdde}
\end{eqnarray}
where $\vec{e}_r$ is the radial unit vector,
$G$ the gravitational constant, 
$c$ the speed of light,
$M_{\star}$ the mass of the star, 
$R_{\star}$ the stellar radius,
$\omega_{\star}$ the rotation frequency of the star,
$m$ the mass of the grain,
$r$ the distance of the grain to the star, 
$\mathbf{v}$ the grain velocity
and $\dot{r}$ the radial velocity, 
$\beta\dma{pr}$ the ratio of the radiation pressure force to the gravitational force.

Other forces, in particular the stellar wind pressure and the Lorentz force, 
are implemented in the code but will not be used in this study.
The pressure due to the stellar wind is comparable to the radiation pressure
only for submicron-sized grains around late-type stars.
As we choose to focus on K-type and earlier stars (Sec.~\ref{sec:star_composition}), 
we will not take into account the stellar wind pressure.
For consistency and simplicity, we will rename the $\beta\dma{pr}$ parameter as \be\
in the following.
The Lorentz force acting on charged grains interacting with the large-scale stellar magnetic field
can also affect the grain dynamics, as evidenced by \cite{Czechowski2012} 
or \cite{Rieke2016}.
We will neither discuss the Lorentz force as it exceeds the scope of this study.

The initial conditions of the simulations depend on the scenario that
is considered. These are detailled in Secs.~\ref{sec:PRdrag} and
\ref{sec:Comets} for the PR-drag pile-up and the cometary delivery
scenarios, respectively. The grain dynamics is computed until one of
following criteria is met:
\begin{itemize}
  \item the grain sublimates completely. This is reached when the
    grain size is below the lower limit of the predefined size grid,
    which in most cases corresponds to a size smaller than 1\,nm.
  \item the grain falls onto the star. This is assumed to happen when
    the distance of the grain to the surface of the star is less than
    0.1\,$R_{\star}$.
  \item the grain is expelled. This is assumed to occur when the
    distance to the star is over 1000\,au.
  \item the grain is too old. This is considered to be the case when
    the integration time is over one million years, meaning the grain
    has not evolved.
\end{itemize}
The integration timestep is taken as a fraction of the local
revolution period (typically a hundredth), to ensure a sufficient
resolution at every distance from the star.  As a test of the code, we
reproduced the results in Figure\,5 of \cite{Krivov1998} with great
fidelity, as shown in Appendix\,\ref{app:Krivov}.
However, for the grains released from parent bodies in a distant belt 
and then migrating inward by PR-drag, 
like in the scenario developed in Sec.\,\ref{sec:PRdrag}, 
this short timestep becomes a numerical limitation. 
Therefore, and as long as the grain remains far from the star, 
we opt in this case for the orbit-averaged prescription of \citeauthor{Wyatt1950} 
(\citeyear{Wyatt1950}, their Eq.\,9)
to evolve the grains by PR-drag. 
This approach allows to save computational time during the less critical evolution stages 
(stage~I as defined in Sec.\,\ref{sec:PRdraggeneralbehaviour}), 
and is similar to the methodology employed by
\cite{Kobayashi2009} and \cite{VanLieshout2014}.
According to \cite{Wyatt1950}, 
the quantity $a e^{-4/5} (1-e^2)$ remains constant during the PR-drag migration, 
where $a$ is the grain semi-major axis and $e$ its eccentricity. 
We use this conservation principle to estimate the semi-major axis and the eccentricity 
as the grain migrates inward until the full, non orbit-averaged simulation is switched on, 
in contrast with what was done in previous studies. 
The switch is done when the grain reaches an equilibrium temperature at periastron 
that is half its sublimation temperature, to prevent sublimation to occur during the orbit-averaged phase. 
We also continuously monitor the evolution of the grain radius due to sublimation during this phase, 
in order to stop the orbit-averaged treatment if the radius is decreased by more than 1\% of its initial value. 
We have checked on a test run that this approach provides the same results as those obtained with the full-simulation. 
It should be noted that, in the cometary scenario developed in Sec.\,\ref{sec:Comets}, 
the whole grain evolution is done using non orbit-averaged equations.

\subsection{Stellar and grain properties}
\label{sec:star_composition}

\begin{table}
\caption{Reference stars used in the code.  Luminosity and mass are
  estimated by the code by interpolating the values computed for
  spectral type.  }
\label{tab:StellarParam}
\begin{tabular}{c c c c c c}
\hline\hline
  Spec.	& Name	& Distance	& V-band	& Luminosity 	& Mass \\
type	& 	& (pc)	&  mag.	& ($L_{\odot}$) 	& ($M_{\odot}$) \\ \hline
A0	& \object{Vega} 			&   7.68$\pm$0.02 \tablefootmark{a}	& 0.03 \tablefootmark{c}	& 57 	& 2.9 \\
F0	& \object{$\rho$ Gem}	& 18.05$\pm$0.08 \tablefootmark{b}	& 4.18 \tablefootmark{c}	& 5.8 		& 1.6 \\
G0	& \object{Iam Ser}		& 11.82$\pm$0.04 \tablefootmark{b}		& 4.42 \tablefootmark{d}	& 2.0 	& 1.05 \\
K0	& \object{54 Psc}			& 11.14$\pm$0.01 \tablefootmark{b}		& 5.88 \tablefootmark{d}	& 0.57 	& 0.79 \\
\hline
\end{tabular}
\tablebib{
\tablefoottext{a}{\cite{vanLeeuwen2007}}
\tablefoottext{b}{\cite{Gaia2016, Gaia2018}}
\tablefoottext{c}{\cite{Vizier}}
\tablefoottext{d}{\cite{VanBelle2009}}
}
\end{table}

In this paper, we consider four different stellar types, ranging from
A0 to K0. For that purpose, we choose four representative nearby
stars, that do not necessarily possess hot exozodiacal dust. Their
properties are summarized in Tab.\,\ref{tab:StellarParam}. 

\begin{table*}
\caption{Grain parameters used in the code.  Sublimation parameters
  $A$ and $B$ refer to those used in \cite{Lebreton2013}.
  Appendix\,\ref{app:TermoProp} provides a comparison with other
  sublimation formulae and notations used in the literature.  }
\label{tab:grainParam}
\centering
\begin{tabular}{l  l l l l c }
\hline\hline
  Name	& Symbol	& Carbon	& Astrosilicate	& Glassy silicate & Reference \\ \hline
Density & $\rho$ (kg.m$^{-3}$)	& 1.78$\times 10^3$	& 3.5$\times 10^3$	& 2.37$\times 10^3$ 	 & 1, 1, 2 \\
Mean molecular mass	& $\mu$ (g.mol$^{-1}$)	& 12.01 	& 172.2	& 67.00 	 & 1, 1, 2 \\
Sublimation temperature & $T\dma{sub}$ (K)	& 2000	& 1200	& 1200 	 & 1 \\
A & (cgs)	& 37215	& 28030	& 24918 	 & 3, 4, 5 \\
B & (cgs)	& 7.2294	& 12.471	& 7.9356 	 & 3,
4, 5 \\
\hline
\end{tabular}
\tablebib{
(1)~Carbonaceous material and Silicates of \cite{Lebreton2013};
(2)~Obsidian of \cite{Lamy1974} ;
(3)~C$\dma{1}$ specy of \cite{Zavitsanos1973}; 
(4)~Astronomical silicate of \cite{Kama2009};
(5)~Silicate of \cite{Kimura1997} }
\end{table*}

We consider three different grain compositions, parameterized by their
physical, optical and thermodynamical properties. In the following,
"carbon" will refer to amorphous carbonaceous grains, "astrosilicates"
and "glassy silicate"
to amorphous silicate grains.
The latter two compositions differ by their optical indexes.  
Optical indexes for carbon grains are taken from \citet[][ACAR sample]{Zubko1996},
while those for astrosilicates are from \citet{Draine2003}.  The
optical indexes for glassy silicates combine measurements for obsidian
from Lake Co. Oregon \citep{Pollack1973,Lamy1978} in the spectral
range 0.1\mum--50\mum, with a constant value for the real part beyond
$\lambda$=50\mum , and a constant value from $\lambda$=50\mum\ to
300\mum\ for the imaginary part, followed by the imaginary part of the
astrosilicates of \citet{Draine2003} beyond $\lambda$=300\mum .  Below
$\lambda$=0.1\mum , both the real and imaginary parts are assumed to
be constant.  This set of optical indexes for the ``glassy silicates''
corresponds to the one used in \citet{Kimura1997} and
\citet{Krivov1998} for silicate grains, with the only addition of the
extension beyond $\lambda$=300\mum\ which is specific to this study.

We employ the Mie theory, valid for hard spheres, to compute the dust
optical properties. These are used to derive the \be\ ratios
\citep[e.g. Eq.\,3 in][]{Sezestre2017} and the radial profiles of the
grain temperature \citep[e.g. Eq.\,4 in][]{Lebreton2013}. Both depend
on the grain size, on the grain composition and on the star that is
considered, as shown in Figs.\,\ref{fig:betaCar}, \ref{fig:betaSi} and
\ref{fig:betaGla} in the case of the \be\ ratios. 

The sublimation prescription is taken from \citet[][their Eqs. 17 and
  18]{Lebreton2013}, and follows the methodology described in
\cite{Lamy1974}. The evolution of the grain size $s$ reads:
\begin{equation}
\frac{\ma{d}s}{\ma{dt}} = - \frac{\alpha}{\rho} \sqrt{\frac{k\dma{B}
    T}{2 \pi \mu m\dma{u}}} \rho\dma{eq} ,
\end{equation}
where $\rho$ is the grain density,
$k_B$ is the Boltzmann constant, 
$T$ is the grain temperature,
$\mu$ is the mean molecular mass of the considered dust composition,
and $m\dma{u}$ is the atomic mass unit.
The equilibrium gas density $\rho\dma{eq}$ around the grain is given by:
\begin{equation}
\log\dma{10} \rho\dma{eq} = B - \frac{A}{T\dma{sub}} - \log\dma{10} T\dma{sub} ,
\end{equation}
with $T\dma{sub}$ being the sublimation temperature of the grain.
We have assumed that the pressure of the gas surrounding the grain
\citep[$\rho\dma{gas}$ in][]{Lebreton2013} is negligible, and the
efficiency factor $\alpha$ to be 0.7 like in \cite{Lamy1974}.  The
thermodynamical properties are documented in
Tab.\,\ref{tab:grainParam} for each of the three compositions
considered in this paper.  The sublimation prescription used here is
similar to the one in \cite{Kobayashi2011}, with the transformations
from one set of thermodynamical parameters to another given in
Appendix\,\ref{app:TermoProp}.  At each timestep in the dynamical
code, the mass lost by a grain due to sublimation is computed, and the
grain size as well as the \be\ value are modified accordingly for the
next dynamical timestep. It is worth noting that the sublimation
timescales can be very sensitive to the composition. In particular,
the behaviour of the glassy silicates is very different from that of
the carbon and astrosilicate grains. For example, while it takes
$2\times 10^6$ and $3\times 10^6$\,s to entirely sublimate a carbon
and an astrosilicate grain of 1\,$\mu$m, respectively, once the
sublimation temperature is reached, a glassy silicate grain of the
same size will sublimate in only $10^2$\,s at its own sublimation
temperature.

\subsection{Synthetic Spectral Energy Distributions}
\label{sec:syntheticobservations}

By combining the different, single-size (single-\be ) grain runs, we
can estimate a density profile, as parameterized by the vertical
optical depth $\tau$, assuming that the grains are produced at steady
state from the parent belt.  The usual method consists in recording
the grains positions at regularly spaced time intervals, and pile-up
these different positions following a procedure similar to
\cite{Thebault2012} until the grain is removed from the system
(ejection, sublimation or fall onto the star). Here, we employ a
different approach to compute $\tau$, described in details in
Appendix\,\ref{app:DensMap}. It combines density profiles derived from
the limited number of test grains for which the dynamics has been
calculated accurately, and timescale estimates for a broader range of
grains sizes, to produce 2-D ($r,s$) density and optical depth maps.
These maps are obtained assuming an initial differential size
distribution proportional to $s^{-3.5}$.

We also developped a Python implemented version of the GRaTeR
radiative transfer code \citep{Augereau1999} that allows to calculate
thermal emission and scattered light maps at any wavelength from the
2-D ($r,s$) maps, as well as spectral energy distributions (SED) of
the exozodis in order to directly compare our numerical results with
the observations.


\section{PR-drag pile-up scenario}
\label{sec:PRdrag}

We consider a set-up similar to the one explored by
\cite{Kobayashi2011} and \cite{VanLieshout2014}, with a population of
small grains assumed to be released by collisions in a Kuiper-belt
like ring (parent bodies located at $r_0=50$\,au), 
whose evolution is then followed,
taking into account PR-drag and sublimation near the star, until the
grains leave the system either by total sublimation, fall onto the
star or dynamical ejection.  We explore 4 stellar types and 3
different grain compositions (see Tables\,1 and 2).  We consider 24
\emph{initial} grain sizes, ranging from 1.7\,nm to 1\,mm, and thus 24
different initial $\beta$ values (vertical tick marks in
Figs.\,\ref{fig:betaCar}, \ref{fig:betaSi} and \ref{fig:betaGla}).  
We consider that the grains are released from parent bodies on circular
orbits at $r_0 =50$\,au, so that the grains' initial orbit is given by $a=r_0 \times
(1-\beta)/(1-2\beta)$ and $e=\beta/(1-\beta)$.

\subsection{Grain evolution}
\label{subsec:GrainEvo}

\begin{figure*}[tp!]
\centering
\hbox to \textwidth
{
\parbox{0.49\textwidth}{
\includegraphics[width=0.5\textwidth, height=!, trim=0 1cm 0 0]{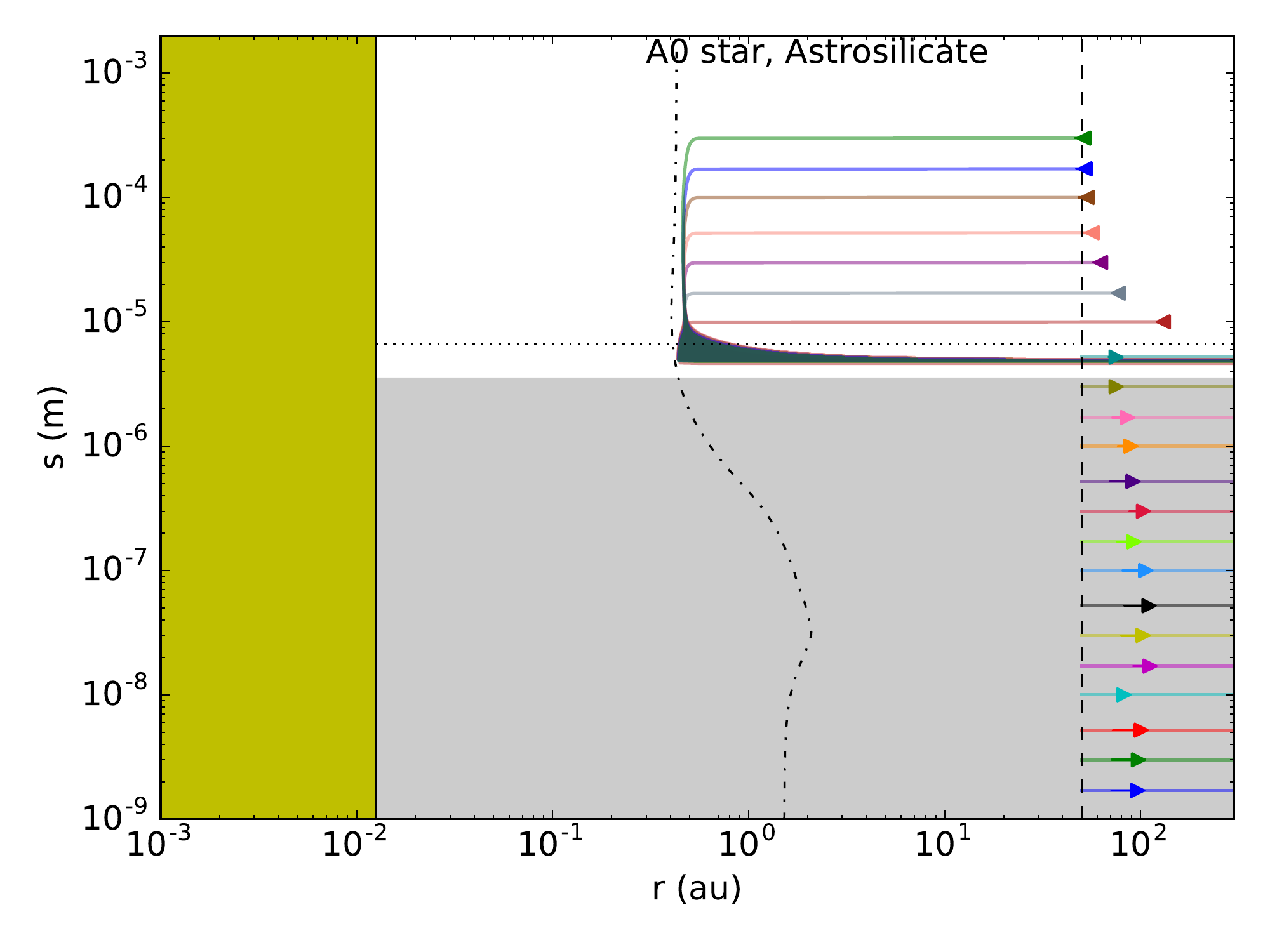}
\subcaption{ }
}
\hfill
\parbox{0.49\textwidth}{
\includegraphics[width=0.5\textwidth, height=!, trim=0 1cm 0 0]{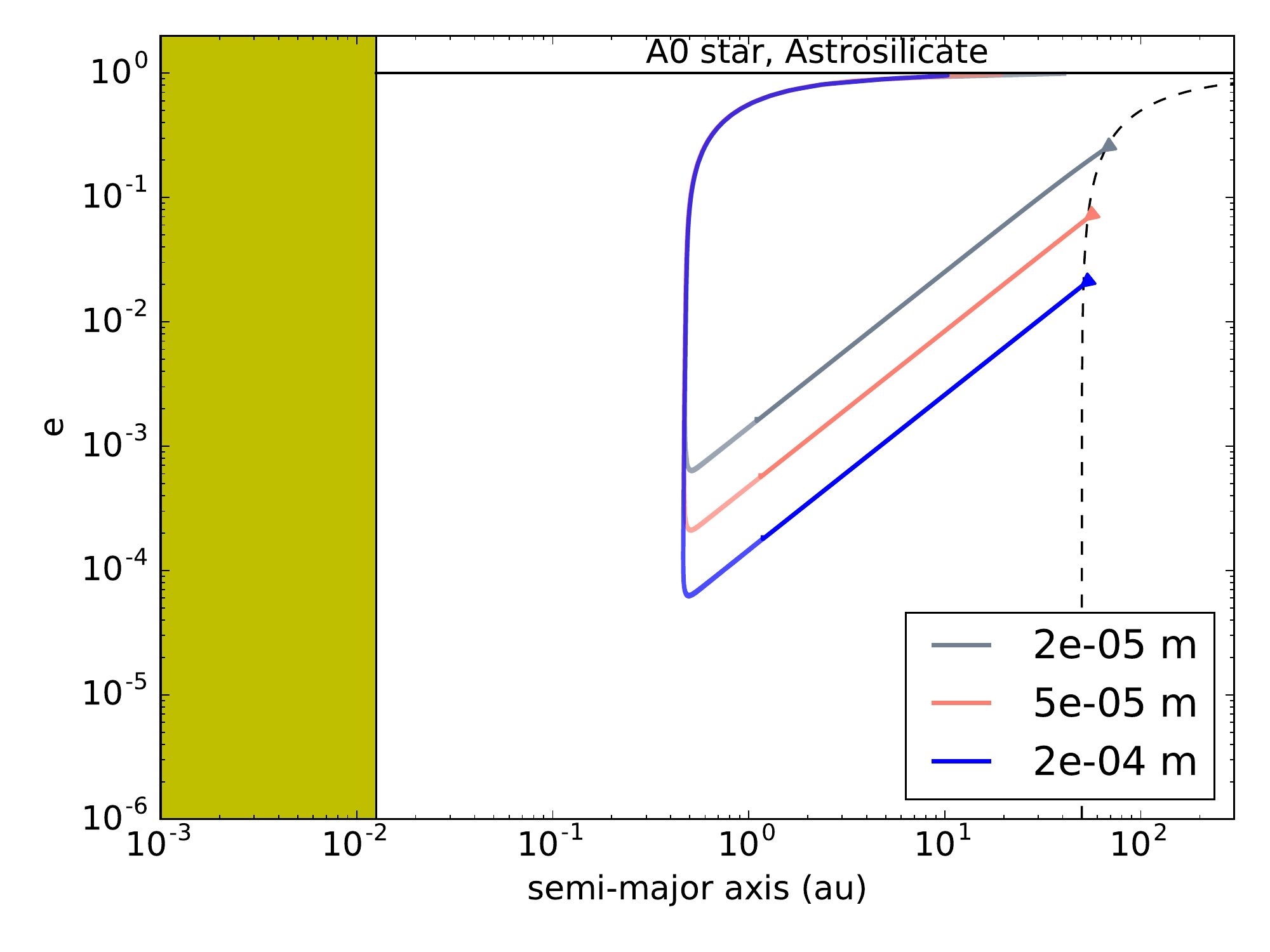}
\subcaption{ }
}
}
\hbox to \textwidth
{
\parbox{0.49\textwidth}{
\includegraphics[width=0.5\textwidth, height=!, trim=0 1cm 0 0]{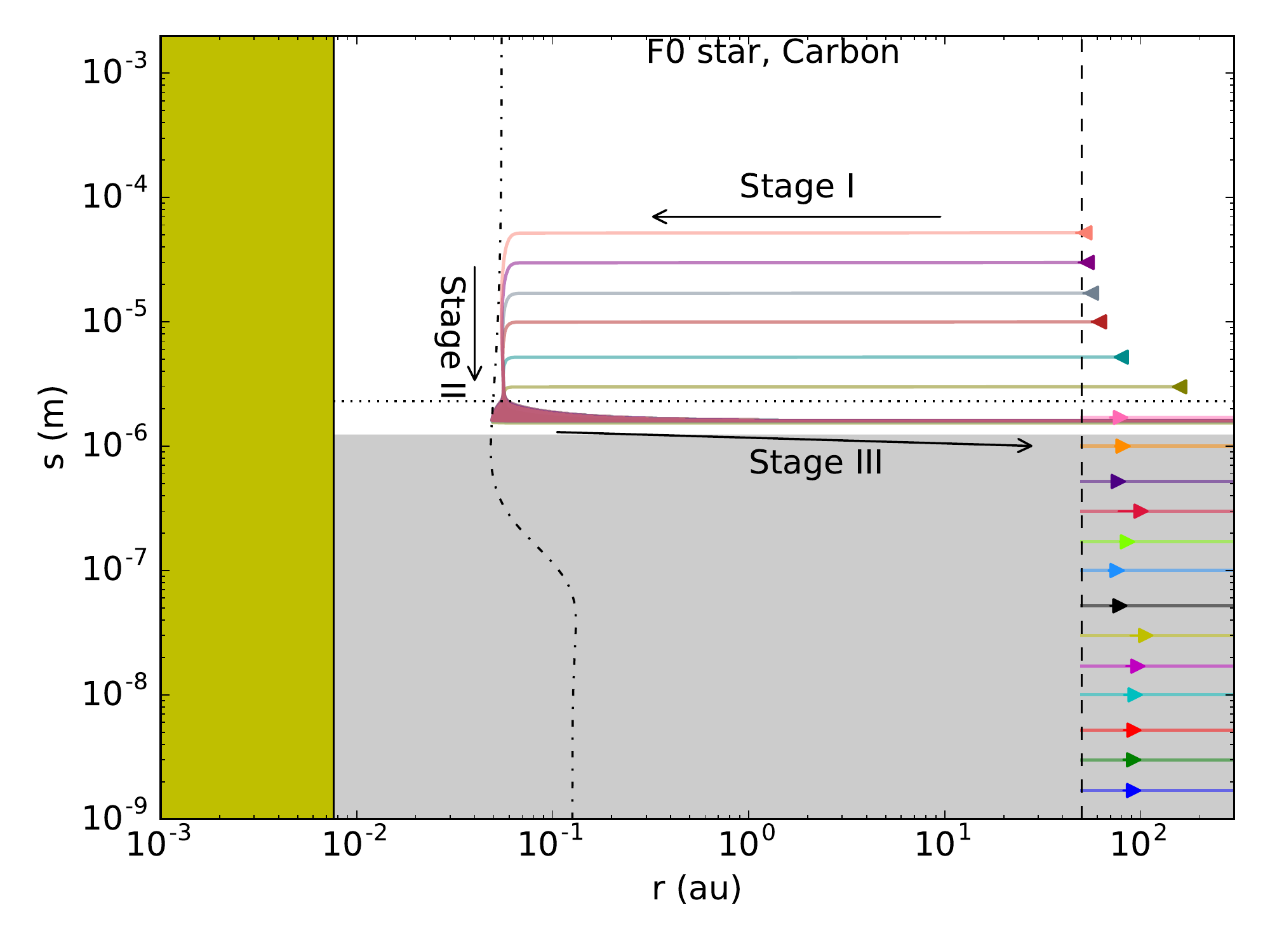}
\subcaption{ }
}
\hfill
\parbox{0.49\textwidth}{
\includegraphics[width=0.5\textwidth, height=!, trim=0 1cm 0 0]{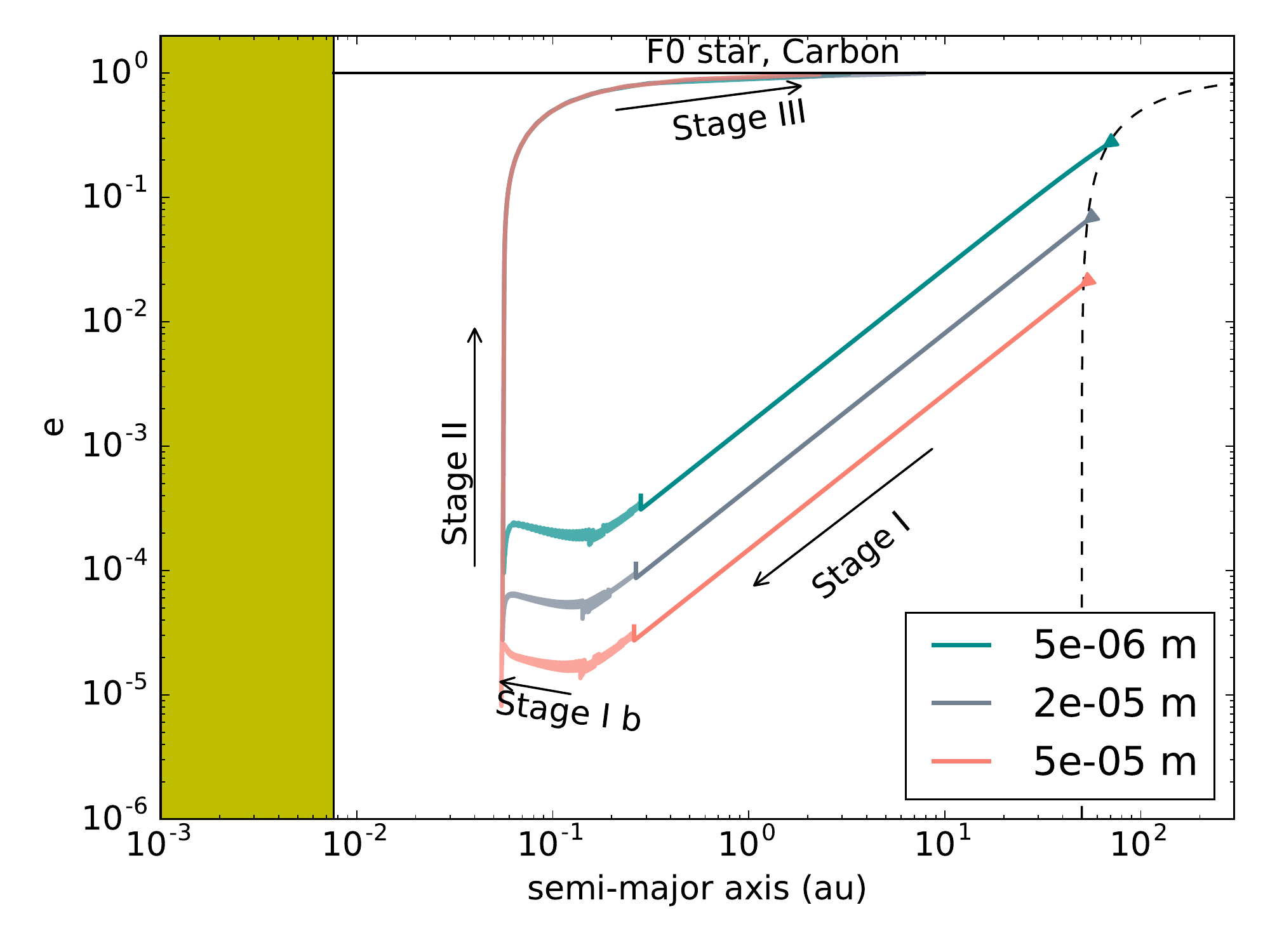}
\subcaption{ }
}
}
\hbox to \textwidth
{
\parbox{0.49\textwidth}{
\includegraphics[width=0.5\textwidth, height=!, trim=0 1cm 0 0]{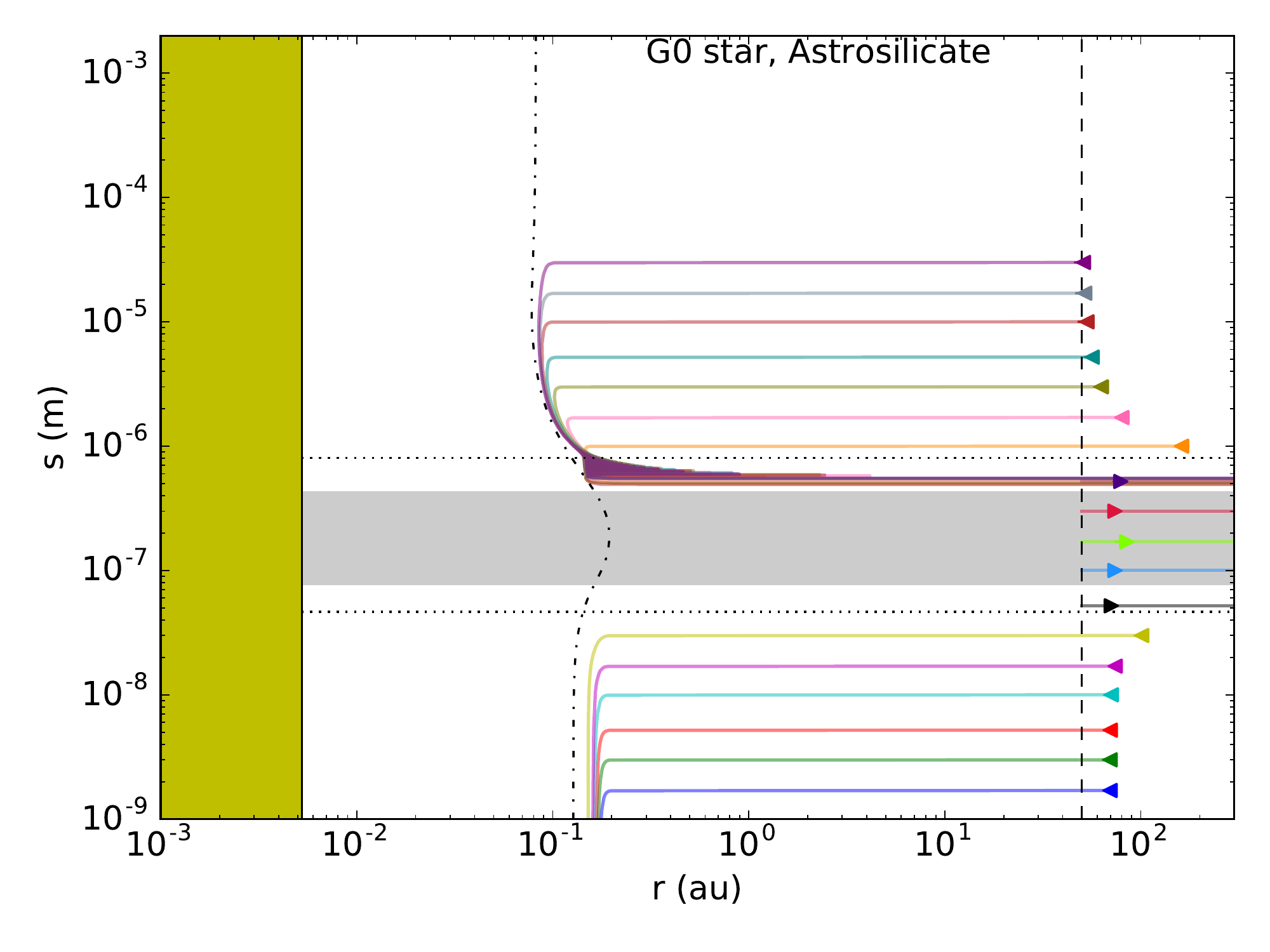}
\subcaption{ }
}
\hfill
\parbox{0.49\textwidth}{
\includegraphics[width=0.5\textwidth, height=!, trim=0 1cm 0 0]{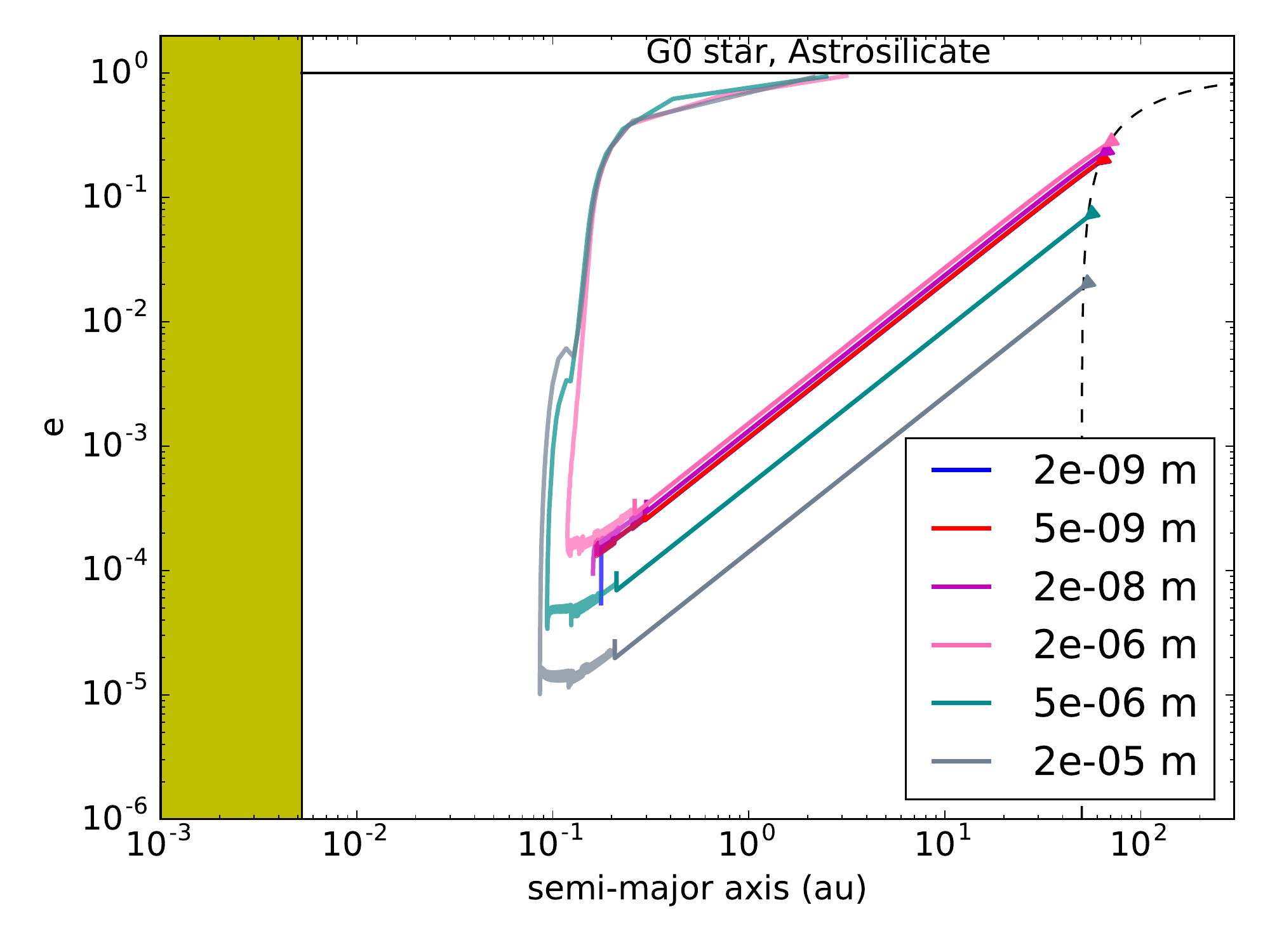}
\subcaption{ }
}
}
\caption{PR-drag pile-up scenario: Grain evolution, as a function of initial grain size, 
  for three different stellar types and 2 grain compositions.
  \textsl{Left panels}: Grain size as a function of stellar distance.
  The arrows denote the temporal evolution.  The dash-dotted line is
  the sublimation distance as a function of grain size, while the
  vertical dashed line indicates the position of the parent bodies.
  The gray horizontal zone identifies the range of grains with
  $\beta > 1$, and the horizontal dotted lines correspond to $\beta =
  0.5$ (see also Fig.\,\ref{fig:BetaAll}).  \textsl{Right panels}:
  Eccentricity as a function of semi-major axis. The horizontal plain
  line represents the $e$=1 limit beyond which particles are on
  unbound orbits.  For both the left and right panels, the left yellow
  area corresponds to the physical location of the star
  and the arrows denote the evolution way. }
\label{fig:PB}
\end{figure*}

\begin{figure*}[tp!]
\label{PBG0Gla}
\centering
\hbox to \textwidth
{
\parbox{0.49\textwidth}{
\includegraphics[width=0.5\textwidth, height=!, trim=0 1cm 0 0]{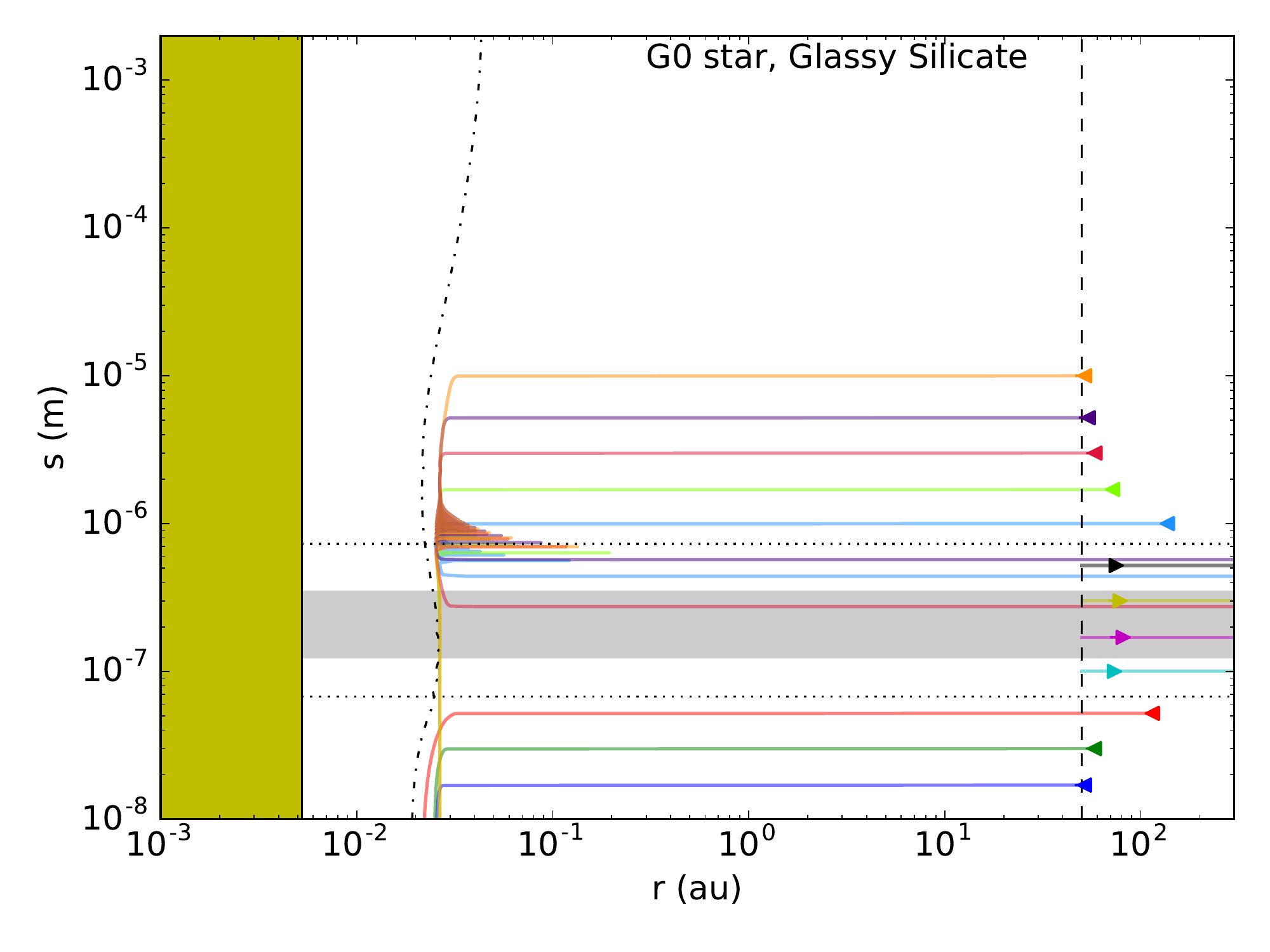}
\subcaption{ }
}
\hfill
\parbox{0.49\textwidth}{
\includegraphics[width=0.5\textwidth, height=!, trim=0 1cm 0 0]{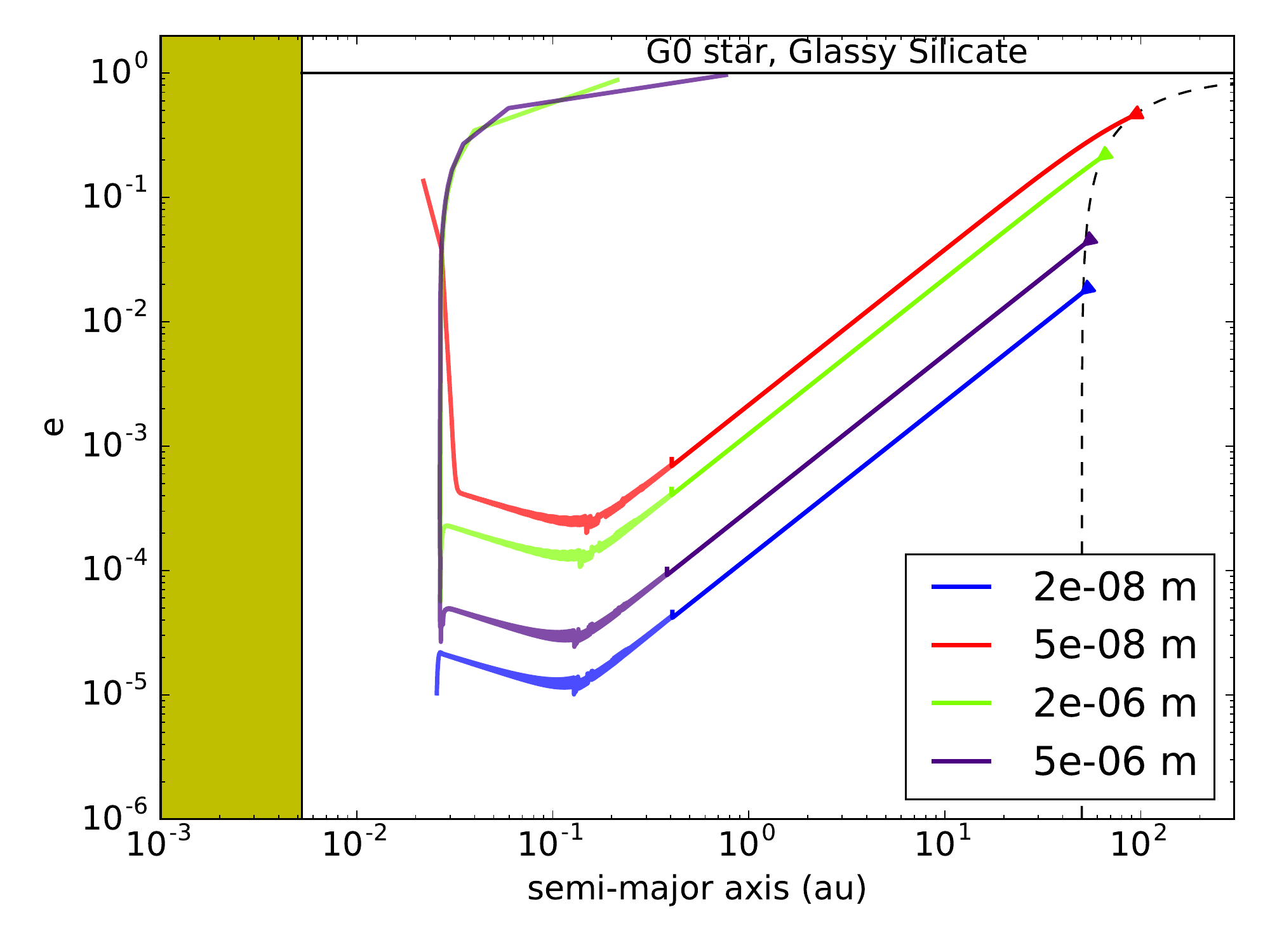}
\subcaption{ }
}
}
\hbox to \textwidth
{
\parbox{0.49\textwidth}{
\includegraphics[width=0.5\textwidth, height=!, trim=0 1cm 0 0]{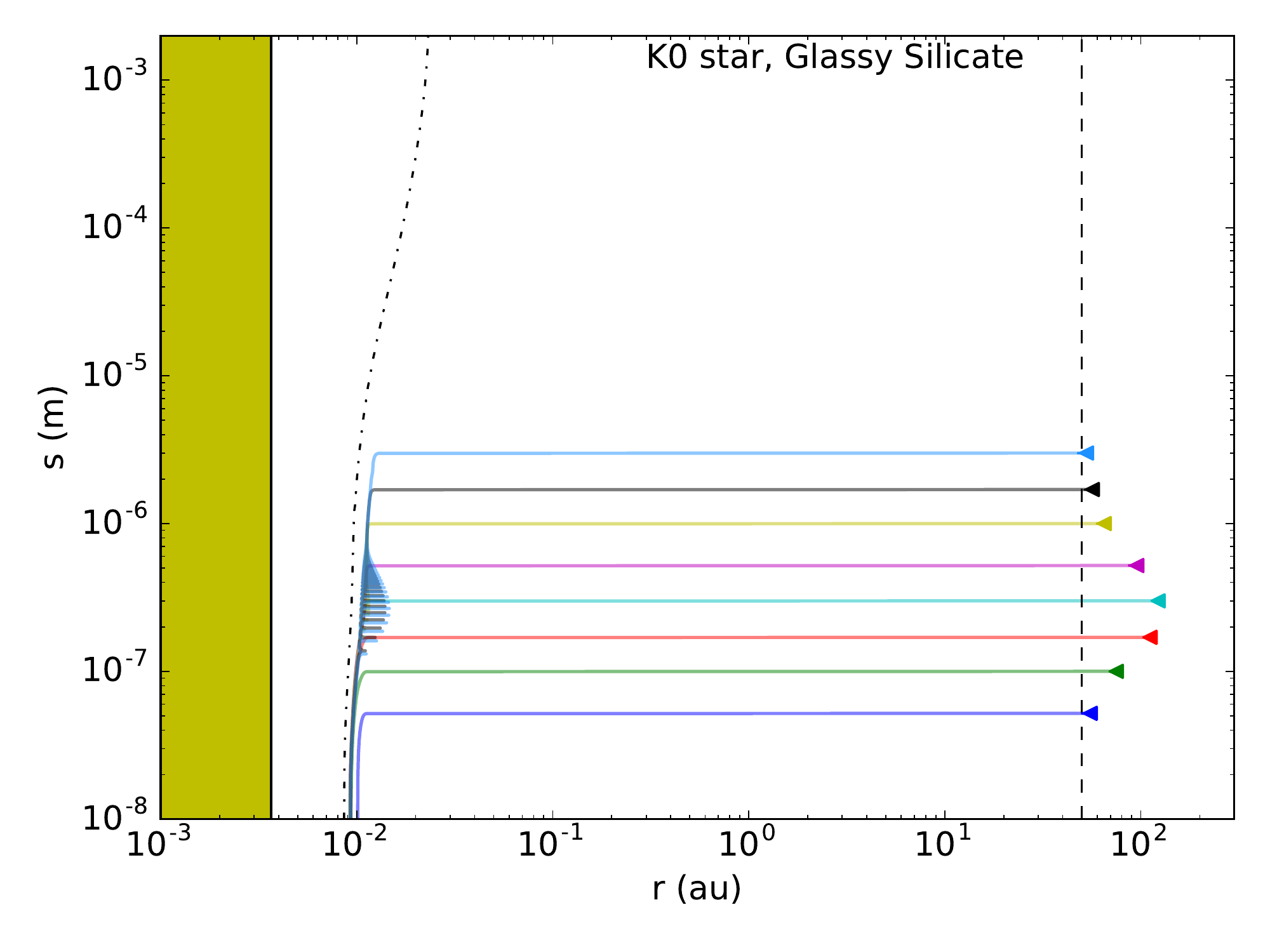}
\subcaption{ }
}
\hfill
\parbox{0.49\textwidth}{
\includegraphics[width=0.5\textwidth, height=!, trim=0 1cm 0 0]{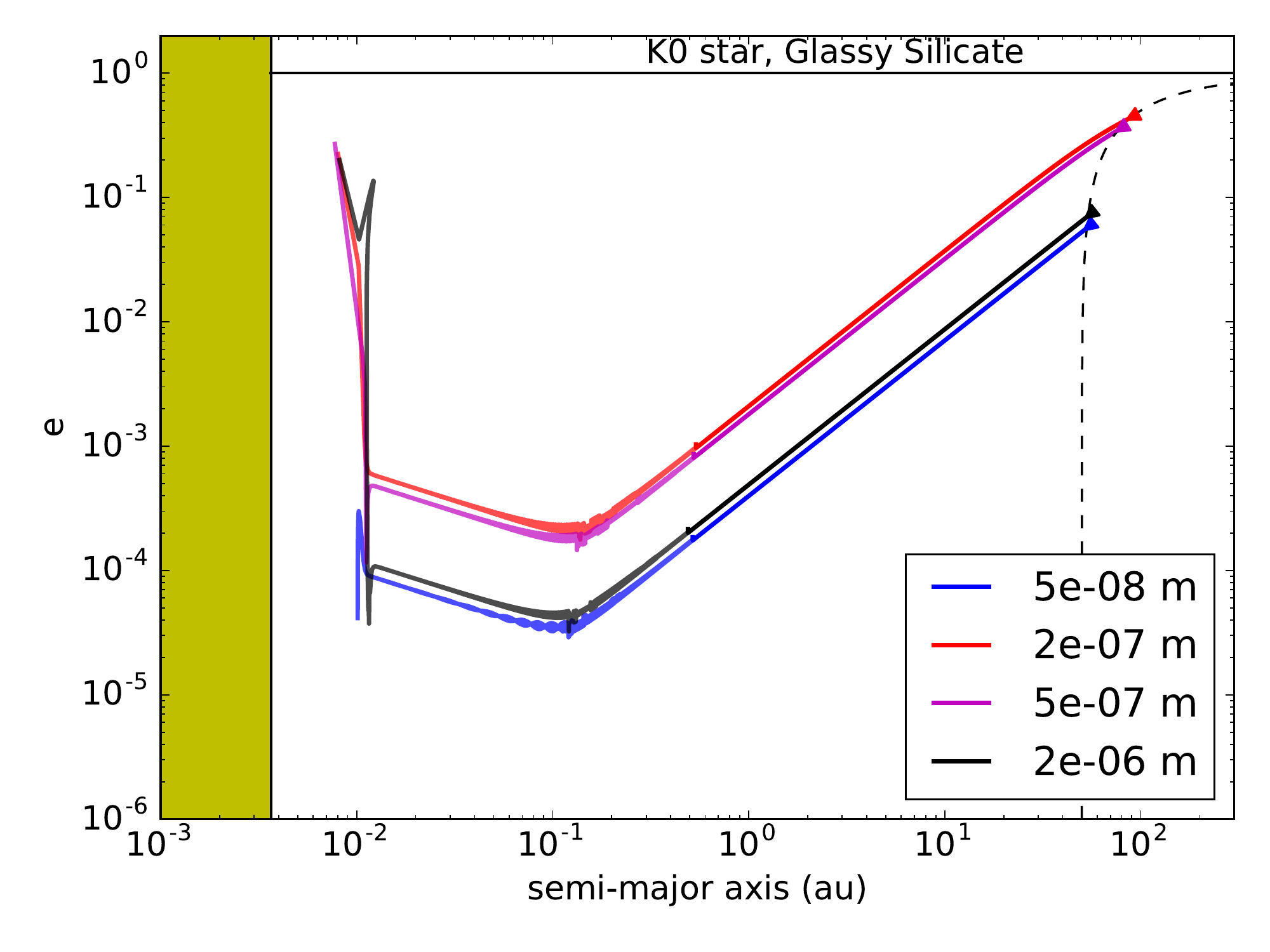}
\subcaption{ }
}
}
\caption{Same as Fig.\,\ref{fig:PB} but for the G0 and K0 stars, and
  glassy silicates.}
\label{fig:PBGlaK0}
\end{figure*}

\subsubsection{General behaviour}
\label{sec:PRdraggeneralbehaviour}

Figs.\,\ref{fig:PB} and \ref{fig:PBGlaK0} present the evolution of
grain sizes and orbital elements for a subset of the explored
parameter space (stellar type, grain composition).

As can be seen, the initial stage \citep[labelled "stage I",
  after][and reproduced in Fig.\,\ref{fig:PB}c and
  \ref{fig:PB}d]{Kobayashi2009} is relatively similar for all cases
and corresponds to the behaviour found by previous studies using
orbit-averaged equations: the grain drifts inward due to PR-drag and
its orbit is progressively circularized, while its size remains
constant because it is too far from the sublimation region. We note,
however, that, contrary to the predictions of orbit-averaged
prescriptions, the eccentricity stops to decrease at a given point and
starts to slowly increase again as its semi-major axis continues to
drop (named Stage Ib in Fig.\,\ref{fig:PB}d). This inflection point
corresponds to a "residual" value below which the osculating
eccentricity of the PR-drag drifting particle cannot fall, which is
due to the intrinsic curvature of the tightly wound spirals that the
grain actually follows as it migrates inward.  The osculating
eccentricity corresponding to these spirals can be approximated, to a
first order, by (d$a$/$a$)$\dma{orb}$, which is the relative variation
of the particle's semi-major axis, due to PR-drag, over one orbital
period as given by the averaged equations used by \cite{Kobayashi2009}
or \cite{VanLieshout2014}. Taking the right-hand term of Eq.\,1 of
\cite{Kobayashi2011} (drift rate due to PR-drag), we get
\begin{equation}
\ma{d}a = \frac{2 \beta G M_*}{ac} T
\end{equation}
where $T$ is the orbital period and $c$ the speed of light.
This leads to a residual eccentricity of the order of
\begin{equation}
e\dma{res} \approx \frac{\ma{d}a}{a} =\frac{4\pi\beta}{c}
\sqrt{\frac{GM_*}{a}},
\end{equation}
which indeed increases with decreasing $a$.  This non-zero
$e\dma{res}$ is always relatively small, less than a few $10^{-3}$,
but it cannot be ignored because even such a small value can make a
difference as to the final fate of a grain as it starts sublimating.

As expected, the situation radically changes as the grains approach
the sublimation region.  As already identified in previous studies, as
the grains start to sublimate, radiation-pressure increases and
eventually halts their inward drift.  During this "stage II" \citep[to
  follow again][]{Kobayashi2009}, the grain shrinks while staying at
its sublimation radius $r\dma{s}$, which does not always correspond to a
constant distance to the star because grain temperatures, and thus
their sublimation distance, depend on their size (see for instance
Fig.\,\ref{fig:PB}e).  In parallel with this size decrease, the
grain's eccentricity increases rapidly. At one point, this
eccentricity becomes significant enough for the particle to spend only
a very small fraction of its orbit in the narrow sublimating region
around $r\dma{s}$.  The grain then enters "stage III", where its
sublimation drastically slows down, only occurring at periastron
passages.  
Its orbital eccentricity continues to increase, 
albeit more slowly than before, 
receiving additional "kicks" at each sublimating-periastron passage.

The final fate of the grain was not investigated in
\cite{Kobayashi2009} or \cite{VanLieshout2014} because it occurs in a
fast-evolving regime where orbit-averaged equations are no longer
valid. The final fate depends on the dust composition and the stellar
type. This is discussed in details below.

\subsubsection{Final fate: ejection}
\label{subsubsec:ejection}

\begin{figure}
\includegraphics[width=0.5\textwidth]{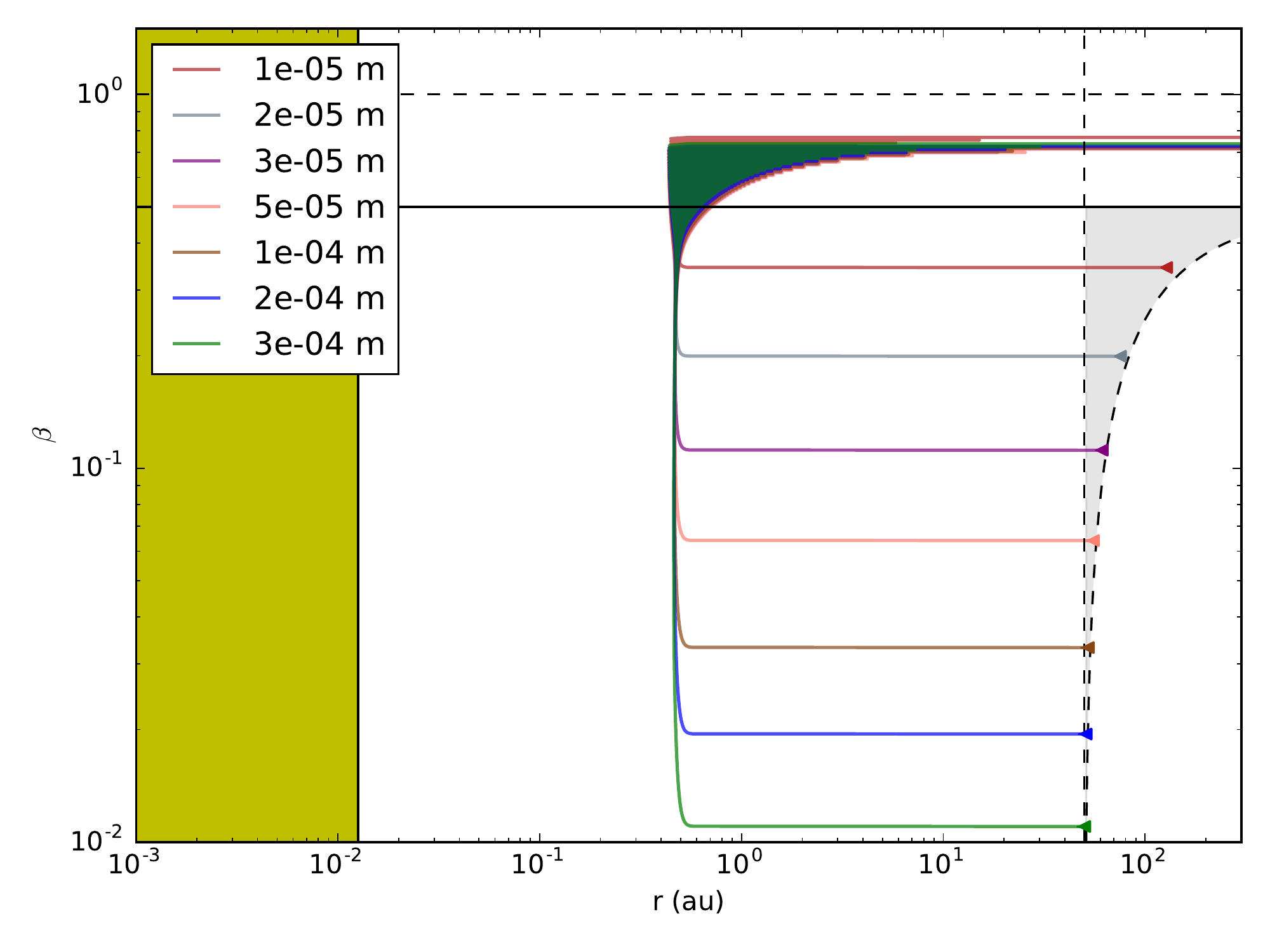}
\caption{PR-drag pile-up scenario: Evolution of \be\ 
  as a function of stellar distance $r$
  for the initially bound astrosilicate
  grains around the A0 star.
  The parent belt is located at 50\,au (vertical dashed line), 
  and the grains are released on increasingly eccentric orbits
  as \be\ raises.
  The arrows represent the inital apoastron of the grains. }
\label{fig:BetaA0}
\end{figure}

In most cases, the dust grain is parked on these ever-more-eccentric
orbits, all having their periastron at $r\dma{s}$, until its eccentricity
reaches 1 and the grain is ejected from the system.  Note that, by the
time it reaches the $e=1$ limit, its $\beta$ is higher than the
classical 0.5 value expected for a grain released from a $\beta=0$
progenitor on a circular orbit (Fig.\,\ref{fig:BetaA0}).  This is
because, when grains start to sublimate, in stage II, PR-drag is still
able to force their eccentricities to low values, lower than the one
they should have according to the canonical $e=\beta/(1-\beta)$
relation.  So that once sublimation gets really intensive and the
grain approaches the $\beta=0.5$ value, its eccentricity is still
relatively small, allowing it to stay on a bound orbit beyond this
critical 0.5 value.  The highest possible $\beta$ value for a grain
reaching $e=1$ is $\beta=1$, obtained for an idealized case where the
$e=1$ grain is produced from a $\beta=0.5$ progenitor on a circular
orbit.  However, in practice, we never obtain $\beta$ values exceeding
0.8-0.85 (see Fig.\,\ref{fig:BetaA0} for the simulation leading to the
highest beta values for $e=1$ particles), which is because the grains'
eccentricities are never exactly zero as they enter stage II.  And
even values as small as a few $10^{-4}$ are enough to prevent the
orbit from reaching $\beta=1$ by the time it reaches $e=1$ (this can
be understood by looking at Equ.\,58 of \cite{Kobayashi2009} and
Equ.\,48 of \cite{VanLieshout2014}, which give the evolution of $e$
during stage II as a function of the initial $e$ it has when it enters
this stage).

The fact that $\beta <0.85$ by the time the grains are ejected has
important consequences.  It means that the DDE always remains
negligible, because its magnitude only becomes significant for $\beta$
values very close to 1.  Taking Eq.\,\ref{eqn:Fpr} and
\ref{eqn:Fdde}, the ratio of the DDE force to the radiation
pressure+PR-drag force along the velocity indeed reads :
\begin{equation}
\frac{F\dma{DDE}}{F\dma{PR}} = \frac{\omega_{\star} R_{\star}^2 }{4
  \sqrt{r G M_{\star} (1-\beta)} }. 
\end{equation} 
For the highest $\beta$ value obtained in our runs, i.e., 0.85 for an
F0 star and carbon grains, we get a maximum $F\dma{DDE}/F\dma{PR}$
value of 0.03.  We can thus safely conclude that DDE only has a very
marginal influence on the grains' evolution, whose effect can be
neglected on the density pile-up of grains close to $r\dma{s}$.

\subsubsection{Final fate: total sublimation}
\label{subsubsec:sublim}

For a small subset of our simulations, the grains' fate is radically
different, as they are removed from the system by total sublimation.
These correspond to the specific cases of a G0 star and glassy
silicates and of a K0 star for both astrosilicate and glassy silicates
(Fig.\,\ref{fig:PBGlaK0}).  The behaviour for the K0 cases is easy to
understand: the maximum possible $\beta$ value is indeed always below
1, regardless of particle sizes (Fig.\,\ref{fig:BetaAll}).  This means
that, as they sublimate during stage II, grains will never reach
$\beta$ values high enough for them to reach the $e=1$ limit.  They
will thus stay on bound orbits all the time, while still sublimating
at their orbital periastron, so that they eventually get fully
sublimated. As a consequence, the grains that reach the maximum
possible \be\ value will survive longer than grains close in size but
with smaller \be\ values.

For a G0 star and glassy silicates, \be\ can exceed one, but only in a
relatively narrow size range (see Fig.1c). As a consequence, as it
sublimates, the radius of a grain can directly cross the whole $\beta
> 1$ size domain, and even sometimes the $\beta > 0.5$ one, before
having had the time to be pushed on an unbound orbit.  The fate of
a grain is very difficult to predict in advance, as it strongly depends
on its orbital location by the time it begins to significantly
sublimate.  All we can safely establish is that there is a significant
fraction of grains that will disappear because of full sublimation
(Fig.\,\ref{fig:PBGlaK0}a).

\begin{figure}[tbp]
\includegraphics[width=0.5\textwidth]{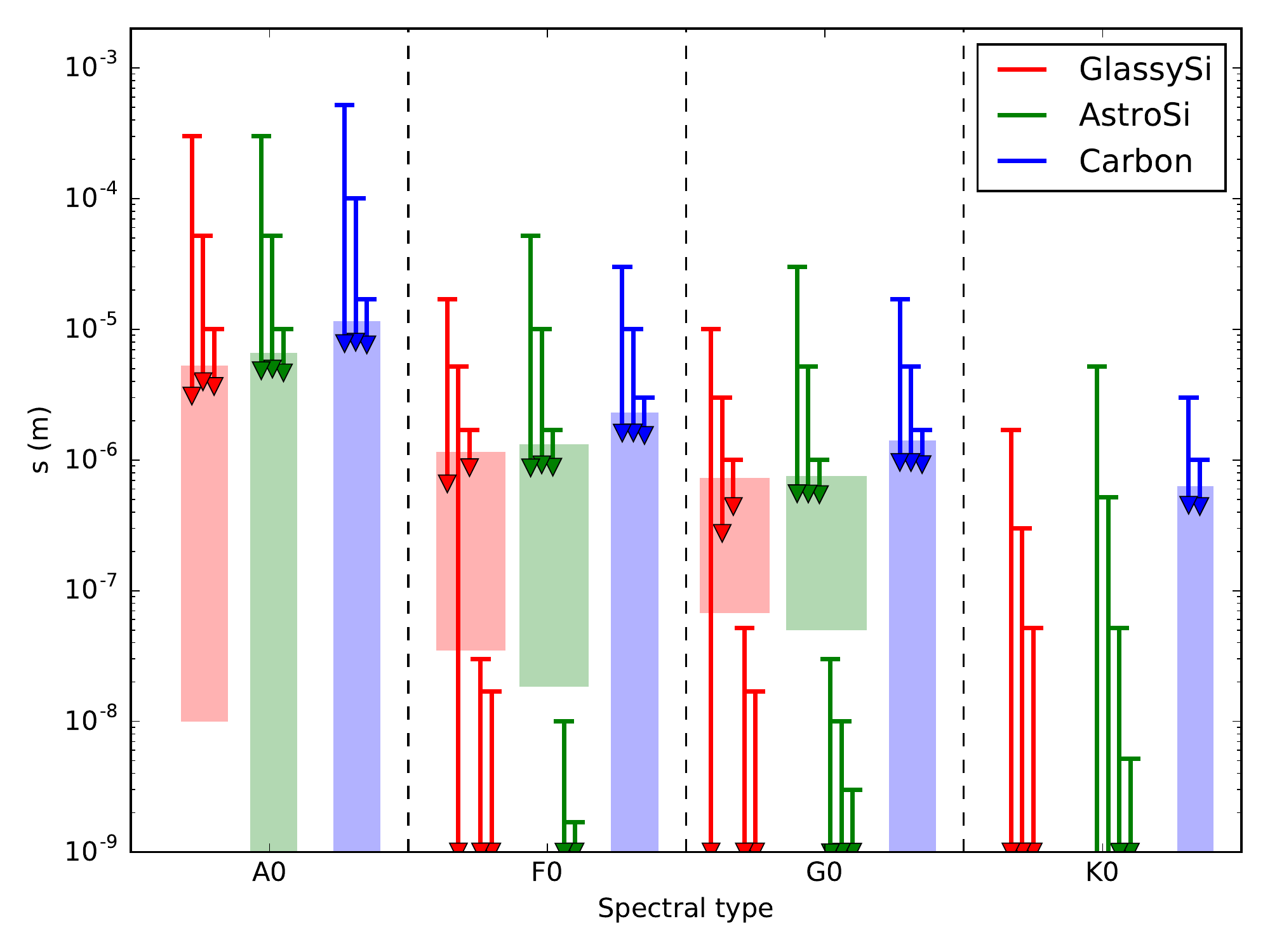}
\caption{Arrows representing the initial to final grain size evolution
  for a sample of simulations.
  The arrow head corresponds to the ejection size, 
  or to total sublimation if it reaches the bottom of the figure.
  The colored rectangles are the range of grain sizes for which $\beta > 0.5$.}
\label{fig:Gsize}
\end{figure}

\subsection{Global disk properties and spectra}
\label{subsec:Spectra}

\subsubsection{Grain size}

\begin{figure*}[tbp]
\begin{center}
\hbox to \textwidth
{
\parbox{0.33\textwidth}{
\includegraphics[width=0.33\textwidth, height=!, trim=0 1cm 0 0]{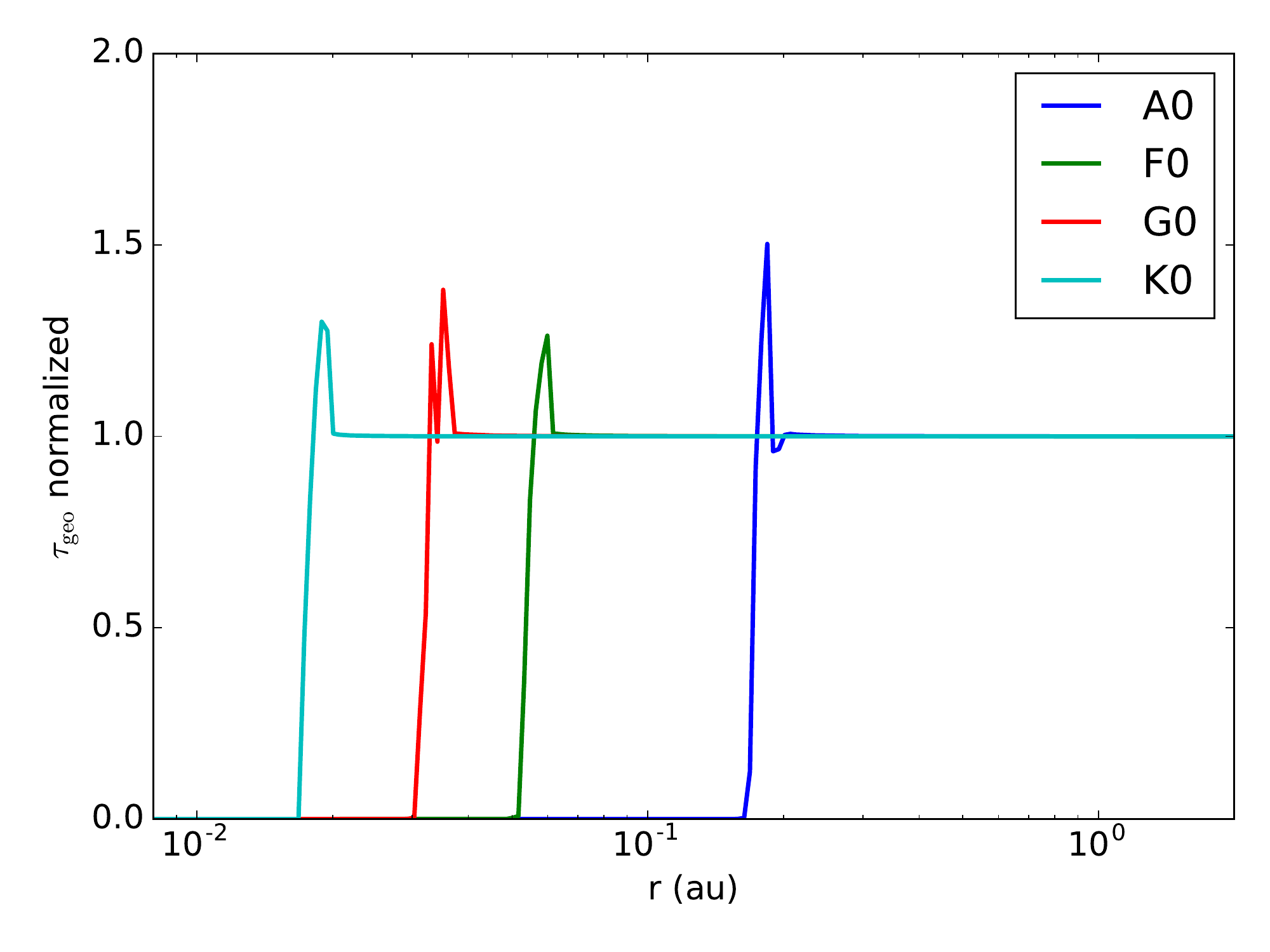}
\subcaption{Carbon}
}
\hfill
\parbox{0.33\textwidth}{
\includegraphics[width=0.33\textwidth, height=!, trim=0 1cm 0 0]{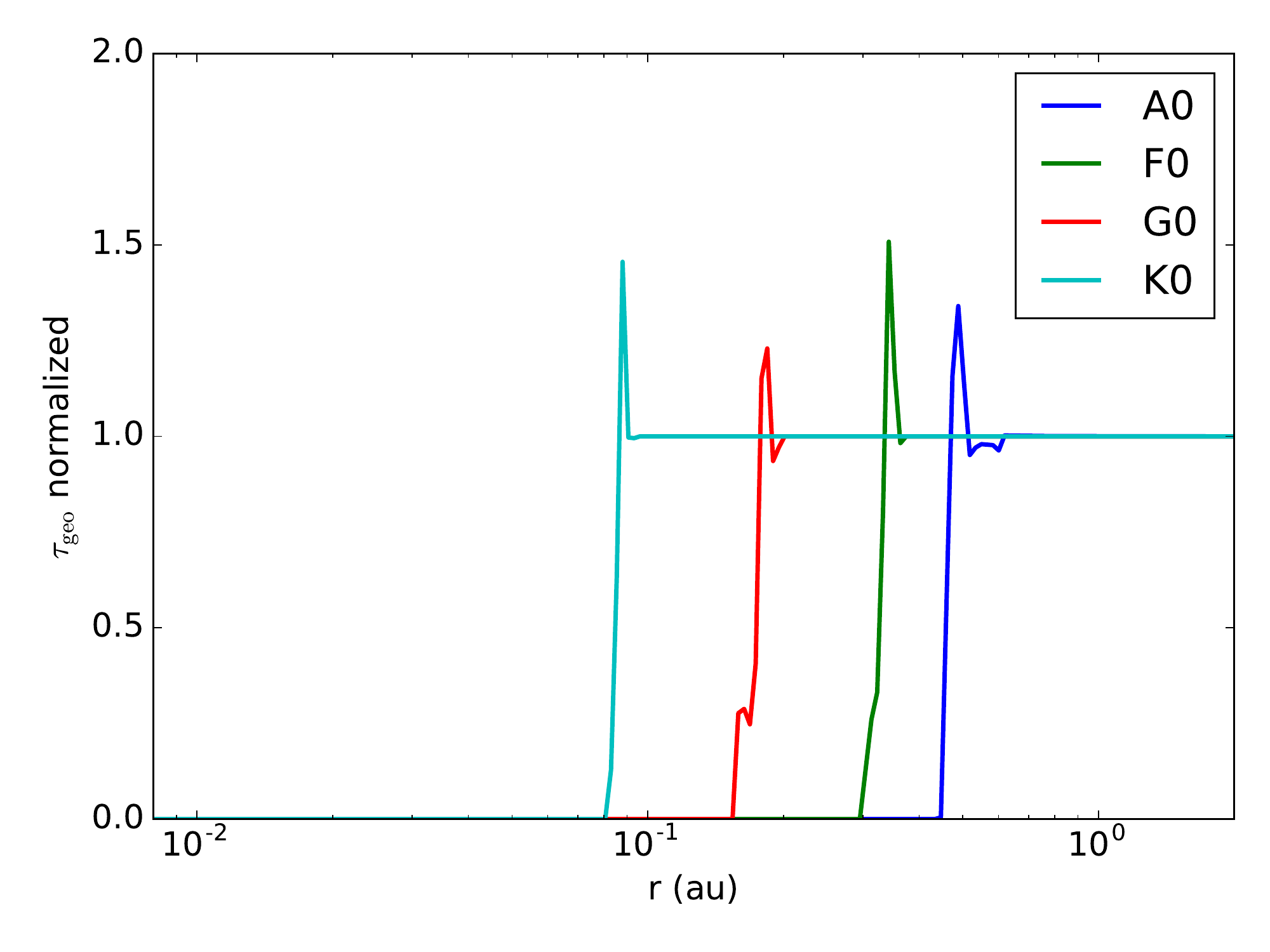}
\subcaption{Astrosilicate}
}
\hfill
\parbox{0.33\textwidth}{
\includegraphics[width=0.33\textwidth, height=!, trim=0 1cm 0 0]{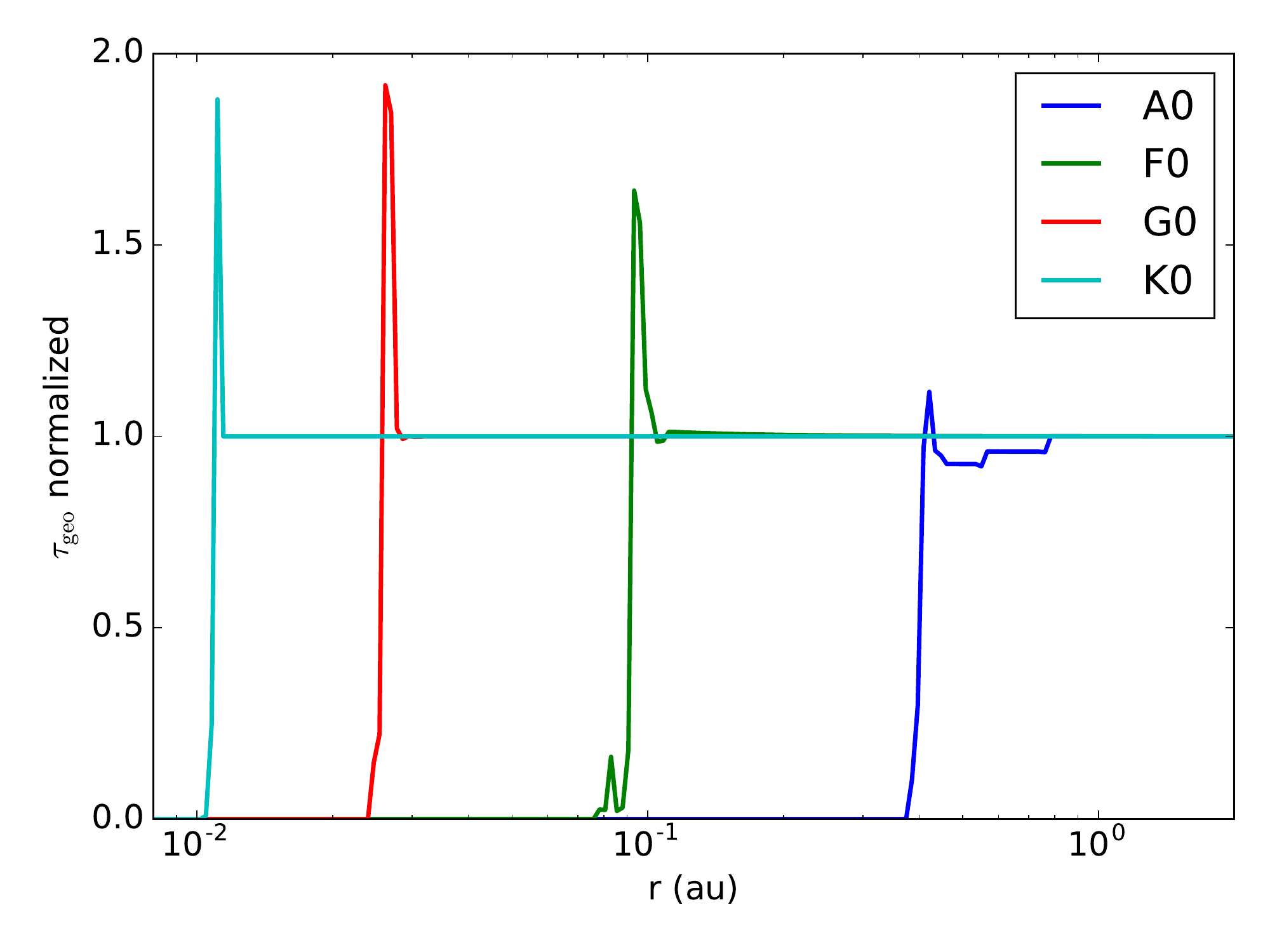}
\subcaption{Glassy silicate}
}
}
\end{center}
\caption{PR-drag pile-up scenario: Radial profiles for the (size-integrated) geometrical optical depth
 for all grain compositions and stellar types.}
\label{fig:IntOptDepth}
\end{figure*}

\begin{table*}
\caption{PR-drag pile-up scenario: Minimal distance and grain size reached for each configuration.  
  A grain size of 0 means that the grain is fully
  sublimated.  If two values are indicated, they correspond to the
  final sizes for the smallest and biggest grains.}
\label{tab:SpTcompo}
\centering
\begin{tabular}{c c c c c c c c c c c c}
  \hline\hline
 & \multicolumn{2}{c}{A0 star} & & \multicolumn{2}{c}{F0 star} & &
  \multicolumn{2}{c}{G0 star} & & \multicolumn{2}{c}{K0 star} \\
  \cline{2-3} \cline{5-6} \cline{8-9} \cline{11-12}
 & $r\dma{min}$ (au) & $s\dma{min}$ ($\mu$m) & & $r\dma{min}$ (au) &
  $s\dma{min}$ ($\mu$m) & & $r\dma{min}$ (au) & $s\dma{min}$ ($\mu$m)
  & & $r\dma{min}$ (au) & $s\dma{min}$ ($\mu$m) \\
Carbon & 0.16 & 7.6 & & 0.05 & 1.5 & & 0.03 & 0.9 & & 0.02 & 0.4 \\ 
Astrosilicate & 0.44 & 4.6 & & 0.14 & 0 / 0.9 & & 0.08 & 0 / 0.5 & & 0.04 & 0 \\
Glassy silicate & 0.38 & 3.1 & & 0.07 & 0 & & 0.02 & 0 & & 0.01 & 0 \\
\hline
\end{tabular}
\end{table*}

Figure\,\ref{fig:Gsize} and Table\,\ref{tab:SpTcompo} provide an
overview of the smallest grain sizes that can be reached for our
sample of stars and grain compositions.  We see that, for early-type
stars, we are unable to produce submicron-sized grains regardless of
the considered grain composition.  This absence of submicron-sized
grains extends also to the case of late-type stars when considering
carbonaceous dust.  These results are in apparent contradiction with
constraints on dominant grain sizes derived from precise spectral
modeling of observed exozodis, which always tend to favour
submicron-sized dust (see Sec.\,\ref{sec:Intro}).

The only cases for which submicron-sized grains are produced are those
where the grain can experience full sublimation, i.e., those for which
the maximum possible $\beta$ value is lower than 1 or barely exceeds
it. As has been discussed in Sec.\,\ref{subsubsec:sublim}, this is
only true for K0 stars (astro- and glassy silicates) and G0 stars
(glassy silicates only). However, even in these cases, the lifetime of
such tiny grains is very short (sublimation being very fast and
efficient), which might not be enough to leave an observable
signature.

\subsubsection{Surface density profiles}
\label{sec:PRdrag_tau}

\begin{figure*}
\begin{center}
\hbox to \textwidth
{
\parbox{0.49\textwidth}{
\includegraphics[width=0.5\textwidth, height=!, trim=0 0.7cm 0 0,clip]{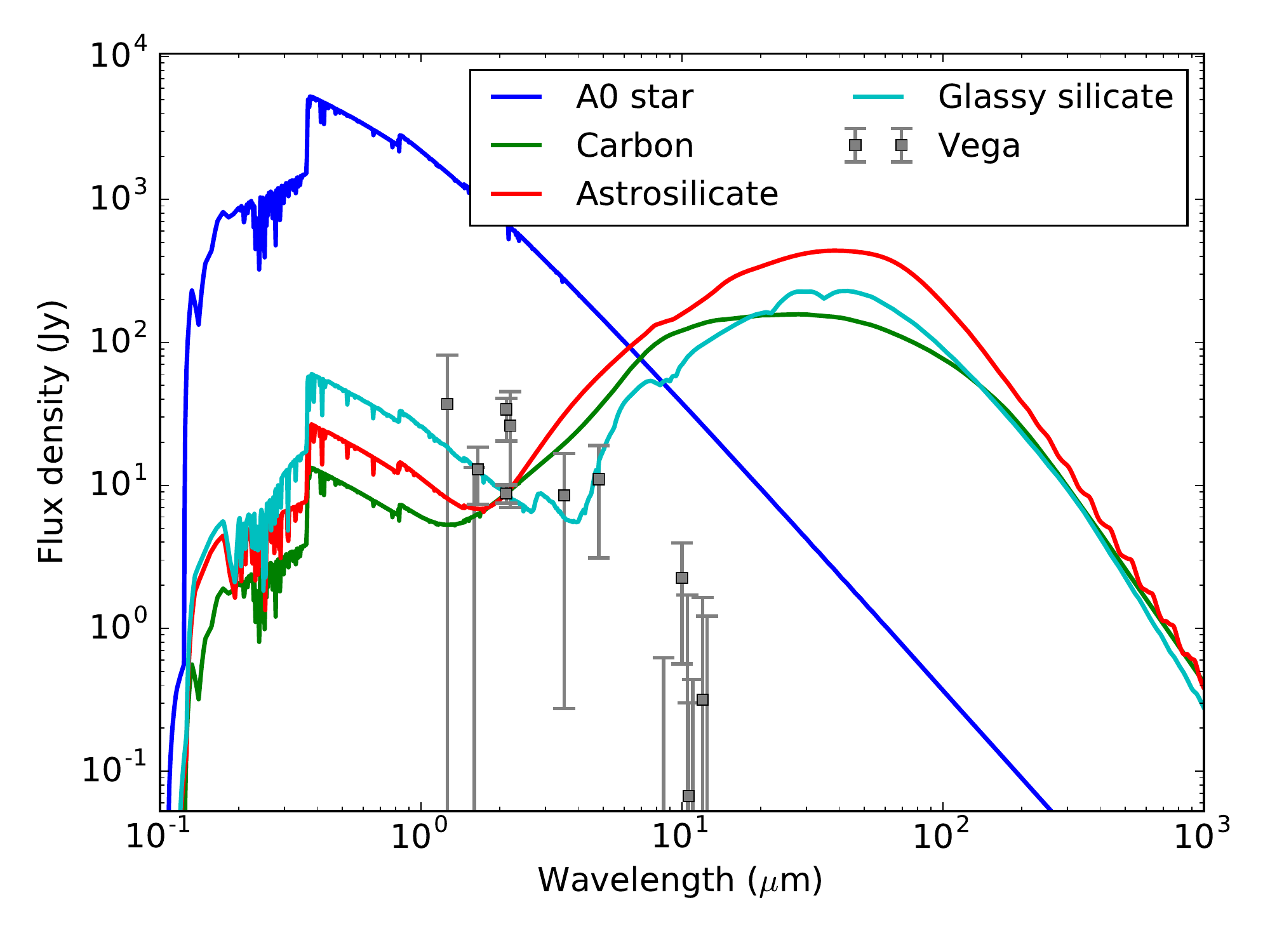}
\subcaption{}
}
\hfill
\parbox{0.49\textwidth}{
\includegraphics[width=0.5\textwidth, height=!, trim=0 0.7cm 0 0,clip]{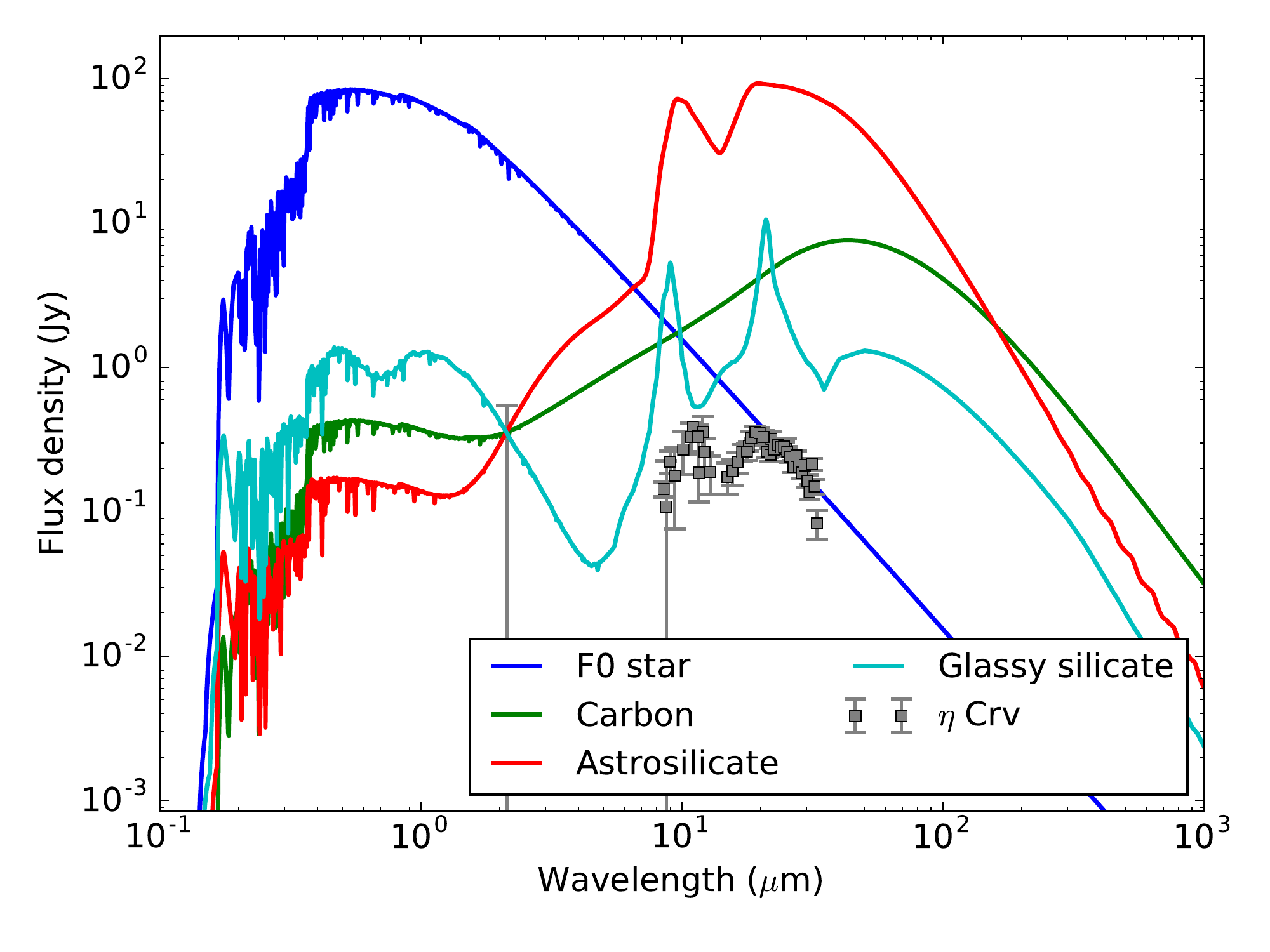}
\subcaption{}
}
}
\hbox to \textwidth
{
\parbox{0.49\textwidth}{
\includegraphics[width=0.5\textwidth, height=!, trim=0 0.7cm 0 0,clip]{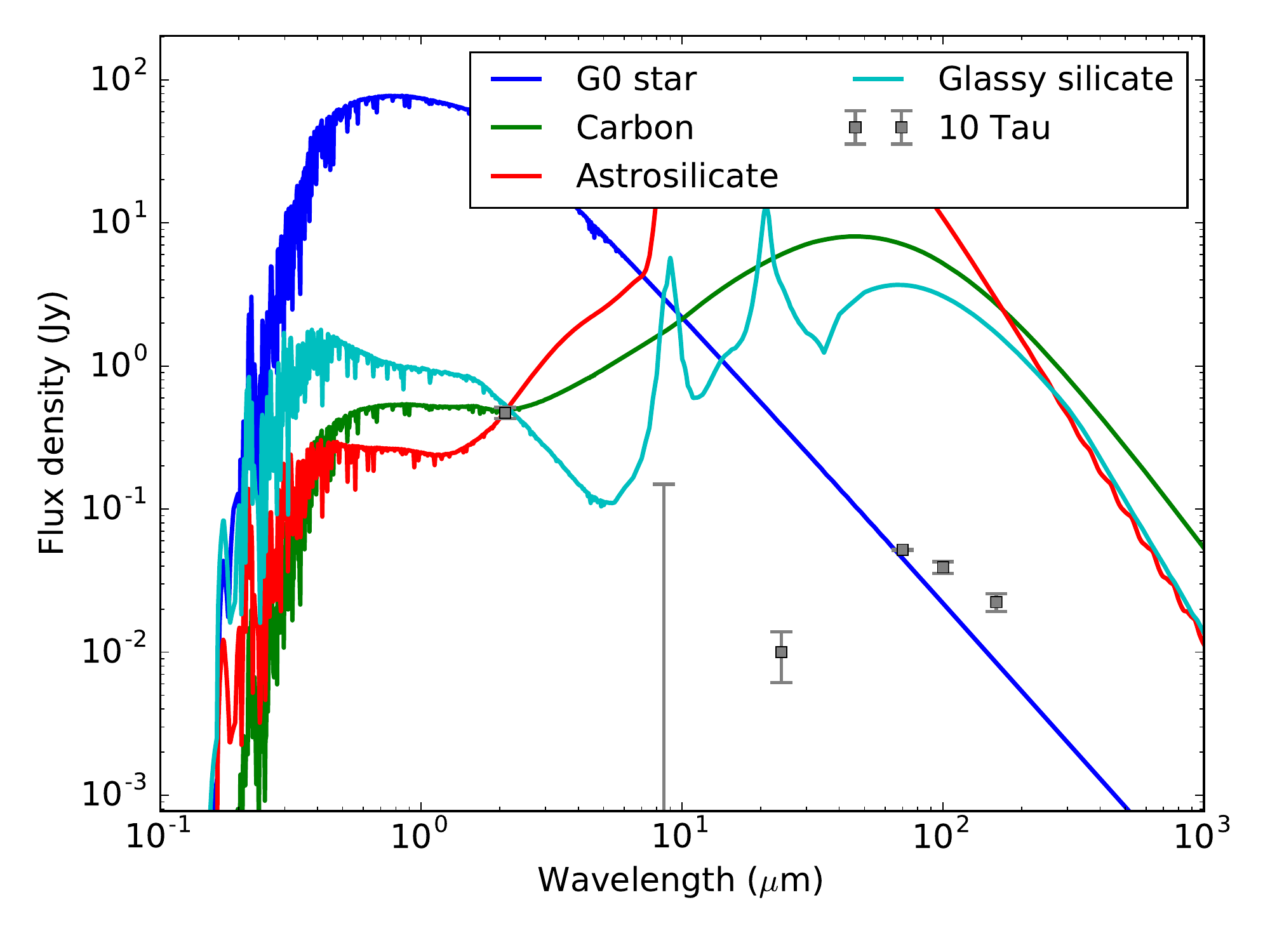}
\subcaption{}
}
\hfill
\parbox{0.49\textwidth}{
\includegraphics[width=0.5\textwidth, height=!, trim=0 0.7cm 0 0,clip]{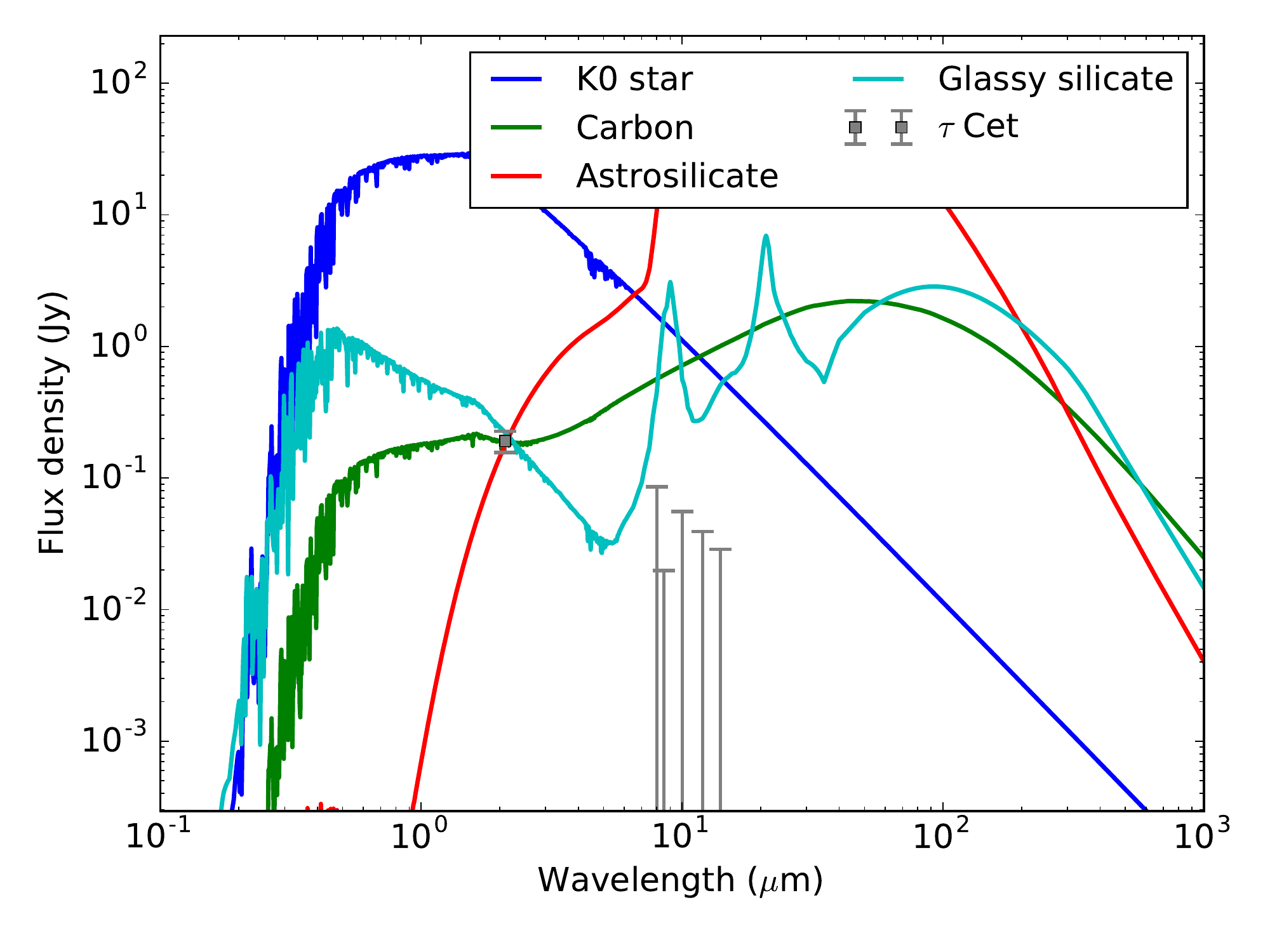}
\subcaption{}
}
}
\includegraphics[width=0.49\textwidth,trim=0.cm 0.7cm 0.cm 0.cm,clip]{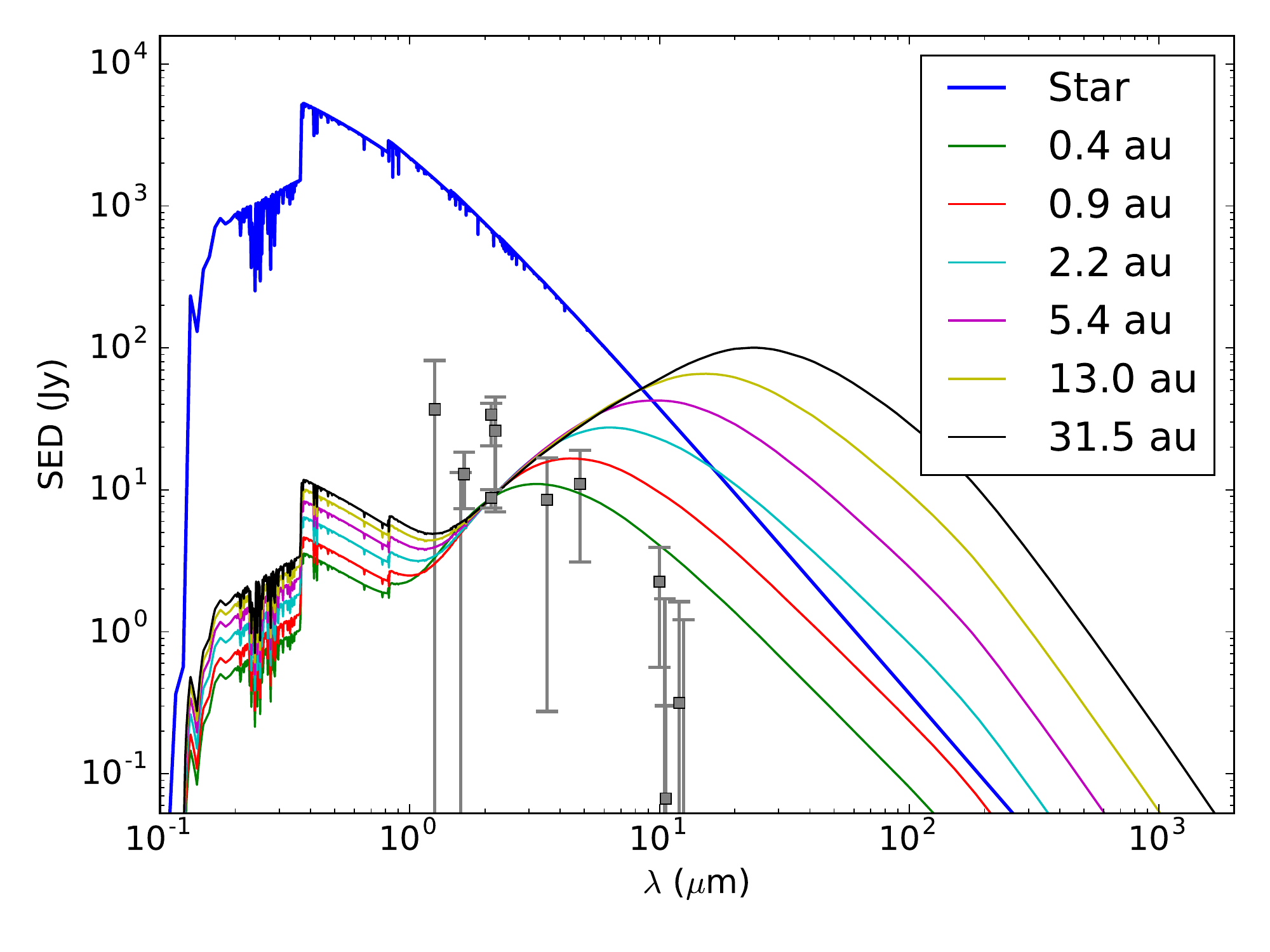}
\subcaption{}
\end{center}
\caption{Panels (a) to (d) : SED for each spectral type and all
  compositions in the PR-drag pile-up scenario.  
  All spectra are normalized to the flux ratio at 2\,$\mu$m, 
  and compared to observed fluxes for exozodis around stars of similar type
  (for the F0 star, for which there is no available observed photometry at 2\mum,
  the value is put to 1\% of the stellar flux). 
  Data points for Vega are
  from \cite{Absil2006}, $\eta$\,Crv are from \cite{Lebreton2016},
  10\,Tau are from \cite{Kirchschlager2017}, $\tau$\,Cet are from
  \cite{difolco2007}. Additional values are from \cite{Absil2013} and
  \cite{Mennesson2014}. Panel (e): 
  SEDs for an A0 star and carbon grains, 
  but only considering the flux within a radial distance
  indicated in the top-right corner, and
  flux-normalized at $\lambda=2.12$\mum\ such that the disk flux
  amounts to 1.29\% of the stellar flux at that wavelength (FLUOR
  excess for Vega).
    }
\label{fig:PB_SED}
\end{figure*}

\begin{figure*}
\begin{center}
\hbox to \textwidth
{
\parbox{0.49\textwidth}{
\includegraphics[width=0.5\textwidth, height=!, trim=0 1cm 0 0]{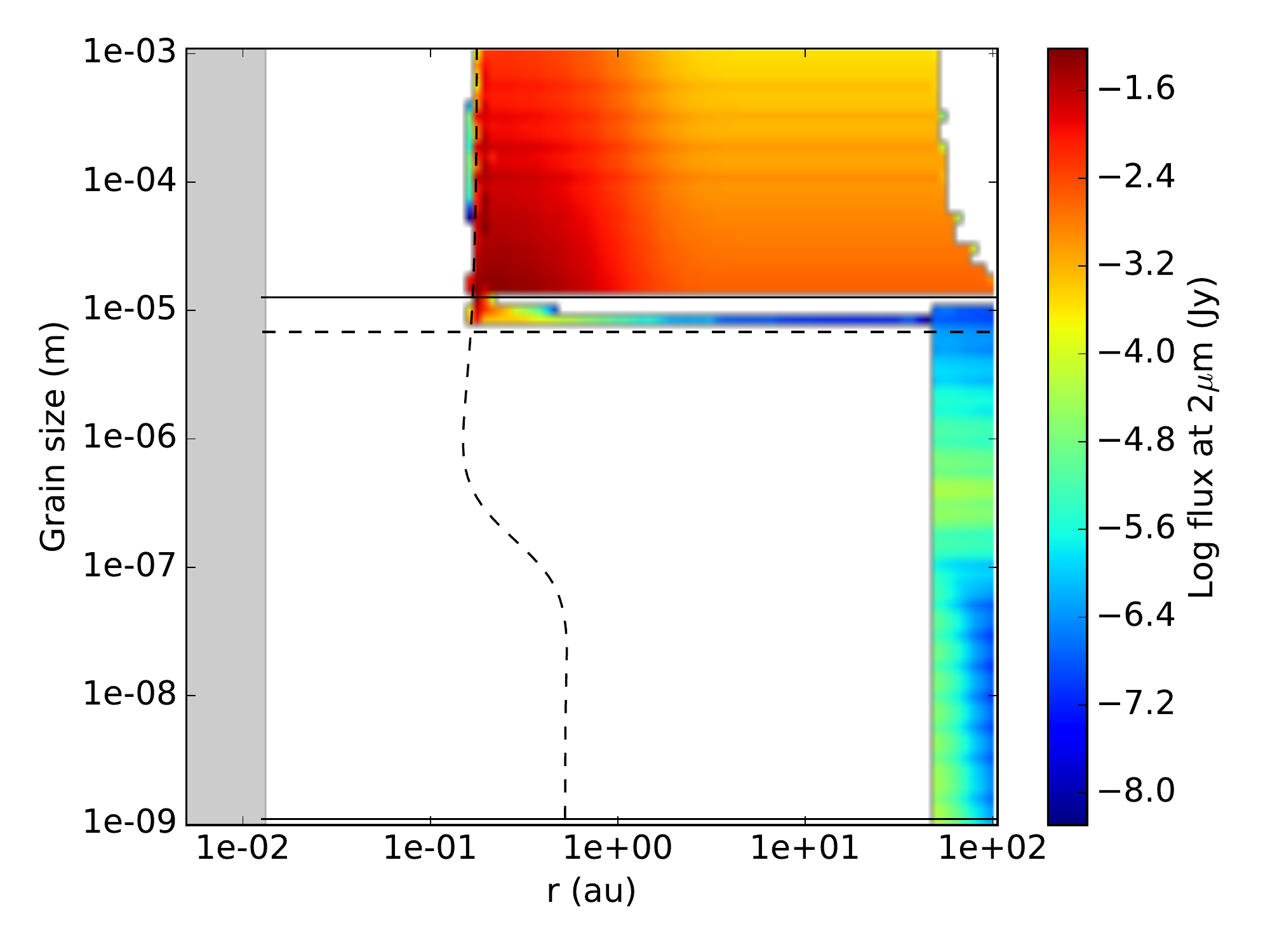}
\subcaption{}
}
\hfill
\parbox{0.49\textwidth}{
\includegraphics[width=0.5\textwidth, height=!, trim=0 1cm 0 0]{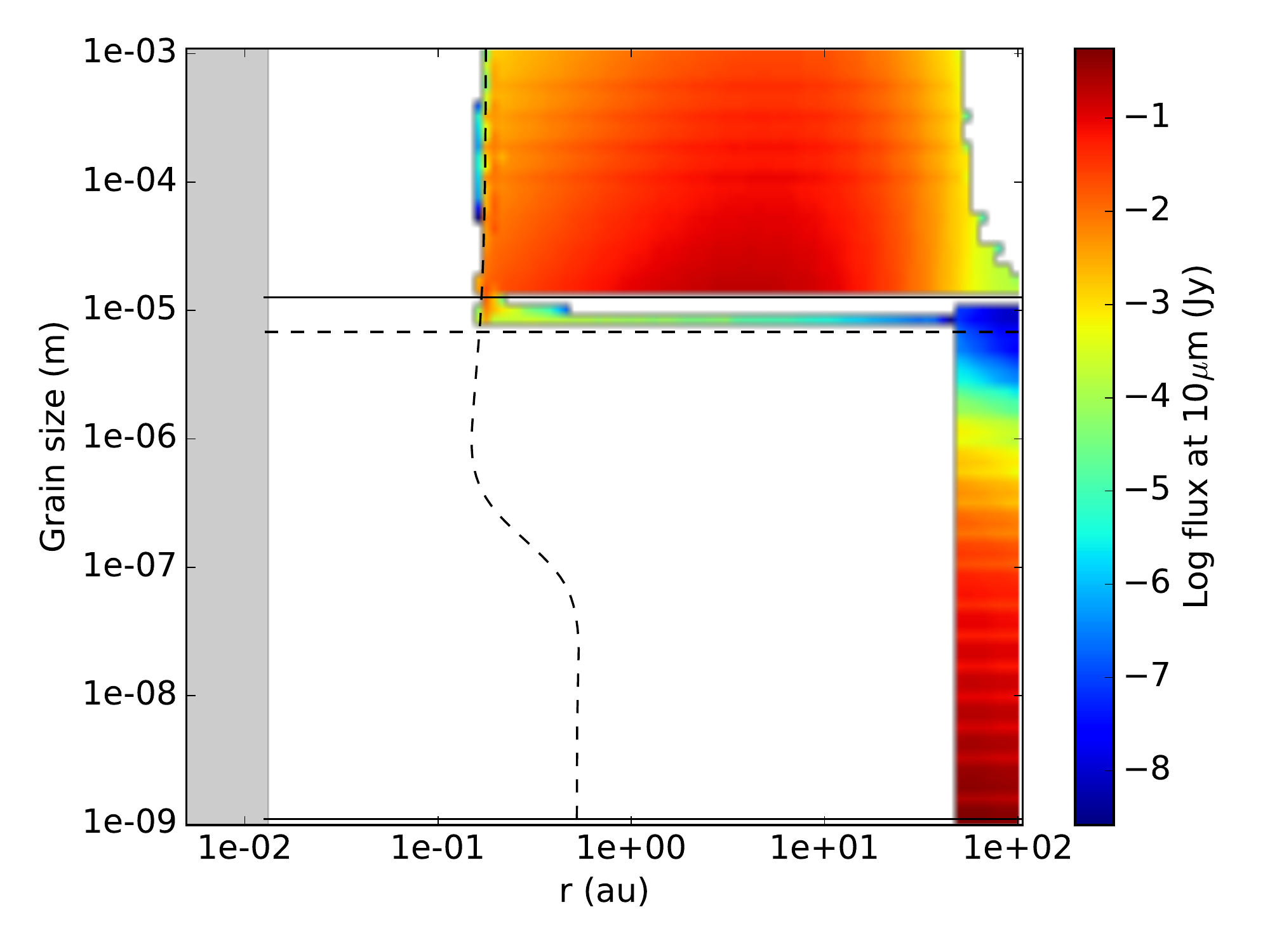}
\subcaption{}
}
}
\hbox to \textwidth
{
\parbox{0.49\textwidth}{
\includegraphics[width=0.5\textwidth, height=!, trim=0 1cm 0 0]{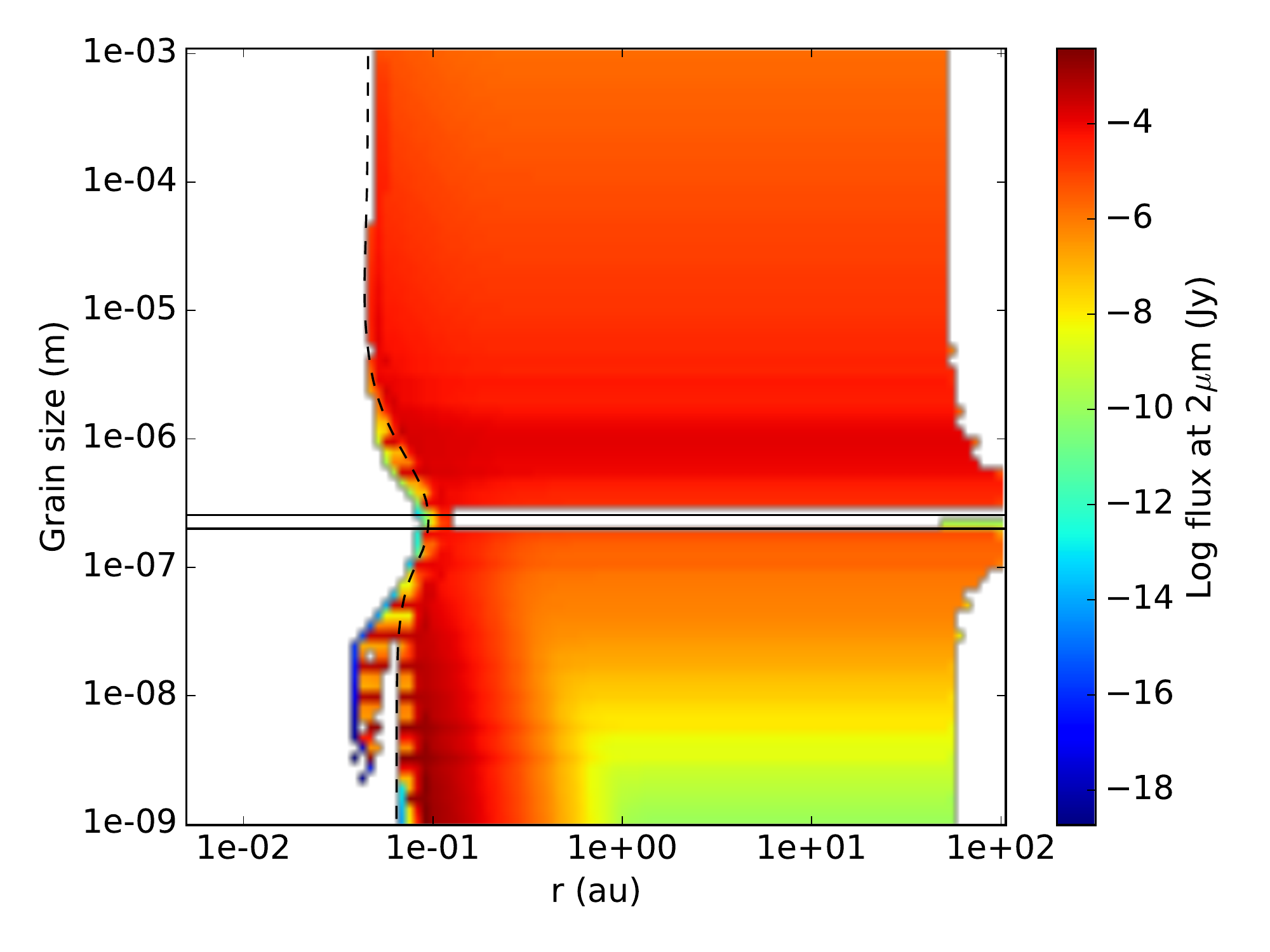}
\subcaption{}
}
\hfill
\parbox{0.49\textwidth}{
\includegraphics[width=0.5\textwidth, height=!, trim=0 1cm 0 0]{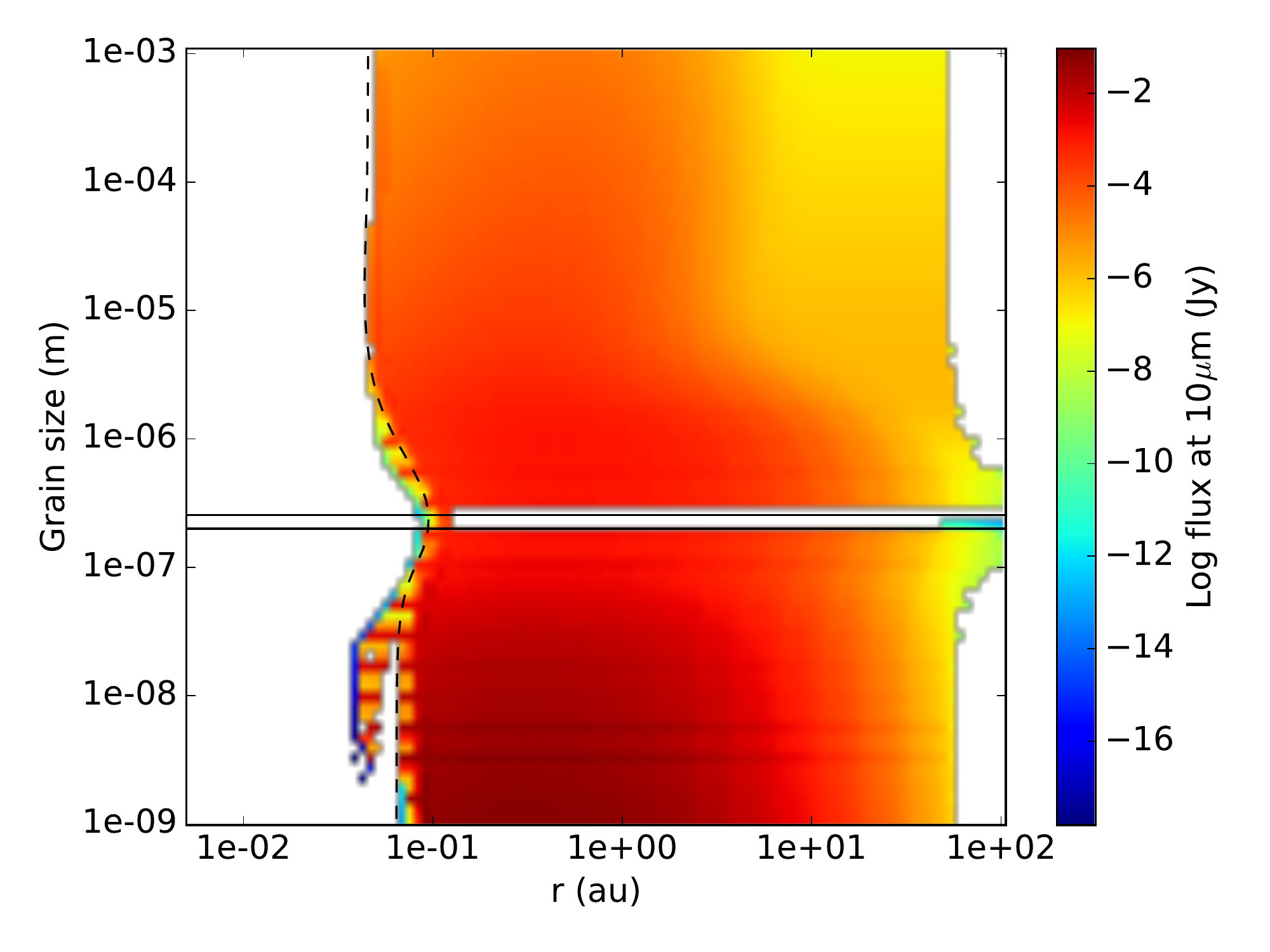}
\subcaption{}
}
}
\end{center}
\caption{PR-drag pile-up scenario: Flux level at 2\,$\mu$m (left) and 10\,$\mu$m (right), for an
  A0 star and carbon grains (top); and K0 star and astrosilicate
  grains (bottom).  See Sec.\,\ref{sec:syntheticobservations} and
  App.\,\ref{app:DensMap} for the methodology.  The verticalish dashed
  line correspond to the location of the sublimation radius $r_{s}$ as
  a function of grain size, while the full and dashed horizontal lines
  indicate the $\beta=0.5$ and $\beta=1$ limits, respectively.}
\label{fig:PBOptMap}
\end{figure*}

Another prediction of radiative transfer models of exozodis is that
the hot dust is expected to be confined close to its sublimation
radius. In order to evaluate the level of dust pile-up at $r\dma{s}$, we
compute the radial distribution of the total geometrical optical
depth, $\tau$, of the dust produced by the PR-drag scenario, following
the approach described in App.\,\ref{app:DensMap}.

As can be seen in Fig.\,\ref{fig:IntOptDepth}, the $\tau(r)$ profiles
are almost flat for most of the domain between the release belt
position down to $r\dma{s}$, which is the expected result for a PR-drag
scenario \citep{Burns1979}. The only departure from the flat profile
occurs close to the sublimation radius, $r\dma{s}$, where we obtain a
dust pile-up generating a density enhancement of a factor of a few at
most (earliest type stars, carbon grains) with respect to the plateau
at larger distance, which is compatible with the values obtained by
\cite{VanLieshout2014}.  This enhancement is due to the biggest grains
sublimating at $r\dma{s}$, all passing through the ($r$,$s$) = (0.2\,au,
10$^{-5}$\,m) bin in the 2D maps in the case of an A0 star and carbon
grains (see also the upper left panel of Fig.\,\ref{fig:PBOptMap})
before being expelled.  
The radial extent of this pile-up is also very narrow, 
and is only marginally resolved in our simulations.  
Thus we can constrain the enhancement to
occur over less than 10$^{-2}$\,au, 
corresponding to a ratio $\Delta r /r$ of around 0.1.  
We note that, for astrosilicates and glassy silicates, the
pile-up is even weaker than for carbon grains. This is because, for
most stellar types, the \be\ ratio is below 1 for the smallest grains of these compositions
(Fig.\,\ref{fig:BetaAll}). This means that, as grains start to
sublimate close to $r\dma{s}$, they will not stay a long time on eccentric
orbits (the reason for the pile-up) before sublimating completely.

\subsubsection{SED}
\label{sec:PRdrag_SED}

As mentioned in the introduction, the best way to evaluate how well
our numerical scenario is able to explain exozodis observations is not
to estimate how it reproduces the \emph{predictions} of radiative
transfer models regarding dust size or pile-up, but rather to evaluate
how well it reproduces the observational constraints themselves.  To
this effect, we use the Python version of the GRaTeR code developed
for this study to generate synthetic SEDs.  As explained in
Sec.\,\ref{subsec:GenePhilo}, because we neglect collisional effects
in the parent belt, we cannot constraint the absolute level of
dustiness, and thus the absolute near-IR fluxes. But we can focus on
the relative balance between the near-IR and mid-IR fluxes as the main
criteria to assess the validity of exozodi producing scenarios. As a
consequence, we chose to rescale all our synthetic SEDs in order for
the emission at 2\,$\mu$m to correspond to the level measured for
4 observationally detected exozodis corresponding to the 4 different
spectral types considered: Vega (A0), $\eta$ Corvi (F0), 10 Tau (G0)
and $\tau$ Ceti (a G8 star that is relatively close to a K0 one). We note that,
for all these cases, the flux excess at $2\,\mu$m is always of the
order of $\sim 1\%$ of the stellar contribution.

The four corresponding synthetic SEDs are shown in
Figs.\,\ref{fig:PB_SED}a to d.  We clearly see that, for all
considered spectral types and grain compositions, the shape of the
synthetic SED contradicts the observational constraints. 
The synthetic
SEDs peak in the far-IR and the flux density in the 10--20\,$\mu$m
domain is always 
much higher than the one found in exozodi observations,
which are only of the order of a few percent of the stellar flux around 10\mum.
This clearly illustrates the fact that the pile-up near
the sublimation region is far from being sufficient to boost the near-IR flux
at the point where it can dominates over the mid-IR flux. 
This mid-IR flux excess is due
to the continuous flow of PR-drag drifting grains in the region
between the production belt and $r\dma{s}$. This can be clearly seen
in Figure~\ref{fig:PBOptMap}, which shows, for the two "extreme" A0
and K0 cases, the contributions of the grains to the fluxes at 2 and
10\,$\mu$m, as a function of their size and spatial location.

We checked if one way to alleviate this problem could be to start with
a parent belt much closer than the considered 50\,au. 
We reproduce this situation in a simple manner,
 without running additional simulations. 
 We consider our original simulations with a parent belt at 50\,au, 
 and integrate the flux coming from the grains within a given distance to the star. 
 That distance is assumed to mimic the new location of the parent belt. 
 In the specific case of Vega, the results are shown Fig.\,\ref{fig:PB_SED}e, 
 where all fluxes have been normalized to the same observed flux at $\lambda = 2.12\,\mu$m.
The "unwanted" mid-IR (10\,$\mu$m) flux falls down to
observation-compatible levels only for an extremely close-in parent belt
located at 0.4\,au. This solution appears highly unlikely given that
such a massive collisional belt would probably not be able to survive
long enough so close to the star to sustain the hot dust for a
duration comparable to the age of the star.

Overall, these conclusions hint at another production process than the
PR-drag mechanism to populate the hot exozodiacal dust systems.

\begin{table*}
\caption{Parameters for the cometary release at perihelion scenario :
  assumed release position ($r\dma{p}$, in au) and resulting comet
  orbit eccentricity ($e$), blowout $\beta$ value ($\beta\dma{blow}$)
  and corresponding blowout grain size ($s_{\mathrm{blow}}$, in
  $\mu$m) (see Sec.\,\ref{sec:Comet_NumSetup} for details on the
  assumptions). }
\label{tab:SizeBoundTan}
\centering

\begin{tabular}{c c c c c c c c c c}
\hline\hline
  & \multicolumn{4}{c}{A0 star} & & \multicolumn{4}{c}{F0 star} \\
    & $r\dma{p}$ & $e$ & $\beta\dma{blow}$ & 
  $s\dma{blow}$ & & $r\dma{p}$ & $e$ & $\beta\dma{blow}$ & $s\dma{blow}$ \\
   \cline{2-5} \cline{7-10}
  Carbon & 0.60 & 0.976 & 1.2$\times 10^{-2}$ & 470 & & 0.16 & 0.994 & 3.1$\times 10^{-3}$ & 310 \\
  Astrosilicate & 2.4 & 0.908 & 4.6$\times 10^{-2}$ & 70 & & 0.43 & 0.983 & 8.5$\times 10^{-3}$ & 65 \\
  Glassy silicate & 1.6 & 0.938 & 3.1$\times 10^{-2}$ & 90 & & 0.11 & 0.996 & 2.1$\times 10^{-2}$ & 210 \\
  \hline
\end{tabular}

\begin{tabular}{c c c c c c c c c c}
  &  \multicolumn{4}{c}{G0 star} & & \multicolumn{4}{c}{K0 star} \\
    & $r\dma{p}$ & $e$ & $\beta\dma{blow}$ & $s\dma{blow}$ & & $r\dma{p}$&
  $e$ & $\beta\dma{blow}$ & $s\dma{blow}$ \\
   \cline{2-5} \cline{7-10}
  Carbon & 0.081 & 0.997 & 1.6$\times 10^{-3}$ & 330 & & 0.038 & 0.999 & 7.6$\times 10^{-4}$ & 240 \\
  Astrosilicate & 0.23 & 0.991 & 4.5$\times 10^{-3}$ & 65 & & 0.11 & 0.996 & 2.1$\times 10^{-3}$ & 51 \\
  Glassy silicate & 0.050 & 0.998 & 9.9$\times 10^{-4}$ & 230 & & 0.027 & 0.999 & 5.3$\times 10^{-4}$ & 120 \\
  \hline
\end{tabular}
\end{table*}


\section{Exocometary dust delivery scenario}
\label{sec:Comets}

Another classic process of exozodiacal dust production is the cometary
grain release very close to the star. In this scenario, an outer mass
reservoir remains necessary, but the dust grains are deposited by
large, undetectable parent bodies in the immediate vicinity of the
place where they are detected. A benefit of this process compared to
the PR-drag pile-up scenario is that it leaves essentially no
observable signature in between the parent belt and the exozodi.

In \citet{Bonsor2012}, we investigated the planetary system
architecture required to sustain an inwards flux of exocomets. In
\citet{Bonsor2013}, we highlighted the importance of
planetesimal-driven migration of the planet closest to the inner edge
of the belt to maintain this flux on sufficiently long timescales
\citep[see also][]{Raymond2014}. In
\citet{Marboeuf2016}, we evaluated the cometary dust ejection rate as
a function of the distance to the star and spectral type, to help
connecting dynamical simulations to exozodi observations in future
studies \citep[e.g.][]{Faramaz2017}. Here, we make an important step
further by discussing the fate of the grains once released by an
exocomet passing close to the star, and by calculating the resulting
emission spectra for a direct comparison to the data.

\begin{figure*}[t]
\centering
\hbox to \textwidth
{
\parbox{0.33\textwidth}{
\includegraphics[width=0.33\textwidth, height=!, trim=0 1cm 0 0]{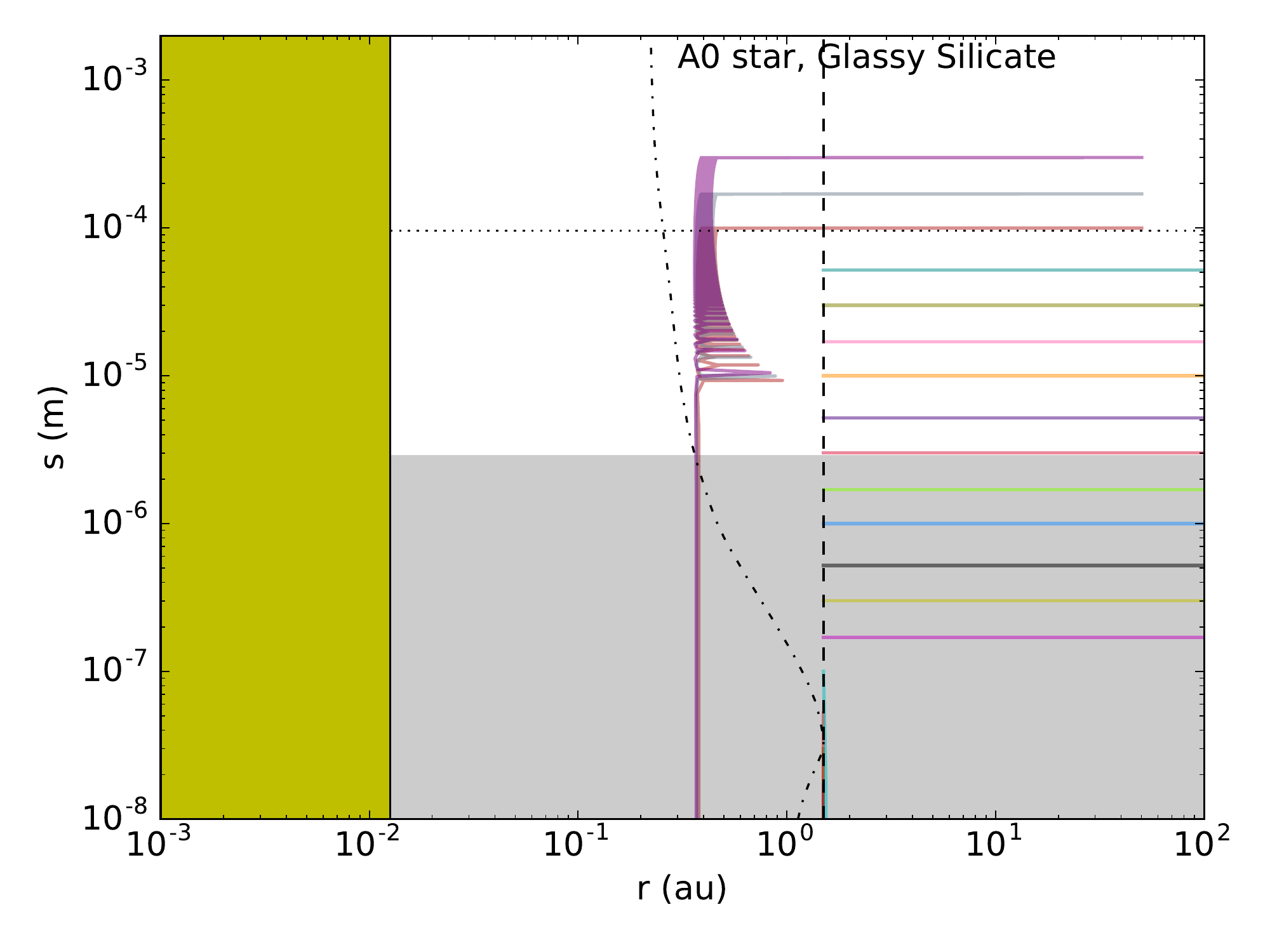}
\subcaption{ A0, glassy silicate}
}
\hfill
\parbox{0.33\textwidth}{
\includegraphics[width=0.33\textwidth, height=!, trim=0 1cm 0 0]{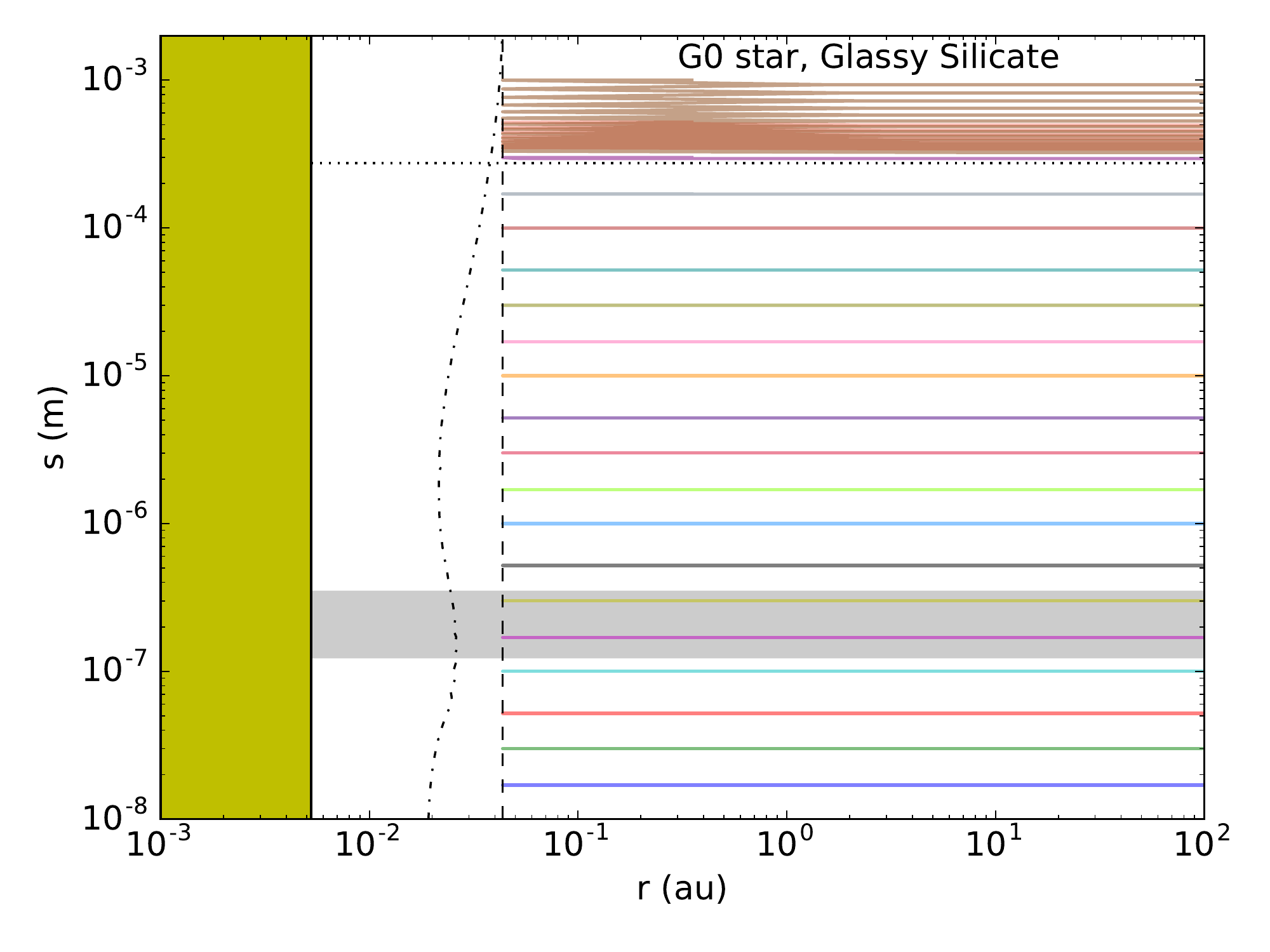}
\subcaption{G0, glassy silicate }
}
\hfill
\parbox{0.33\textwidth}{
\includegraphics[width=0.33\textwidth, height=!, trim=0 1cm 0 0]{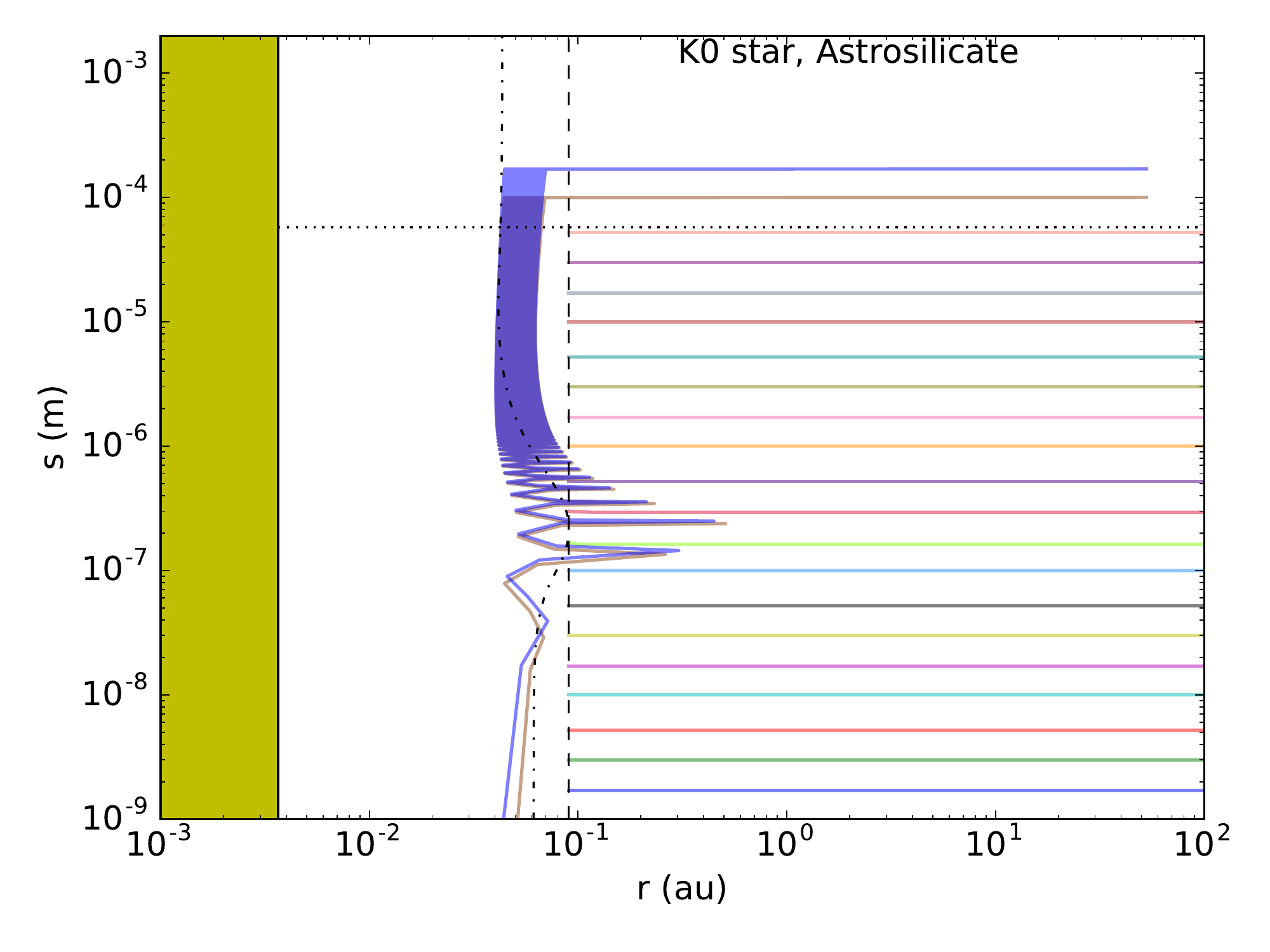}
\subcaption{K0, astrosilicate }
}
}

\hbox to \textwidth
{
\parbox{0.33\textwidth}{
\includegraphics[width=0.33\textwidth, height=!, trim=0 1cm 0 0]{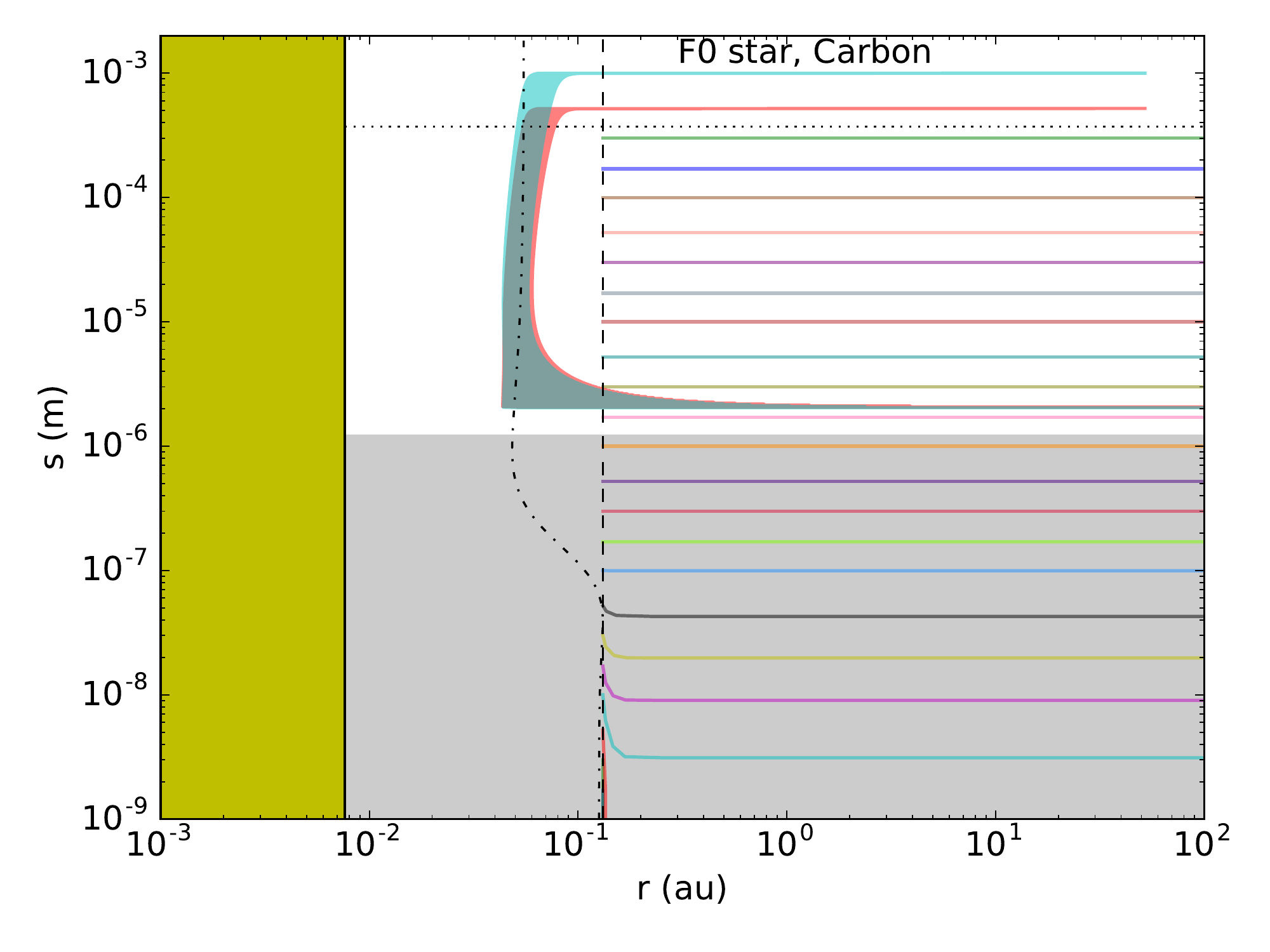}
\subcaption{F0 star, Carbon} } \hfill
\parbox{0.33\textwidth}{
\includegraphics[width=0.33\textwidth, height=!, trim=0 1cm 0 0]{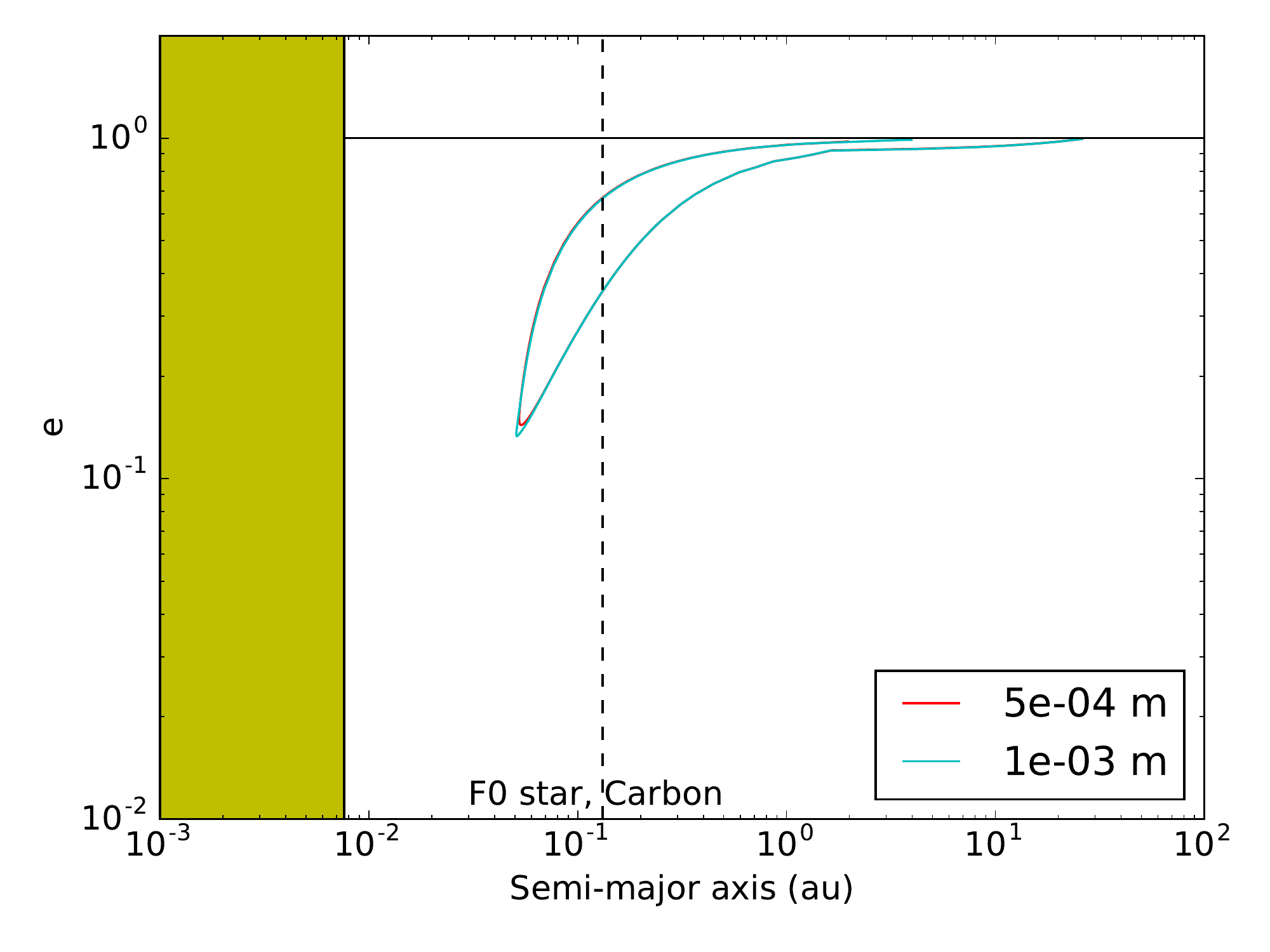}
\subcaption{F0 star, Carbon}
}
\hfill
\parbox{0.33\textwidth}{
\includegraphics[width=0.33\textwidth, height=!, trim=0 1cm 0 0]{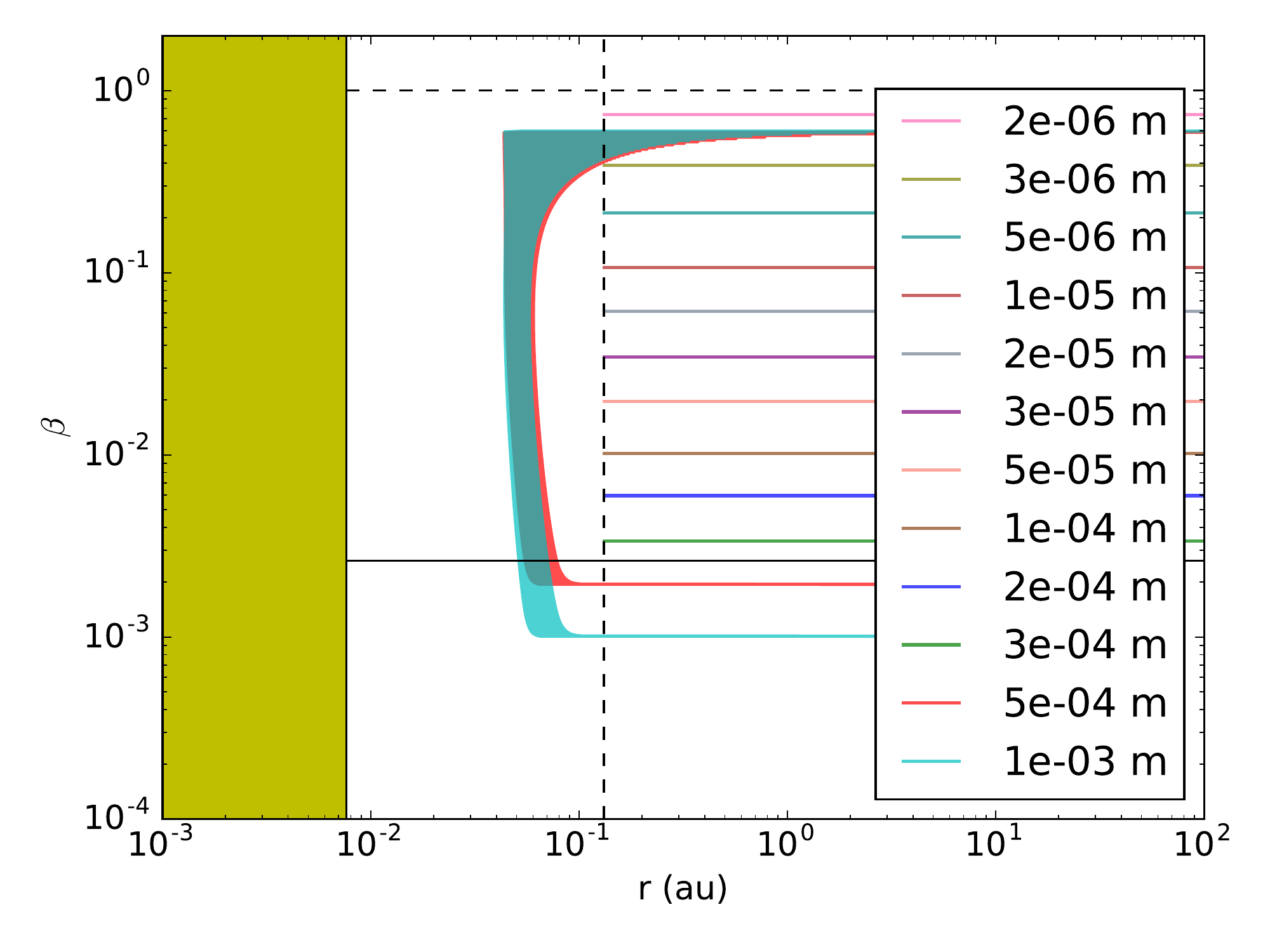}
\subcaption{F0 star, Carbon}
}
}
\caption{Panels (a) to (c) : grain size as function of the distance
  for all simulations ran in the exocometary dust delivery at
  perihelion scenario (the yellow area, the dotted and
  dash-dotted lines have the same meaning as in Fig.\,\ref{fig:PB}, 
  while dashed line corresponds to the limit between initially bound and unbound grains sizes). 
  Panels (d) to (f) : specific case of an F0 star
  and carbon grains, for which the evolution of the grain size, the
  eccentricity and the \be\ value as a function of distance is
  displayed. In the case of eccentricity, the two curves overlap.}
\label{fig:ComTanSR}
\end{figure*}

\subsection{Numerical setup}
\label{sec:Comet_NumSetup}

In order to compare the outcome of our cometary model to the results obtained for the PR-drag pile-up
scenario (Sec.\,\ref{sec:PRdrag}), we consider  a reservoir of exocomets that have their
aphelion at a fixed distance of 50\,au and a perihelion  $r\dma{p}$ just outside the sublimation limit,
which, for each composition and spectral type, 
we define as the largest sublimation distance of the considered grain sizes
(often corresponding to the smallest grains,
see e.g. vertical dashed lines in Fig.\,\ref{fig:ComTanSR}a to
\ref{fig:ComTanSR}d).

\begin{figure}[tp!]
\includegraphics[width=0.49\textwidth]{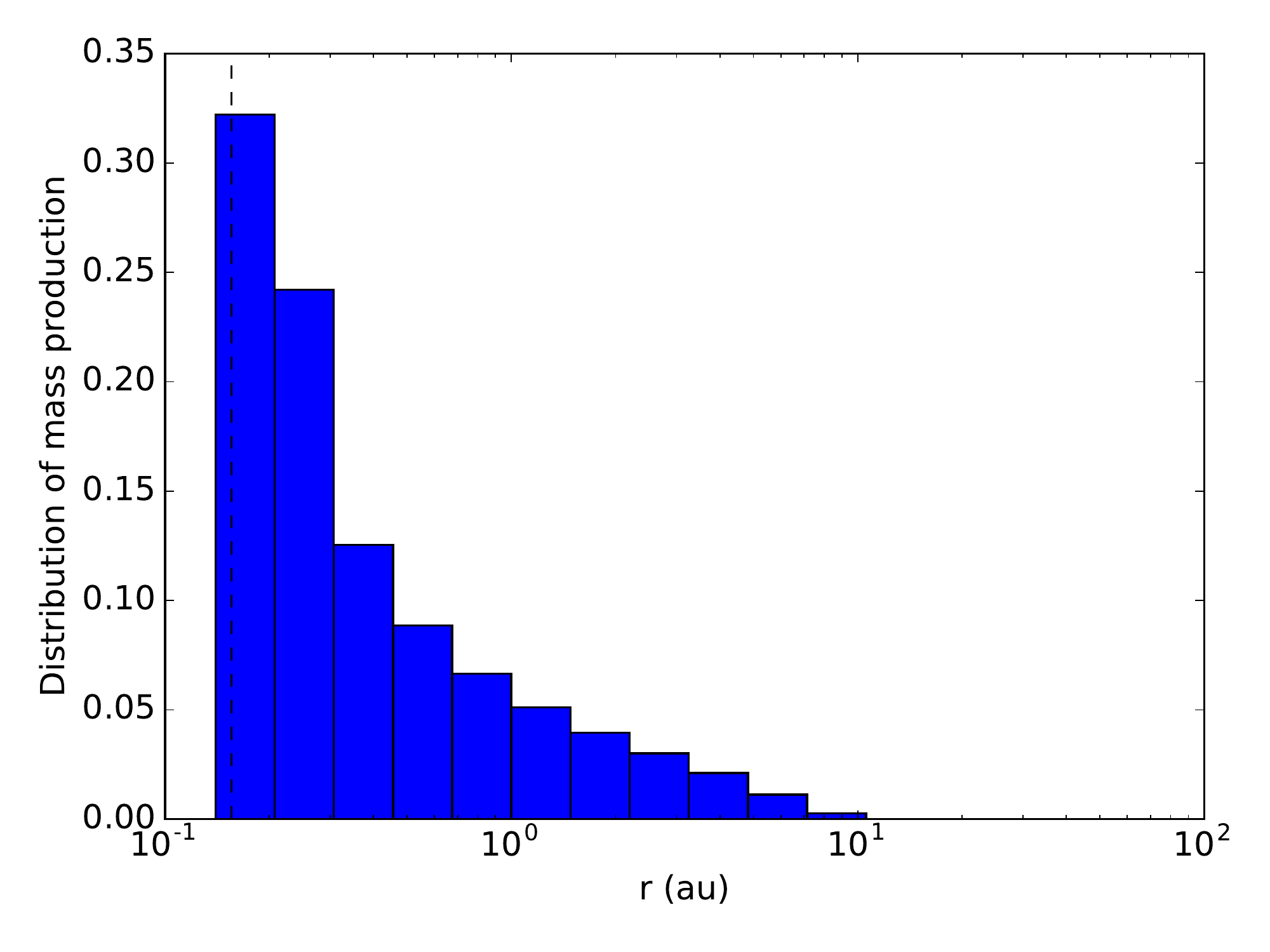}
\caption{Fraction of the surfacic mass loss for an exocomet orbiting
  an F0 star as a function of radial distance.  
  The comet has an apoastron of 50\,au,
  and an eccentricity of 0.994.  
  Based on the comet evaporation prescription of
  \citet[][their Eq.\,17]{Marboeuf2016}.
  }
\label{fig:ComFracMass}
\end{figure}

We assume that all grains leaving the comet are produced when the comet
passes at perihelion $r\dma{p}$. This is a simplifying assumption because
grains should be dragged from the comet by the evaporation of volatiles, which
should happen over a large fraction of its orbit. However, as shown by \citet{Marboeuf2016}, the
volatile and dust production rate strongly increase with decreasing distance to the star
(see Equ.17 of that paper), so that most of the mass loss happens in a narrow region
close to the comet's perihelion, as is clearly illustrated by Fig.\ref{fig:ComFracMass}.

Grains produced at perihelion have the highest possible speed once released from the comet and are thus less
prone to remain bound. More precisely, the blowout limit
in term of \be\ is lowered
compared to the PR-drag pile-up scenario and can be expressed using
\citep[e.g.][]{Murray1999}:
\begin{equation}
\beta\dma{blow} = \frac{1}{2} \left( \frac{1-e_0^2}{1+e_0 \cos \Phi} \right) .
\label{eqn:Bcom}
\end{equation}
where $\Phi$ is the longitude of the release position on the cometary
orbit, and $e_0$ the parent body eccentricity. This reduces to
$\beta\dma{blow} = (1-e_0)/2$ for a release at perihelion.  For the
grain compositions and spectral types explored in this study
(Sec.\,\ref{sec:star_composition}), the release distance varies
between 2\,au and $\sim$\,0.02\,au, corresponding,
for an apoastron of 50\,au, to parent body
eccentricities varying between 0.923 and 0.999, and $\beta\dma{blow}$
values between $3.8 \times 10^{-2}$ and $5.0 \times 10^{-4}$,
respectively.  These low $\beta\dma{blow}$ values translate into large
grain sizes, of several tens to several hundred of $\mu$m, 
for the limiting blow-out size.
The release distances,
orbit eccentricities, blowout $\beta$ values and grain sizes
$s_{\mathrm{blow}}$ are documented in Tab.\,\ref{tab:SizeBoundTan} for
the four spectral types and three compositions investigated in this
study.

\subsection{Grain evolution}

\subsubsection{Carbon and astrosilicate grains}

In our model, the carbon and astrosilicate bound grains
($\beta<\beta_{\mathrm{blow}}$, i.e. $s>s_{\mathrm{blow}}$) are
delivered by an exocomet a little beyond their sublimation distance,
leaving room for dynamical evolution. The fate of these grains shows
similarities with that discussed in the case of the PR-drag pile-up
scenario. Their semi-major axis and eccentricity both decrease by
PR-drag, until they are totally sublimated
(Sec.\,\ref{subsubsec:sublim}) or expelled from the system by
radiation pressure due to partial sublimation
(Sec.\,\ref{subsubsec:ejection}), as illustrated in
Figs.\,\ref{fig:ComTanSR}c and d.

There are, however, important differences between the cometary and
PR-drag pile-up scenarios. In particular, the high orbital
eccentricity of the grains inherited from the comet implies that those
grains spend only a very small fraction of their orbital period
(typically less than a day) close to the sublimation zone in the early
phases of their evolution. Therefore, these grains see their size
remaining essentially constant while migrating inward by PR-drag, like
in the stage\,I of the PR-drag pile-up scenario
(Sec.\,\ref{sec:PRdraggeneralbehaviour}). Their orbit is getting
circularized until sublimation becomes significant, thereby moving
into the stage\,II phase. However, they enter this phase with a
significant residual eccentricity (of the order of 0.1,
e.g. Fig.\,\ref{fig:ComTanSR}e), much larger than in the case of the
PR-drag pile-up scenario. As a consequence, the carbon and
astrosilicate grains are expelled when they reach a \be\ value of
about 0.6 (Fig.\,\ref{fig:ComTanSR}f), to be compared to the 0.8 accessible in the PR-drag scenario. 
The only exception is for
astrosilicate grains around the K0-type star, for which the \be\ = 0.6
value is never reached accross all grain sizes, meaning that the bound
grains end up completely sublimated. Another consequence of the
significant residual eccentricity during stage II is that, here again, the grains
do not reach \be\ values close enough to 1 for the DDE mechanism to
operate.

\subsubsection{Glassy silicate grains}

The behaviour of glassy silicate grains is quite different. The
sublimation timescale of these grains at the sublimation temperature 
is four orders of magnitude lower than for the other compositions
considered here (Sec.\ref{sec:star_composition}). In this case, the
sublimation time becomes comparable to the time spent by the grain
close to the sublimation limit in only one perihelion flyby. For the
A- and F-type stars, this results in a complete sublimation of the
bound grains after the circularization of stage\,II.

For later type stars, the fate of the bound grains is affected by the
unusual fact that, at a given distance, large glassy silicate grains
are hotter than smaller ones (see e.g. the almost vertical dash-dotted
lines in Figs.\,\ref{fig:PBGlaK0} and \ref{fig:ComTanSR}b). Therefore,
the large bound glassy silicate grains are released at, or very close
to, their sublimation distance around late-type stars in our
model. Their high temperature, combined with the intrinsic sublimation
efficiency of glassy silicates compared to carbon and astrosilicate
grains, implies that their sublimation timescale is in this case lower
than the PR-drag timescale. 
As consequence, these bound grains become small enough to
be expelled before their orbits can be circularized.

\subsection{Global disk properties and spectra}
\label{sec:comets_diskspectra}

\subsubsection{Surface density profiles}
\label{sec:comets_tau}

\begin{figure}
\centering
\includegraphics[width=0.49\textwidth]{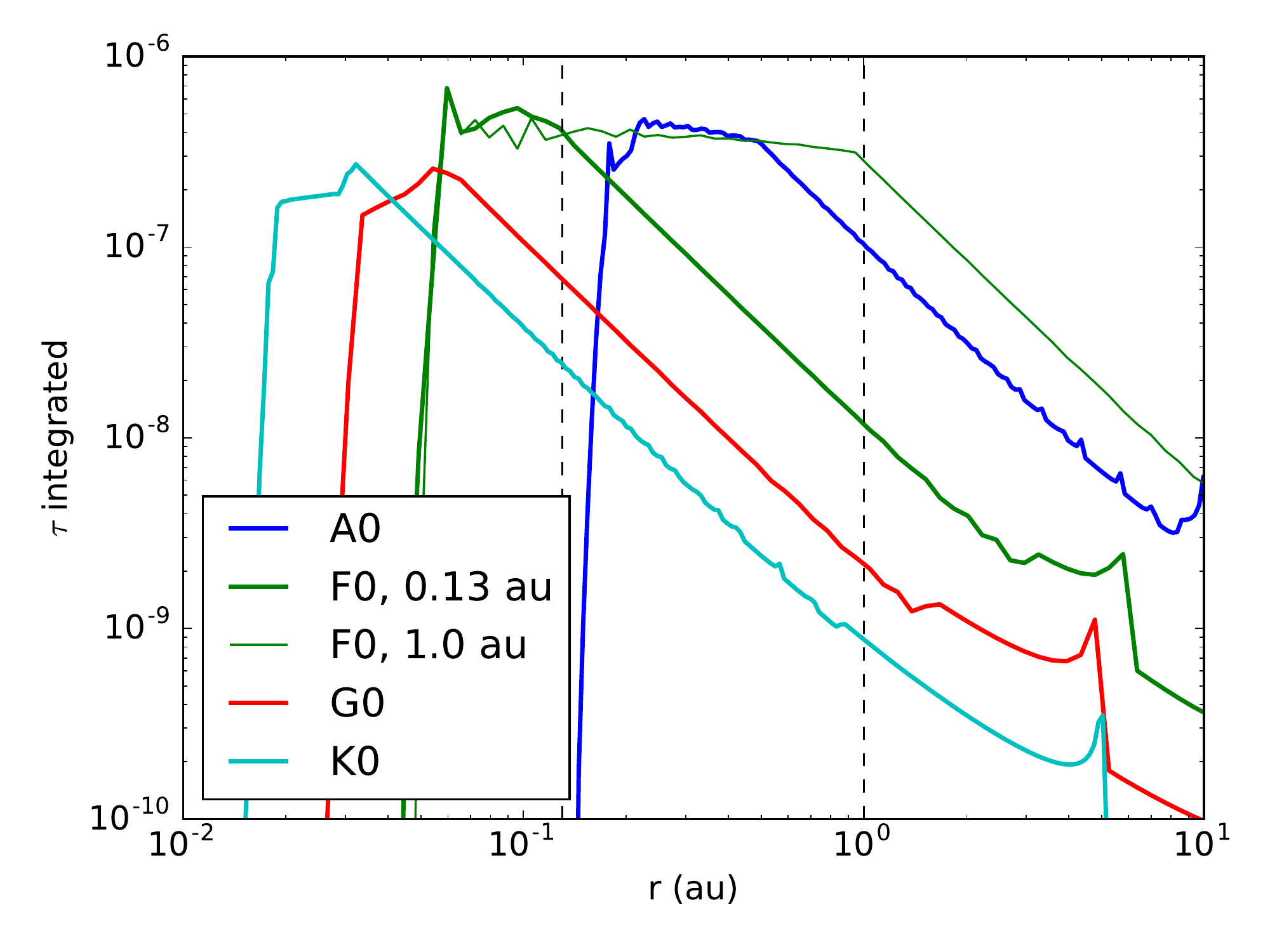}
\caption{Total geometrical optical depth ($\tau$) in the exocometary
  dust delivery at perihelion scenario, assuming carbon grains. For
  illustrative purpose, a model with an exocomet perihelion larger
  than in the nominal case is shown for the F0 star.}
\label{fig:ComTau}
\end{figure}

\begin{figure*}
\begin{center}
\hbox to \textwidth
{
\parbox{0.49\textwidth}{
\includegraphics[width=0.5\textwidth, height=!, trim=0 0 0 0]{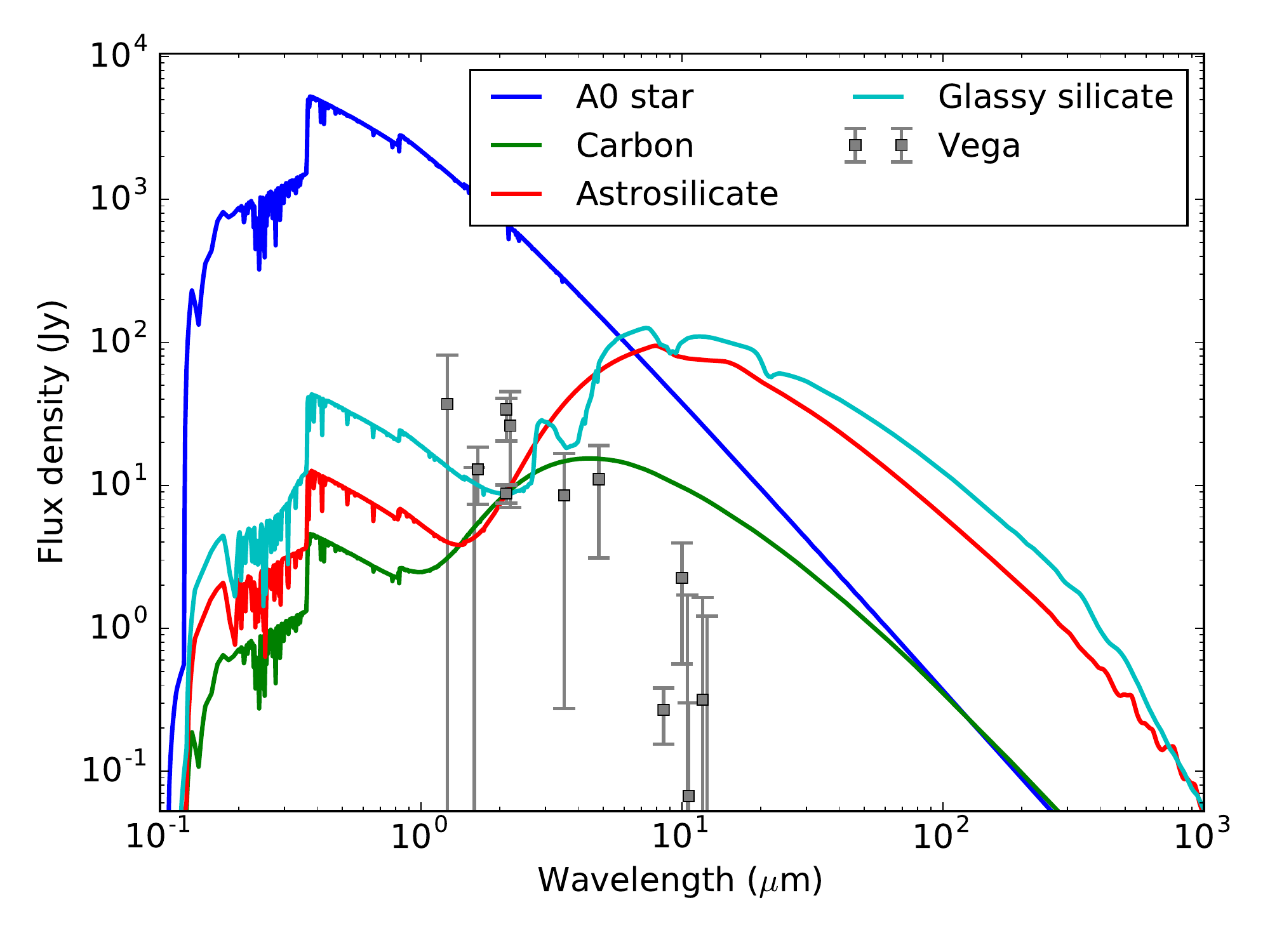}
\subcaption{}
}
\hfill
\parbox{0.49\textwidth}{
\includegraphics[width=0.5\textwidth, height=!, trim=0 0 0 0]{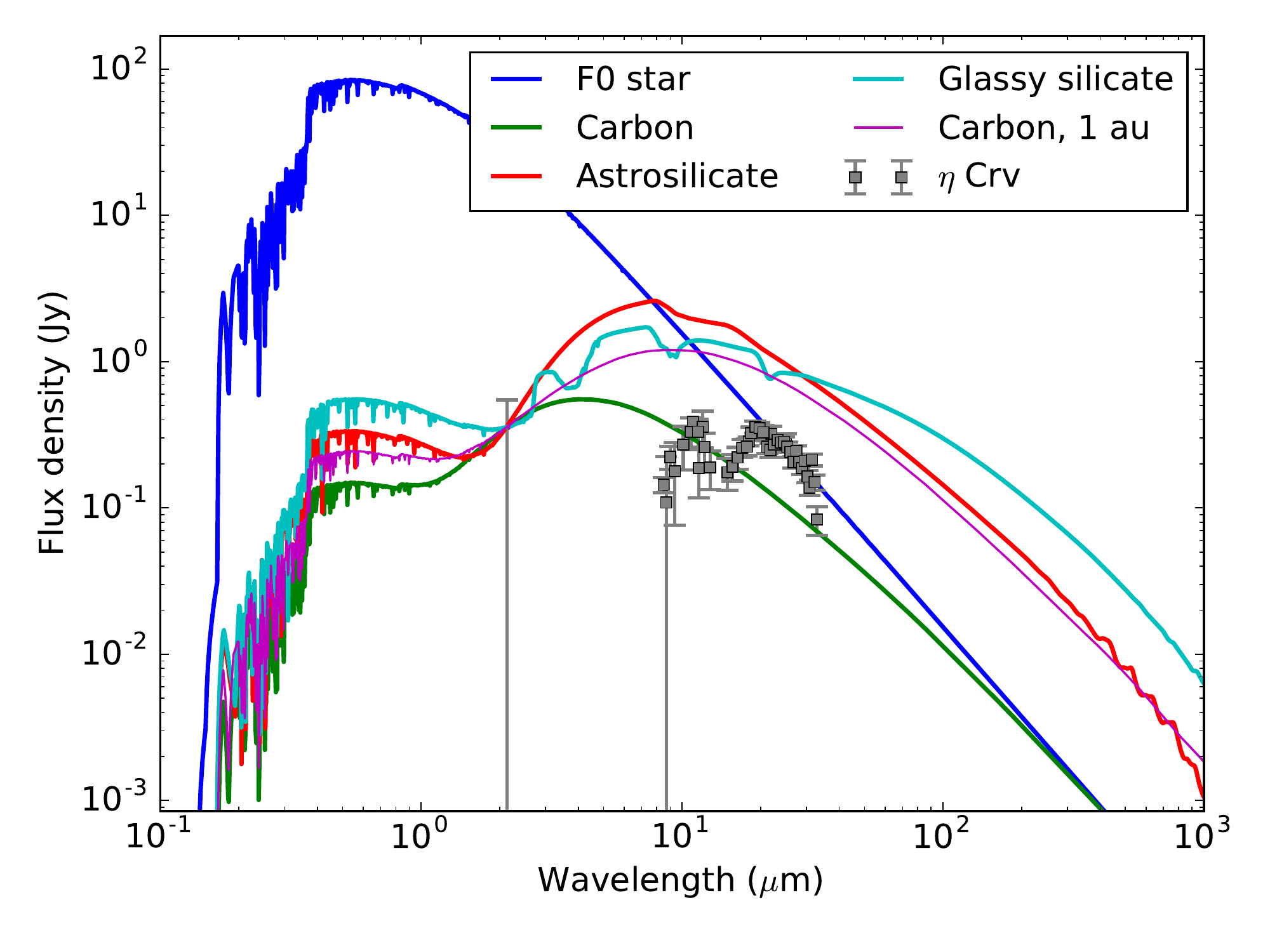}
\subcaption{}
}
}
\hbox to \textwidth
{
\parbox{0.49\textwidth}{
\includegraphics[width=0.5\textwidth, height=!, trim=0 0 0 0]{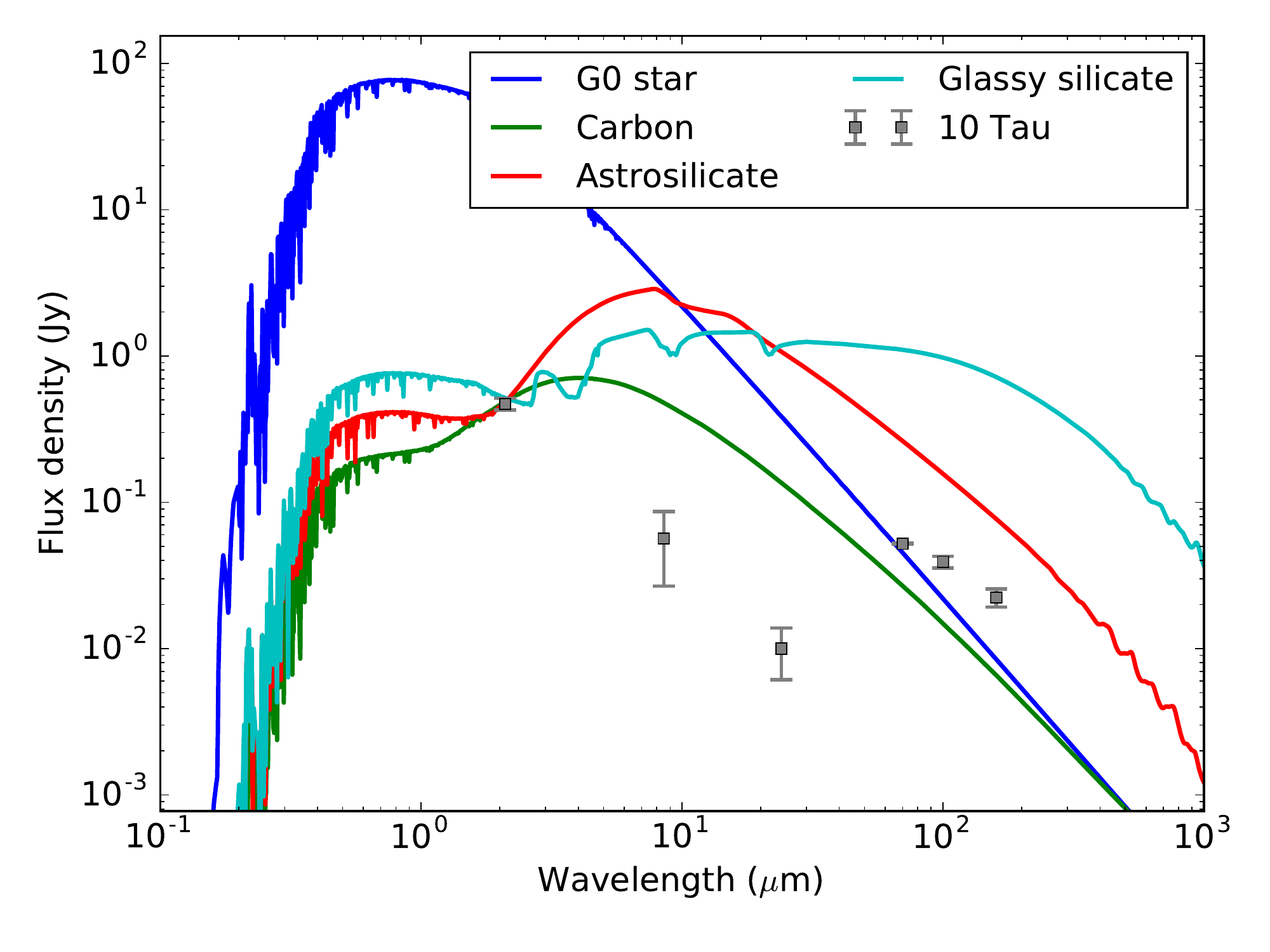}
\subcaption{}
}
\hfill
\parbox{0.49\textwidth}{
\includegraphics[width=0.5\textwidth, height=!, trim=0 0 0 0]{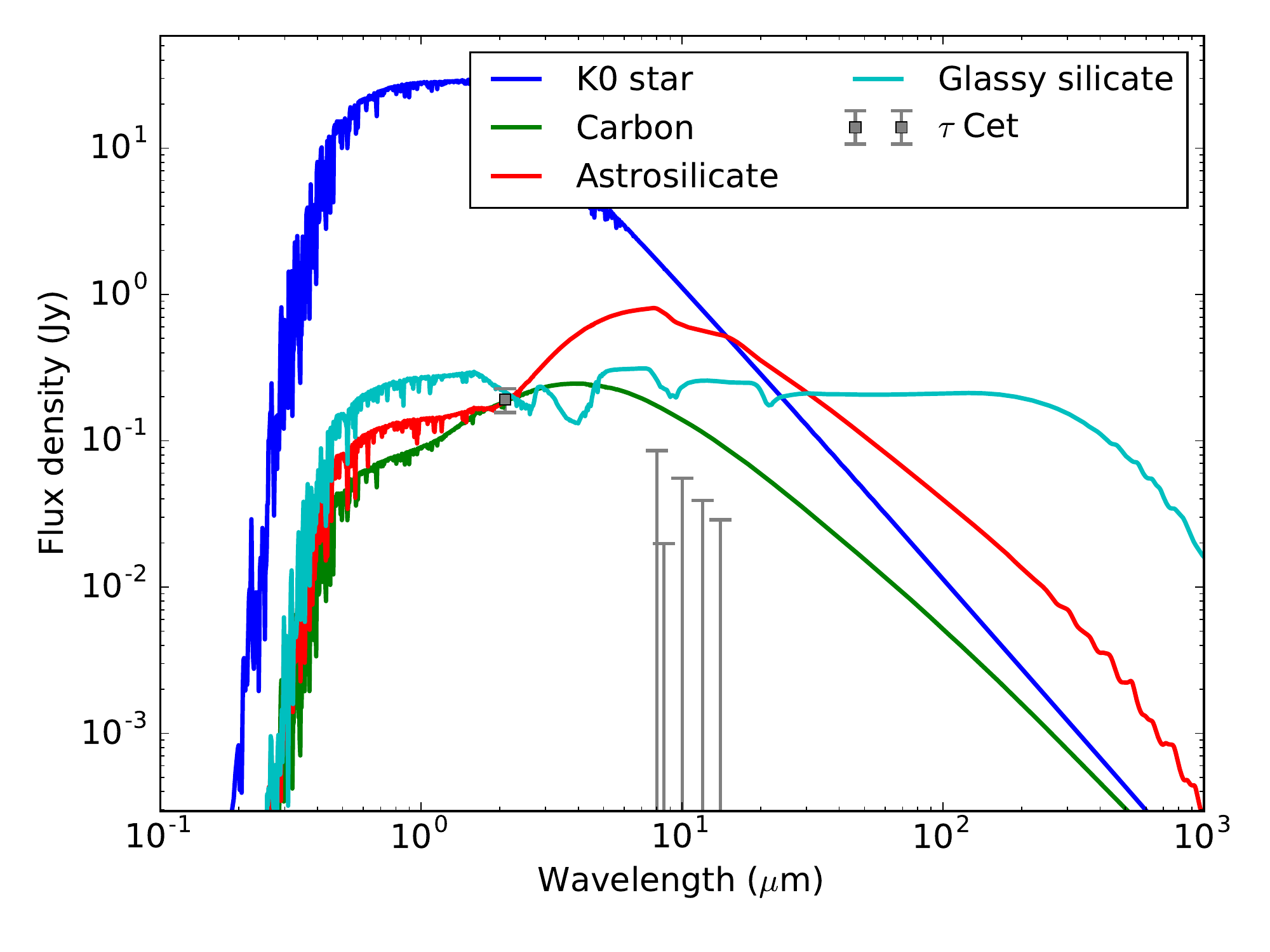}
\subcaption{}
}
}
\end{center}
\caption{Same as Fig.\,\ref{fig:PB_SED} but for the exocometary dust
  delivery at perihelion scenario. 
  }
\label{fig:ComTan_SED}
\end{figure*}

We calculate the radial distribution of the total geometrical optical
depth, $\tau$, of the dust produced in the comet-release scenario,
following the approach described in App.\,\ref{app:DensMap}, and
already used in the case of the PR-drag pile-up scenario (see
Sec.\,\ref{sec:PRdrag_tau}). As illustrated in Fig.\,\ref{fig:ComTau}
for carbon grains, for a release at a perihelion $r\dma{p}$ that is close to 
the sublimation distance $r\dma{s}$, the radial density profile peaks
at the release location and decreases further out with the distance to the
star as $r^{-1.7}$.  

To check the importance of the position of the periastron with respect to
the sublimation distance, we performed an additional simulation, for an F0 star
and carbon grains, for which the dust is released at 1\,au instead of 0.13\,au in the nominal case.
As can be seen in Fig.\,\ref{fig:ComTau}, we obtain a flat $\tau(r)$ profile in between $r\dma{s}$ and
1\,au, followed by a $r^{-1.7}$ profile at larger distances. 
The plateau between $r\dma{p}$ to $r\dma{s}$ results from the inward migration and
circularization of the orbit of the bound grains by PR-drag, as
already observed for the PR-drag pile-up scenario
(Fig.\,\ref{fig:IntOptDepth}), while the high-eccentricity bound
grains populate the regions outside that distance.
We conclude that the exocometary dust
delivery position can essentially be regarded as playing the same role
as the parent belt distance in the PR-drag pile-up scenario, as far as
the shape of the $\tau(r)$ profile between that reference
position down to the sublimation distance is concerned. Numerically, the PR-drag
pile-up scenario can be considered as an extreme case of the cometary
scenario, with exocomets having zero eccentricity.

\begin{figure*}
\centering
\hbox to \textwidth
{
\parbox{0.49\textwidth}{
\includegraphics[width=0.49\textwidth]{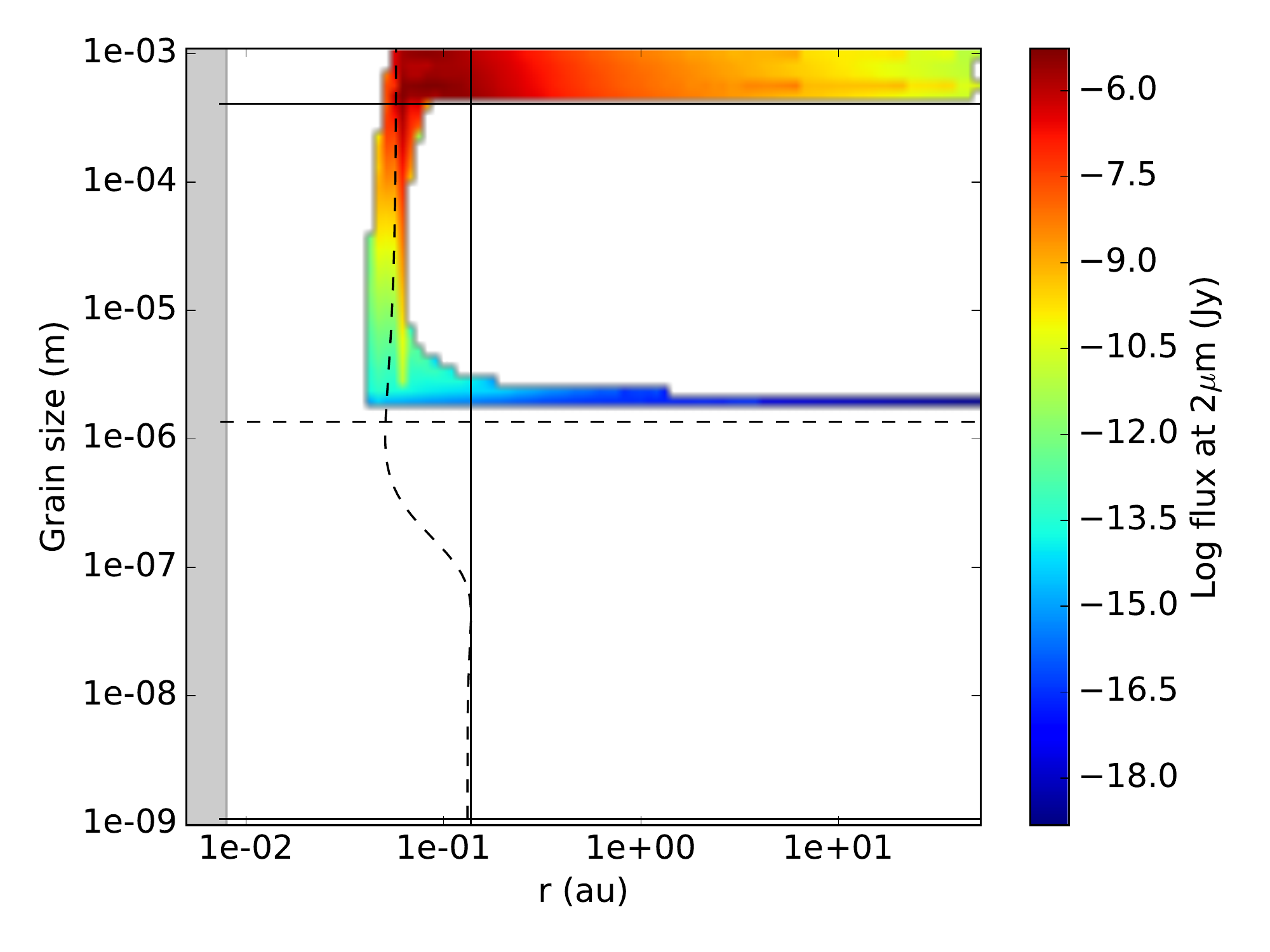}
\subcaption{}
}
 \hfill
\parbox{0.49\textwidth}{
\includegraphics[width=0.49\textwidth]{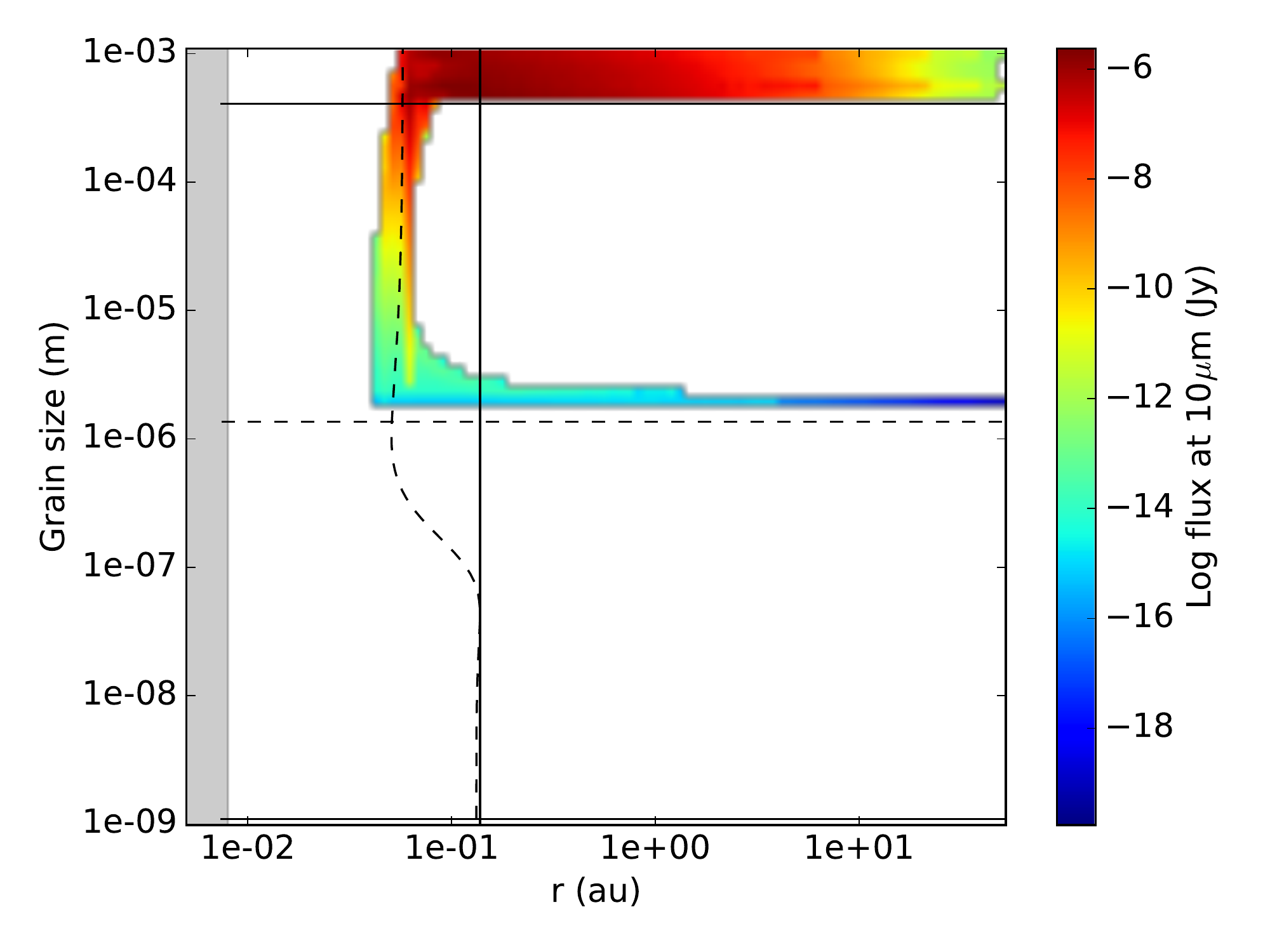}
\subcaption{}
}
}
\hbox to \textwidth
{
\parbox{0.49\textwidth}{
\includegraphics[width=0.49\textwidth]{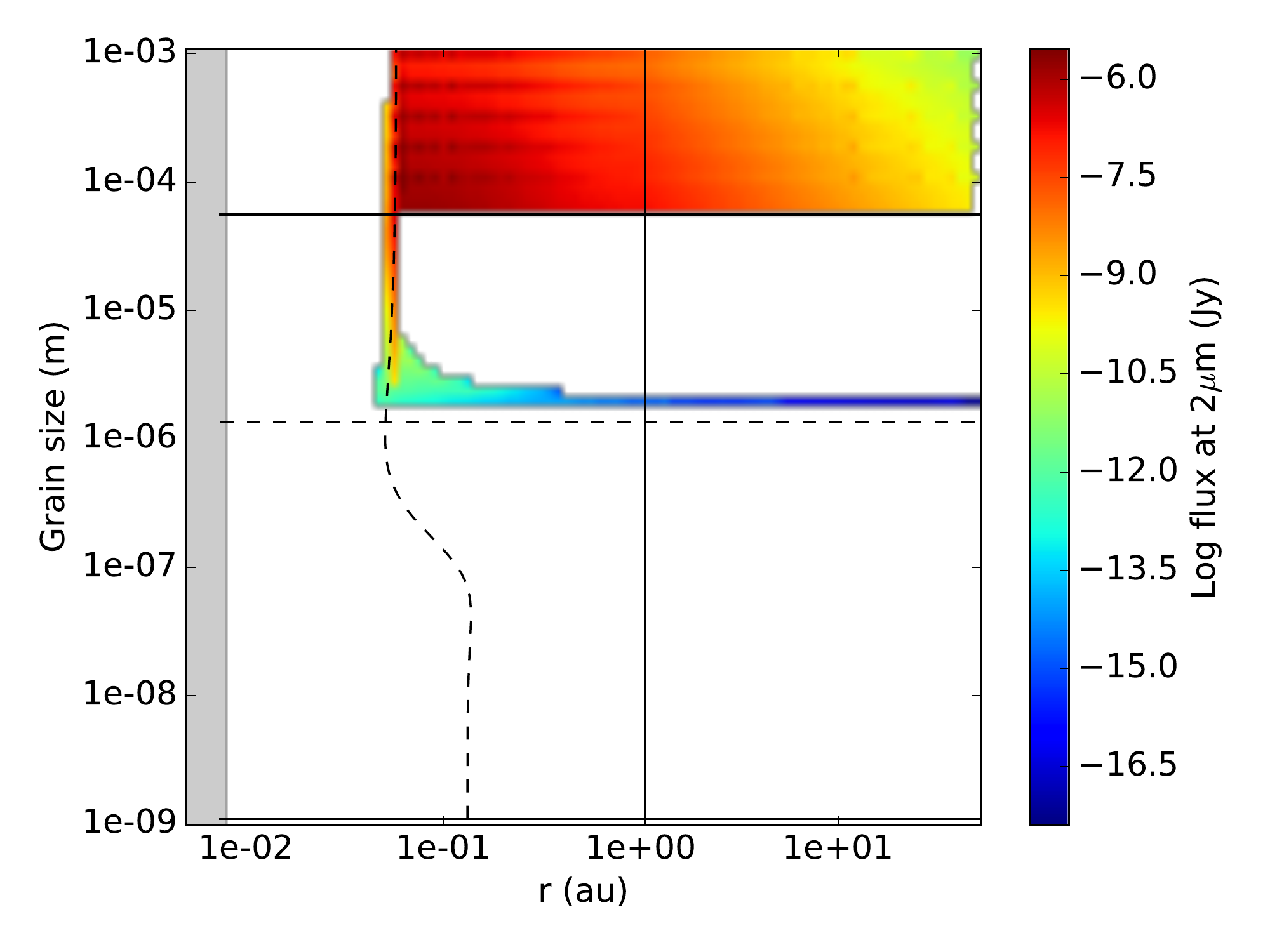}
\subcaption{}
}
 \hfill
\parbox{0.49\textwidth}{
\includegraphics[width=0.49\textwidth]{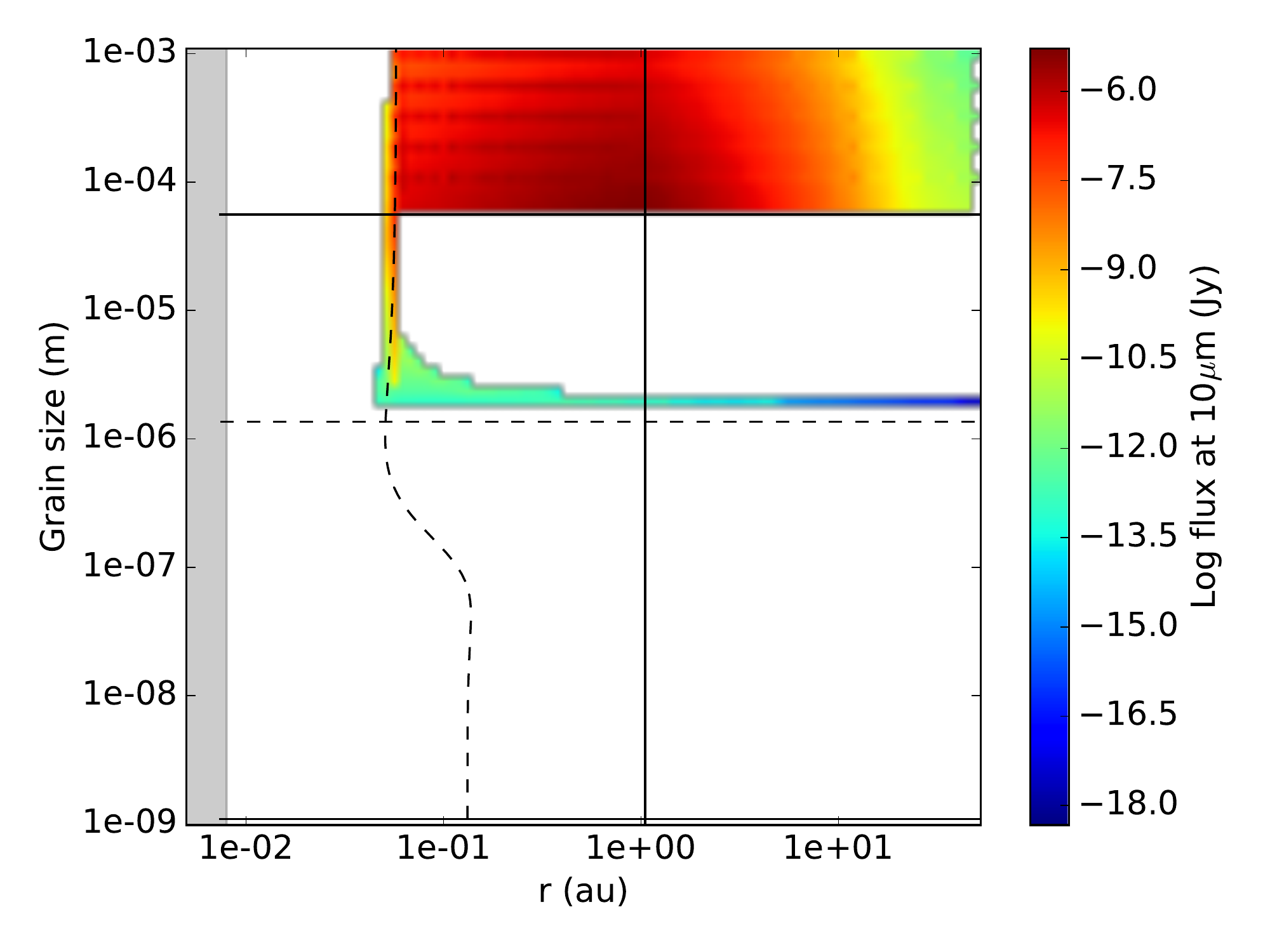}
\subcaption{}
}
}
\caption{Contribution of each size and distance bin to the flux at
  2\,$\mu$m (left) and 10\,$\mu$m (right) in the case of an F0 star
  and carbon grains. The grains are released by a exocomet at 0.13\,au
  (top) and 1\,au (bottom). }
\label{fig:ComFluxMap}
\end{figure*}

\subsubsection{SED}
\label{sec:comets_SED}

We use the same computing and normalization approach as in
Sec.\,\ref{sec:PRdrag_SED} to evaluate the SEDs resulting from the
exocometary dust production scenario. A notable difference with the
PR-drag pile-up scenario, is that it is possible here to connect the
absolute exozodiacal dust level to a number of exocomets passing close
to the star, as discussed further in Sec.\,\ref{sec:comets_flux}.

The synthetic SEDs for the exocometary scenario are displayed in
Fig.\,\ref{fig:ComTan_SED} for the four spectral types and three dust
compositions considered in this study, and are compared to the same
reference observed systems as in Fig.\,\ref{fig:PB_SED}. Overall, the mid- to
far-infrared excess is significantly reduced compared to the PR-drag
pile-up scenario. This is mainly because the grains are
directly deposited next to the sublimation zone, mitigating the amount
of dust beyond the grain release position (comet's perihelion). The
peak of the SED is accordingly shifted to much shorter wavelengths
compared to the PR-drag pile-up scenario. 

In this respect, the carbon grain case is the most favourable one. 
Regardless of the spectral type, the
exozodiacal emission of cometary carbon grains peaks at 3--5\,$\mu$m.
The flux at 10\,$\mu$m is always significantly smaller than the
stellar flux at the same wavelength, by a factor of about five, 
even though it is still not enough to fully fit the data. 
The models with astrosilicate and
glassy silicate grains, on the other hand, predict much too large
emissions in the mid- and far-infrared compared to the observations of
hot exozodis.

In Fig.\,\ref{fig:ComTan_SED}b we also display the SED obtained for the case with a comet
perihelion at $r\dma{p}=1\,$au, far away from the sublimation distance $r\dma{s}$. As can be clearly seen,
the fit to the observed data is much poorer because of the excess mid-IR flux due to PR-drag drifting grains in the
flat region between $r\dma{p}$ and $r\dma{s}$ that was identified in Fig.\ref{fig:ComTau}. In essence, we are here
close to the cases explored in the PR-drag pile-up scenario (Sec.\ref{sec:PRdrag}).

The main contributors to the flux at different wavelengths can be explored
using 2D emission maps as a function of the grain size and distance to
the star. Figures\,\ref{fig:ComFluxMap}a to d show examples for our best case, i.e., an F0
star and a carbon dust composition. We see that the near- and mid-IR emissions
essentially come from the same relatively large grains (several tens to hundreds
of micrometers), which are the smallest bound grains released by the
exocomets (Tab.\,\ref{tab:SizeBoundTan}).
At 2\,$\mu$m, most of the flux
originates from the smallest bound grains just outside the sublimation
distance, while the 10\,$\mu$m emission comes from similar grains in
size but distributed in a broader region centered around the dust
release position (Fig.\,\ref{fig:ComFluxMap}b). At that wavelength,
the drop of the surface density beyond the dust delivery position
(0.13\,au) contributes to moderating the emission from the distant
regions, as was demonstrated in the PR-drag pile-up
scenario when we schematically simulated much closer-in parent belts
(Fig.\,\ref{fig:PB_SED}). These behaviours at 2 and 10\,$\mu$m are
emphasized when the exocomet's perihelion is arbitrarily moved to
1\,au, as shown in Fig.\ref{fig:ComFluxMap}c and d. Finally, we note
that a direct consequence of the fact that the emission originates
from large grains is that the strong silicate features seen in case of
the PR-drag pile-up scenario are here essentially absent.

Overall, it is noteworthy that two key features of the classical
radiative transfer models of exozodis \citep[e.g.][and following
  studies]{Absil2006} are reproduced with the exocometary dust
delivery scenario, namely an accumulation of the grains very close to
the sublimation zone and a preference for carbon-rich
dust. Nevertheless, it should also be noted that the carriers of the
exozodi emission in that case are several orders of magnitude larger
in size than the grains usually required by the classical radiative
transfer models.

\subsection{Inward flux and size of the exocomets}
\label{sec:comets_flux}

As already mentioned, in the previous sections the flux was arbitrarily rescaled in order
to match the observed excess levels in the near-IR.
Now, in order to discuss further the relevance of the exocometary dust delivery scenario,
we aim to evaluate the size and number of exocomets required to physically reach
these observed exozodiacal flux level. For that purpose,
we employ the cometary dust ejection prescription of
\citet{Marboeuf2016}, in particular their equation 17, together with
the orbital parameters documented in Table~\ref{tab:SizeBoundTan}, to
quantify the mass of grains released by a comet over an orbital
period. This also allows to estimate, for a given comet mass, the exocomet lifetime as well as
the number of orbits before complete erosion. These quantities are
used to assess the absolute flux density at 2 and 10\,$\mu$m resulting
from the evaporation of an exocomet of a certain size. The different
steps of the adopted methodology are detailled in
Appendix~\ref{app:FluxCom},
and the results for a typical 10\,km-sized
exocomet are summarized in Tables~\ref{tab:ComFlux1} and
\ref{tab:ComFlux2}.

In the case of carbon and astrosilicate grains, the flux produced at
2\,$\mu$m by one 10\,km-sized exocomet amounts to a few $10^{-5}$ to a
few $10^{-4}$ the stellar flux at the same wavelength. Therefore, a
few tens to a few hundred of active 10\,km-radius exocomets on a
similar orbit are required to reach the observed level of $\sim$1\%
of the stellar flux at 2\,$\mu$m. At 10\,$\mu$m, the dust to star flux
ratio is always larger than at 2\,$\mu$m, by a factor of about twenty
for the carbon grains, and a factor of about 200 for the astrosilicate
grains, in agreement with the results in
Sec.\,\ref{sec:comets_SED}. 
The required number of
active exocomets can be mitigated if their initial size is larger. 
Indeed, because we consider the total flux resulting from the complete
erosion of the exocomet, it scales directly with the exocomet mass and
hence with the exocomet radius to the cube. Therefore, a single active
40 to 80\,km-sized exocomet would be enough to produce a 1\% excess at
2\,$\mu$m.
Glassy silicates fluxes are significantly lower due to the short sublimation lifetime 
of such grains. Around FGK stars, these grains sublimate significantly at each perihelion passage,
and disapear in a few orbits.

Several assumptions enter into the calculation of the number and size
of the exocomets required to reproduce the observations 
(see Sec.\,\ref{sec:Comet_NumSetup} and App.\,\ref{app:FluxCom}),
and the above
values should therefore be taken with caution.  It is also useful
to remind that our conclusions are valid for the specific orbital parameters
summarized in Table~\ref{tab:SizeBoundTan}. Nevertheless, the results
are encouraging in the sense that the sizes and number of exocomets
appear reasonable, thereby providing additional support to the
exocometary dust delivery scenario explored in this study.


\section{Summary and conclusion}

By investigating, with non orbit-averaged equations, the fate of dust
around several different stellar types, and considering different grain compositions,
we are able to draw some conclusions on the properties of exozodis
emission depending if these grains come from an outer parent belt and drift
inward by PR-drag (Sec.\ref{sec:PRdrag}), or if they have an
exocometary origin (Sec.\,\ref{sec:Comets}). We show that : \\

\noindent $\bullet$ in the case of the PR-drag pile-up scenario :
\begin{itemize}
\item for early-type stars, significant amounts of sub-\mum\ sized grains 
  cannot be produced in the inner disc regions, 
  because grains are blown-out by radiation pressure before sublimating down to these sizes. 
  The only case for which sub-\mum\ grains are obtained is that of silicate grains around later-type stars,
\item dust pile-up close to the sublimation radius is moderate, generating
  a density enhancement of a few at most with respect to an otherwise
  flat surface density profile,
\item the near-IR excess is always associated to mid-IR excess at
  the same level, or even much higher. This latest behaviour 
  cannot explain the numerous near-IR excesses without mid-IR excess detections.
\end{itemize}
\noindent $\bullet$ in the case of the Exocometary dust delivery
scenario :
\begin{itemize}
\item a narrow ring forms close to the sublimation zone, near the comet's periastron. 
  This ring is predominantly populated with large grains, a few tens to a few
  hundred $\mu$m in radius depending on the composition, which are the
  smallest bound grains produced at the comet's perihelion,
\item compared to the PR-drag pile-up scenario, the near-IR excess is
  associated with a much smaller mid-IR excess, in better agreement
  with the data, although not totally fitting them. Carbon-rich grains provide the best results,
\item the near-IR excess can be reproduced assuming a realistic inward
  fluxes of exocomets and reasonable exocomet sizes.
\end{itemize}
In addition, we find that the DDE mechanism, which could in principle help forming a
dense dust ring close to the sublimation radius, 
has a very limited effect in both scenarios,
because particles never reach \be\ values close enough to 1 by the time they are blown out.

Therefore, based on simulations performed with the new numerical model
developed in the context of this study, we conclude that the PR-drag
pile-up scenario is unlikely to produce the hot exozodis observed with
near-IR interferometry. The exocomet-release at perihelion scenario,
on the other hand, provides a very promising theoretical framework
that should be explored further. For that purpose, future developments of
our numerical model shall include a post-processing treatment of the
collisions in the dust ring close to the sublimation zone, building up
on the DyCoSS collisional model used in the context of debris disks
\citep[e.g.][]{Thebault2012,Thebault2014}. It should also include a detailed
calculation of the grain charging, following for instance the
prescription of \citet{Kimura2018}, and a careful consideration of the
magnetic topology, in order to thoroughly discuss the efficiency of
magnetic trapping and its impact on the near- and mid-IR emissions.

\begin{acknowledgements}

We thank Hiroshi Kimura for providing the optical constants for glassy
silicates, and for the discussions about the DDE mechanism.
We also thank Rik van Lieshout for useful discussion concerning sublimation processes. We
acknowledge the financial support from the Programme National de
Plan\'etologie (PNP) of CNRS-INSU co-funded by the CNES. 
Our code uses the NumPy \citep{Numpy} and the Matplotlib libraries
\citep{Matplotlib}.
This work has
made use of data from the European Space Agency (ESA) mission {\it
  Gaia} (\url{https://www.cosmos.esa.int/gaia}), processed by the {\it
  Gaia} Data Processing and Analysis Consortium (DPAC,
\url{https://www.cosmos.esa.int/web/gaia/dpac/consortium}). Funding
for the DPAC has been provided by national institutions, in particular
the institutions participating in the {\it Gaia} Multilateral
Agreement.
  
\end{acknowledgements}

\bibliographystyle{aa} \bibliography{library}

\begin{appendix}

\section{Reproducing \cite{Krivov1998}}
\label{app:Krivov}

\begin{figure}[h!]
\begin{center}
\includegraphics[width=0.45\textwidth, height=!]{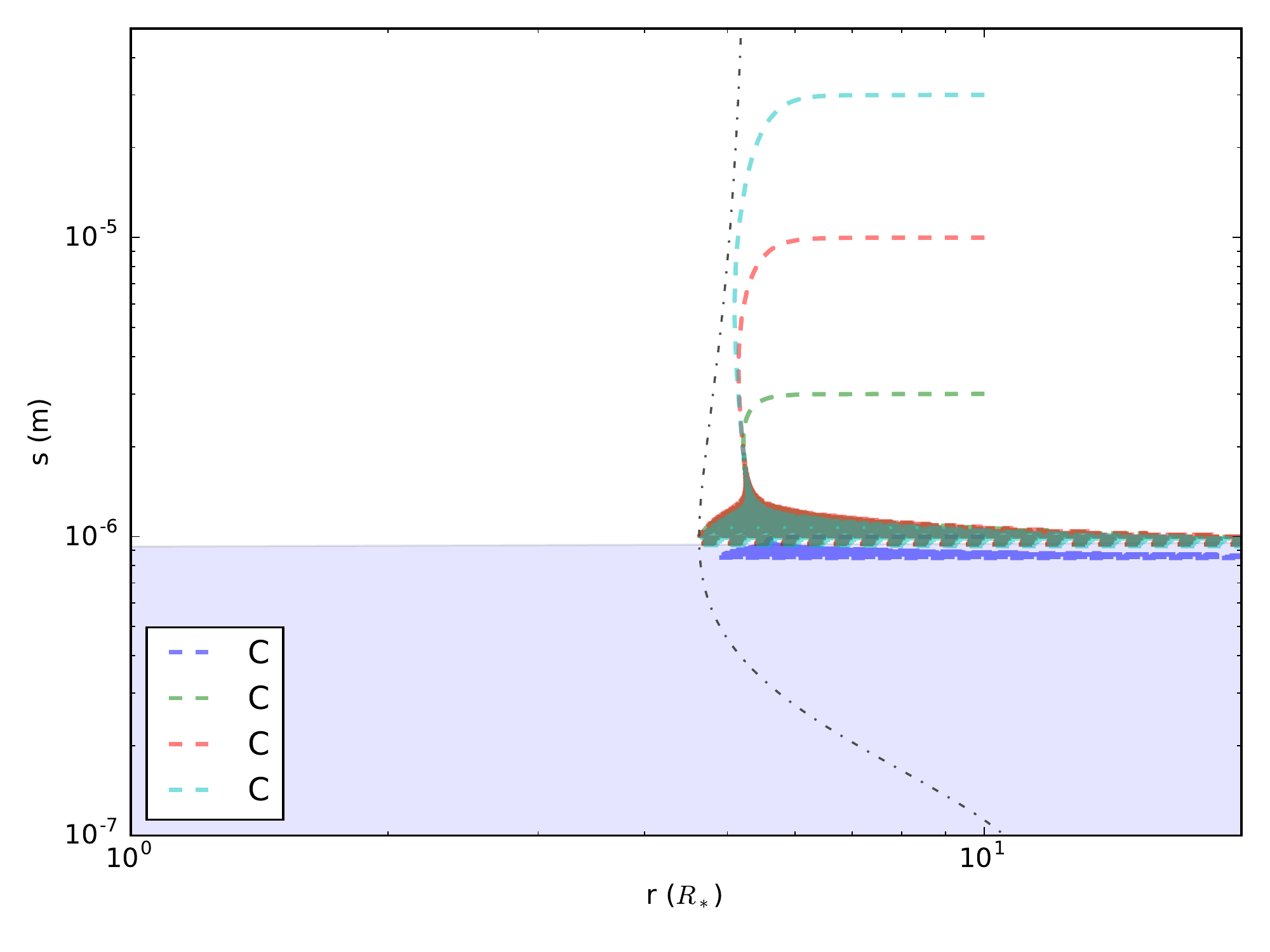}
\includegraphics[width=0.45\textwidth, height=!]{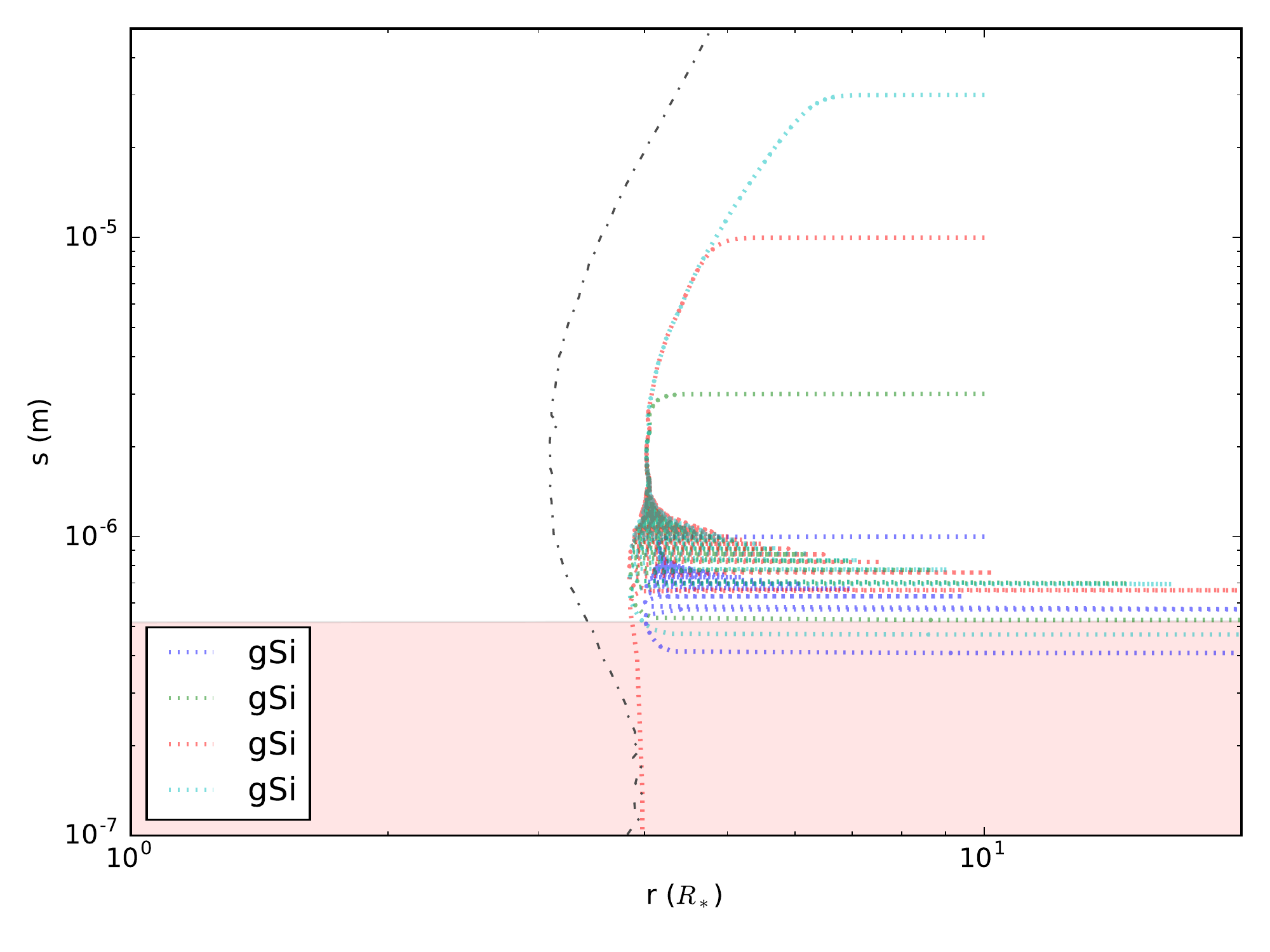}
\caption{Top : evolution of the grain size as a function of the
  distance to the Sun for carbon grains of initial sizes: 1; 3; 10;
  and 30\mum.  Bottom : same for the glassy silicate. These plots
  reproduce the results presented in Fig.\,5 of \cite{Krivov1998}.}
\end{center}
\label{fig:Kri98}
\end{figure}

\section{Thermodynamical properties}
\label{app:TermoProp}

The literature is rich in 2-parameter formula for describing the
sublimation process of dust grains, which are in essence similar but
with different notations, bringing some confusion. Here, we propose a
summary of the equations to transform the two parameters in four
different papers in the literature to the ($A,B$) parameters of
\cite{Lebreton2013} used in this study:
\begin{itemize}
\item ($A_Z,M_Z$) parameters in  \cite{Zavitsanos1973}, table\,3:
\begin{eqnarray}
A & = & M_Z \\
B & = & A_Z + 6 + \log_{10} \frac{k_B}{\mu m_u}
\end{eqnarray}
The value of 6 in the expression of B, coming from the unit system, is
lacking in \cite{Lebreton2013}, explaining the difference with our derived values.

\item ($H,P$) parameters in \cite{Kobayashi2009}:
\begin{eqnarray}
A & = & \frac{\mu m_u H}{k_B \ln 10}  \\
B & = & \log_{10} \frac{\mu m_u P}{k_B}
\end{eqnarray}

\item ($A_C,B_C$) parameters in \cite{Cameron1982}:
\begin{eqnarray}
A & = & \frac{10^4}{B_C} \\ 
B & = & \frac{A_C}{B_C} + 6 - \log_{10} \frac{k_B}{\mu m_u}
\end{eqnarray}

\item ($A,B_L$) parameters in \cite{Lamy1974}:
\begin{eqnarray}
B & = & B_L - \log_{10} \frac{k_B}{\mu m_u \times 1.33322 \cdot 10^3 }
\end{eqnarray}

\end{itemize}

\section{Geometrical optical depth map computation}
\label{app:DensMap}

\begin{figure}
\centering \includegraphics[width=0.49\textwidth, height=!, trim=0 0cm
  0 0]{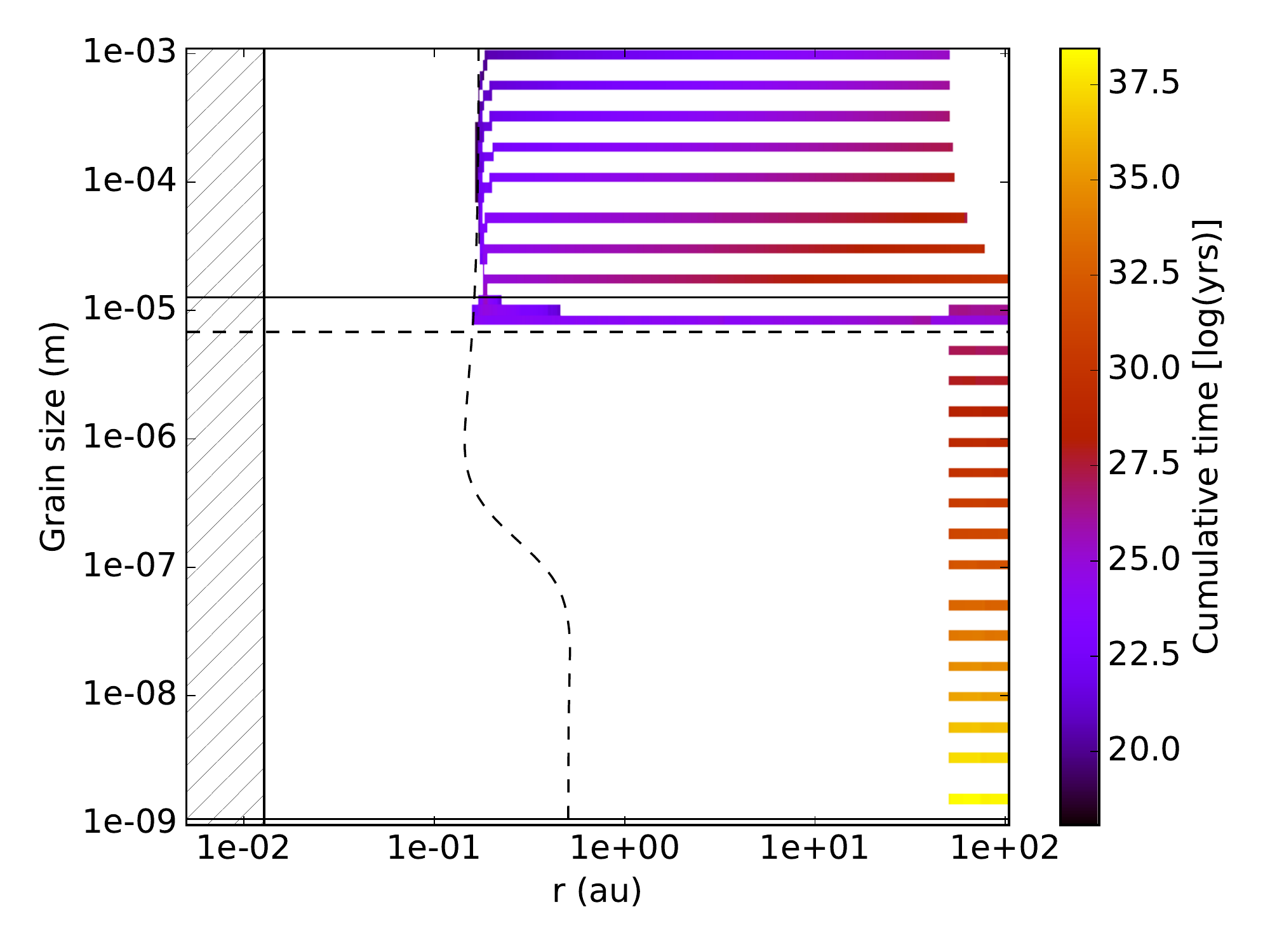}
\caption{Cumulative time map derived from the simulations for the A0,
  carbon case.}
\label{fig:A0CarDensSimu}
\end{figure}

Our goal is to produce density and optical depth maps from
trajectories independently computed for individual grains with our
dynamical code (that includes sublimation). In a first step, we define
a 2D grid of logarithmically-spaced distances to the star ($r$) and
grain sizes ($s$).  Then, for each single-grain size simulation, we
sum up the times spent by the grain during its lifetime in each bin
of the 2D grid, correcting this time by the initial differential size
distribution assumed to be proportional to $s^{-3.5}$ (collisional
erosion). This yields a first 2D map of cumulative times, that is
proportional to a density map if the system is assumed to be at
steady-state. As illustrated in figure~\ref{fig:A0CarDensSimu}, the
limited number of grain sizes for which the dynamics has been computed
leaves, however, empty grain size bins in the 2D map (empty
lines). This leads us to develop a complementary approach to fill the
holes in this map.

\begin{figure}
\centering \includegraphics[width=0.49\textwidth, height=!, trim=0 0cm
  0 0]{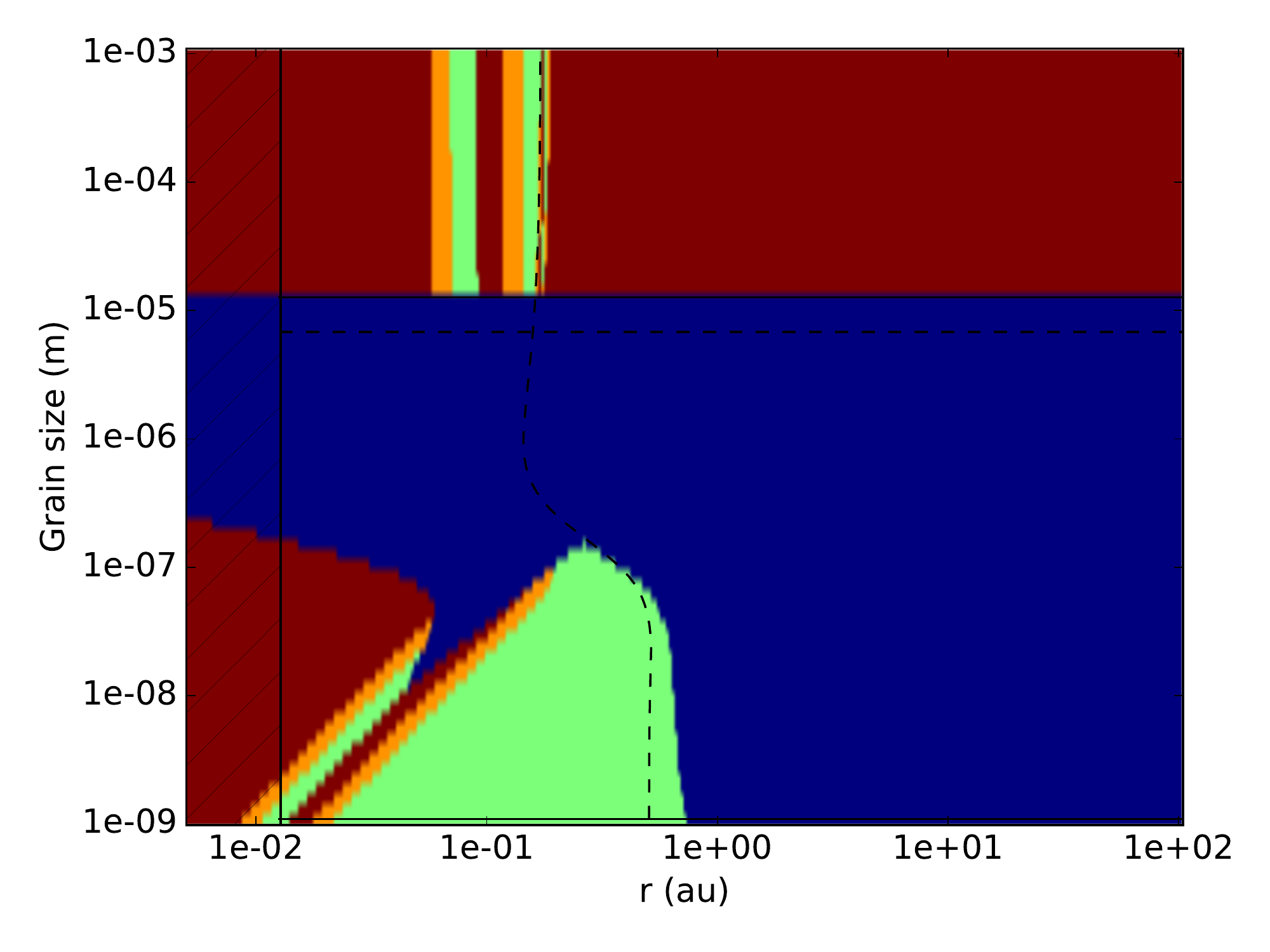}
\caption{Dominant physical processes driving the evolution of a grain,
  as a function of the distance to the star and grain size, with red
  being inward migration by PR-drag, green being sublimation,
  orange being a combination of both and blue
  being ejection by radiation pressure (see text for more detail on how
  the dominant process is estimated).}
\label{fig:A0CarEvo}
\end{figure}

\begin{figure}
\centering \includegraphics[width=0.49\textwidth, height=!, trim=0 0cm
  0 0]{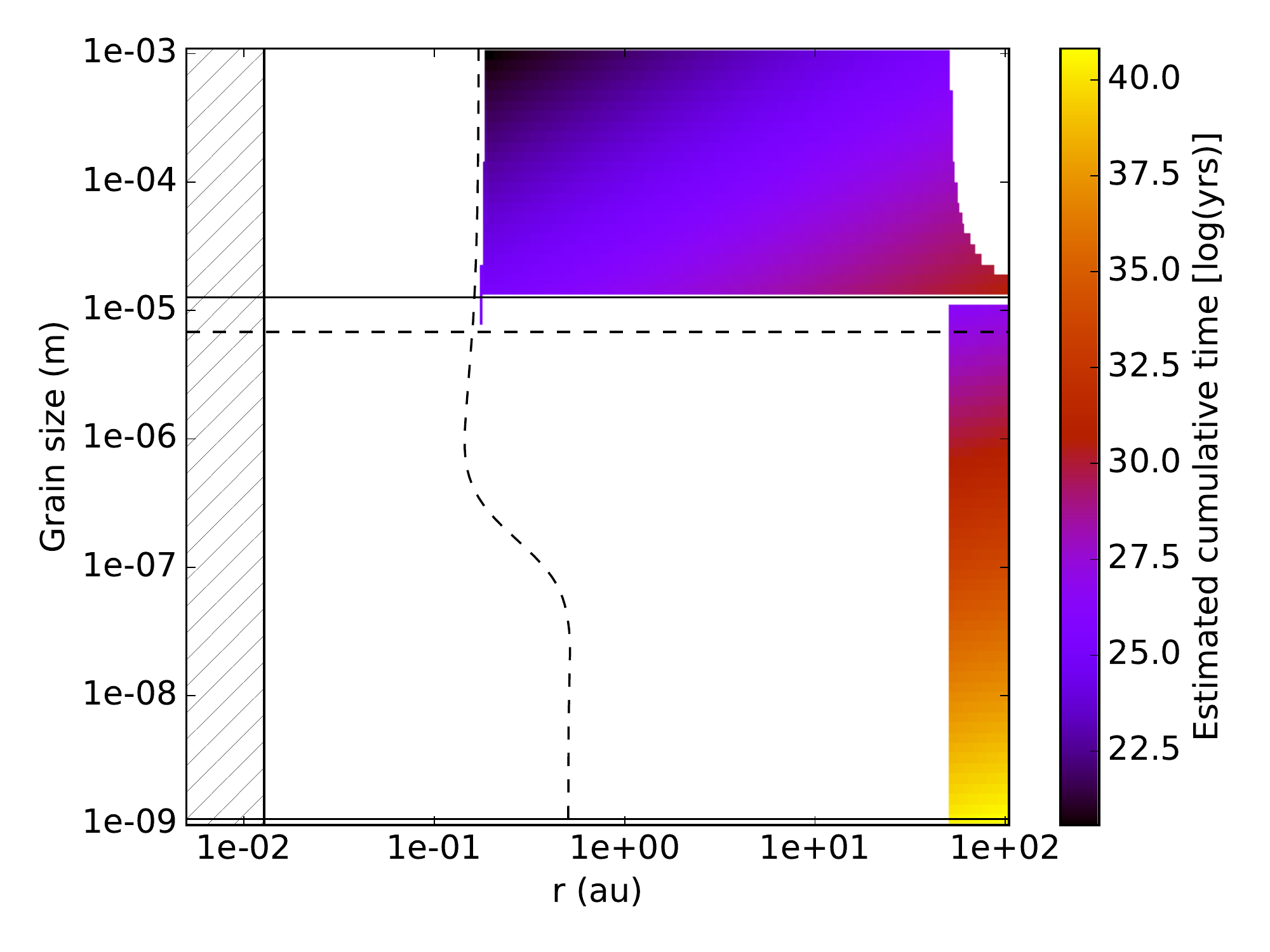}
\caption{Cumulative time map derived from the estimated timescales for
  the A0, carbon case.}
\label{fig:A0CarDensEst}
\end{figure}

We define a second 2D ($r$,$s$) map, of the same size and same bin
values as the first 2D map. We then estimate three timescales for each
bin in the 2D ($r$,$s$) map, to qualitatively evaluate the ability of
a grain to move to a nearby bin due to sublimation, ejection
(radiation pressure) or inward migration (PR-drag). The sublimation
timescale to move from ($r_i$,$s_i$) to ($r_i$,$s_{i-1}$) and the
ejection timescale to move from ($r_i$,$s_i$) to ($r_{i+1}$,$s_i$) are
both taken from \cite{Lebreton2013}. The PR-drag timescale to move
from ($r_i$,$s_i$) to ($r_{i-1}$,$s_i$) is taken from
\cite{Burns1979}. In
each bin, the shortest timescale gives the dominant physical process
and this is used to predict the trajectory of a grain in the 2D
map, allowing a jump from position ($r_i$,$s_i$) to ($r_{i-1}$,$s_{i-1}$)
when sublimation and PR-drag migration timescales are comparable. 
This is illustrated
in figure\,\ref{fig:A0CarEvo}, where we can see that sublimation is
the dominant physical process close to the sublimation distance for
the smallest grains, while inward migration is the dominant physical
process for the biggest grains, and ejection dominates over the other
processes otherwise.

This second 2D map provides crude evolution tracks for the grains that are
used to populate another cumulative time map similar to the one shown
in figure~\ref{fig:A0CarDensSimu}. We proceed as follows. For each
size bin, we populate the initial position bin given the value of \be,
thereby accounting for the eccentricity of the orbit
while the apoastron is fixed by the parent belt position.  Then, we
follow each synthetic grain along the $(r,s)$ plane, with a path
determined by the smallest timescale in each of the successive bins
through which the grain is passing. Each local, smallest timescale is
stored after being multiplied by the initial dust differential size
distribution proportional to $s^{-3.5}$, and summed up to give a
complete map of cumulative times, covering all grain sizes. This is
shown in figure~\ref{fig:A0CarDensEst} and can be directly compared to
the map obtained with the dynamical code
(Fig.\,\ref{fig:A0CarDensSimu}).

We see that the 2D ($r$,$s$) map obtained with the dynamical code
better follows the impact of changing \be\ values due to sublimation,
especifically it better captures the dynamics of the grains just above
the blow-out size. These grains produce the so-called (and
well-documented) pile-up close to the sublimation distance, that is
however not recover in the 2D map built using typical timescales. The
two cumulative time maps are then combined to produce a single, smooth
map, where averaged values are taken when the bins are non-zero in
both maps. This combined map can be regarded as a density map.

To get an optical depth map, the density map is multiplied by the
geometrical cross-section in each bin, and divided by the distance to
the star. This optical depth is thus vertically and azimuthally integrated. 
This can be used to construct
the optical depth radial profile by integrating over all grain sizes
for example, or as a pre-requisite to evaluate the flux in scattered
light and thermal emission (e.g. Figure\,\ref{fig:PBOptMap}). The 2D
maps can also be used to truncate the disk, to estimate the emission
within a certain radius for example.

For the sake of comparison, in all the paper the maps shown are
normalized to get a dust disk to star flux ratio of 1\% at $\lambda =
2\,\mu$m.

\section{Flux level produced by an evaporating exocomet}
\label{app:FluxCom}

Here, we describe the methodology employed to compute the amount of
dust released by the exocomets and the resulting flux
density. The successive steps are the following :

\begin{enumerate}

\item for each set of exocometary orbital parameters (see
  Sec.\,\ref{sec:Comet_NumSetup} and Tab.\,\ref{tab:SizeBoundTan}), we
  calculate the total mass of dust per unit of exocomet surface (in
  kg/m$^2$) released by the exocomet in one complete revolution. This
  is done by integrating equation 17 of \citet{Marboeuf2016} along the
  exocomet orbit over one orbital period. 
  The exocomet is supposed composed at 50\% of dust.  
  The results are independent
  of the actual size of the exocomet, but depend on the assumed grain size
  distribution. In the model of \citet{Marboeuf2016}, the mass is
  calculated assuming a differential grain size distribution
  proportional to $s^{-3.5}$ between $s_{\mathrm{min}} = 1\,\mu$m and
  $s_{\mathrm{max}} = 1$\, mm.

\item an exocomet radius is assumed to get the exocomet surface and
  hence the total mass of dust ejected from the exocomet in one orbit
  using the result of the previous step. The released dust masses are
  documented in the 6th row of Tables\,\ref{tab:ComFlux1} and
  \ref{tab:ComFlux2}, assuming an initial exocomet radius of 10\,km.

\item we make the approximation that all the dust ejected by the
  exocomet during one orbit is released at perihelion. Although this
  may appear a crude approximation, Fig.\,\ref{fig:ComFracMass} shows
  that this remains reasonable because a large fraction of the mass is
  produced close to the star, due to the high eccentricities
  considered in our study, and because the mass loss rate decreases as
  the distance squared to the star in the innermost regions
  \citep[inside the water ice sublimation distance in the model
    of][see their Eq.\,17]{Marboeuf2016}.

\item this mass is distributed over the grain size bins of our model
  grid (see App.\,\ref{app:DensMap}) between 1\,mm downto the smallest
  grain size in our simulation, assuming a differential grain size
  distribution proportional to $s^{-3.5}$.  This step allows each size
  bin to be populated with an absolute number of grains. Note that the
  smallest size of our grid is lower than the $s\dma{min} = 1\,\mu$m
  lower limit adopted by \citet{Marboeuf2016}, but the fraction of the
  total mass contained in the smallest grains is negligible with the
  adopted size distribution. Moreover, only the bound grains are kept
  in the next steps.

\item this setup is used to produce a 2D map of cumulative times as
  described in Appendix~\,\ref{app:DensMap}, weighted by the absolute
  number of grains of each size obtained at the previous step. The 2D
  map is obtained by evolving the grains in position and size over the
  largest lifetime of the biggest grains (3rd row of
  Tables\,\ref{tab:ComFlux1} and \ref{tab:ComFlux2}), see
  Appendix~\,\ref{app:DensMap} for the methodology. This step assumes
  in essence a system at steady state and is equivalent to populating
  the orbits to produce a density map. The 2D map is normalized by the
  time over which the simulation was evolved (3rd row of
  Tables\,\ref{tab:ComFlux1} and \ref{tab:ComFlux2}) to obtain a mean
  2D number density map equivalent to a density map assuming a
  constant dust production process at perihelion. This procedure is
  valid if the grains released at the comet's perihelion which dominate the flux
  survive long enough for their positions to be randomized in longitude
  along their orbits.
  We have checked that, in the case of the A0 star and carbon grains,
  the orbital periods vary by a factor five within the considered grain range.
  This guaranties randomization in longitude within a few orbits,
  i.e. $\sim$1000 years in this case,
  to be compared with the grain lifetime which is typically two orders of magnitude larger.

\item the mean 2D number density map is used to directly compute the
  absolute scattered light and thermal emission at the wavelength of
  $2\,\mu$m, providing an estimate of the mean flux density resulting
  from the first passage of a exocomet (7th row in
  Tables\,\ref{tab:ComFlux1} and \ref{tab:ComFlux2}). As time goes
  one, the exocomet radius shrinks. The number of exocomet orbits
  before complete sublimation and the exocomet lifetimes are
  documented in the 5th and 4th rows of Tables\,\ref{tab:ComFlux1} and
  \ref{tab:ComFlux2}, respectively. The total flux density at
  $2\,\mu$m, produced by all the grains released by the exocomet over
  its lifetime, is then obtained by summing up the contributions at
  each successive perihelion passage until complete erosion of the
  exocomet (8th row of Tables\,\ref{tab:ComFlux1} and
  \ref{tab:ComFlux2}). The flux at $10\,\mu$m (9th row of
  Tables\,\ref{tab:ComFlux1} and \ref{tab:ComFlux2}) is computed in a
  similar way.

\end{enumerate}

\begin{table*}
\caption{Parameters and flux for the A0 and F0 stars, resulting from
  the exocomet evaporation model, assuming a 10\,km-sized
  exocomet. See Appendix~\ref{app:FluxCom} for details.}
\label{tab:ComFlux1} 
\begin{center}
\begin{tabular}{c c c c c c c c }

\hline\hline
Star & \multicolumn{3}{c}{A0} &	 &  \multicolumn{3}{c}{F0}  \\ 
Composition & Carbon & Astrosilicate & Glassy silicate 
& & Carbon & Astrosilicate & Glassy silicate \\
\cline{2-4} \cline{6-8}
Maximum grain lifetime (yr) 
	& $6.3\times 10^4$ & $1.8\times 10^5$ & $5.5\times 10^{5}$ & 
	& $9.9\times 10^4$ & $1.4\times 10^5$ & $2.1\times 10^{5}$ \\
Exocomet lifetime (yr) 
	& $3.5\times 10^3$ & $8.6\times 10^3$ & $6.9\times 10^{3}$ & 
	& $2.0\times 10^4$ & $3.5\times 10^4$ & $1.7\times 10^{4}$ \\ 
Number of exocometary orbits 
	& 48 & 117 & 95 &
	& 204 & 357 & 169 \\
Mass released in one orbit (kg) 
	& $1.5\times 10^{14}$ & $6.0\times 10^{13}$ & $7.4\times 10^{13}$ &
	& $3.5\times 10^{13}$ & $2.0\times 10^{13}$ & $4.4\times 10^{13}$ \\
Flux ratio at 2\,$\mu$m (first orbit) 
	& $2.0\times 10^{-6}$ & $1.2\times 10^{-6}$ & $3.6\times 10^{-7}$ &
	& $3.1\times 10^{-6}$ & $2.2\times 10^{-6}$ & $4.2\times 10^{-9}$ \\
Total flux ratio at 2\,$\mu$m 
	& $3.3\times 10^{-5}$ & $4.8\times 10^{-5}$ & $1.1\times 10^{-5}$ &
	& $2.1\times 10^{-4}$ & $2.7\times 10^{-4}$ & $2.2\times 10^{-7}$ \\ 
Total flux ratio at 10\,$\mu$m 
	& $8.4\times 10^{-4}$ & $1.0\times 10^{-2}$ & $2.6\times 10^{-3}$ &
	& $4.0\times 10^{-3}$ & $3.4\times 10^{-2}$ & $1.6\times 10^{-5}$ \\
\hline
\end{tabular}
\end{center}
\end{table*}

\begin{table*}
\caption{Same as Tab.\,\ref{tab:ComFlux1} for the G0 and K0 stars.}
\label{tab:ComFlux2} 
\begin{center}
\begin{tabular}{c c c c c c c c }

\hline\hline
Star &  \multicolumn{3}{c}{G0} &  & \multicolumn{3}{c}{K0} \\
Composition & Carbon & Astrosilicate & Glassy silicate &
	& Carbon & Astrosilicate & Glassy silicate \\
\cline{2-4} \cline{6-8}
Maximum grain lifetime (yr) 
	& $1.8\times 10^5$ & $2.0\times 10^5$ & $3.3\times 10^{4}$ &
	& $1.8\times 10^5$ & $1.7\times 10^5$ & $6.6\times 10^{5}$ \\
Exocomet lifetime (yr) 
	& $4.3\times 10^4$ & $7.3\times 10^4$ & $3.0\times 10^{4}$ &
	& $9.6\times 10^4$ & $2.0\times 10^5$ & $7.0\times 10^{4}$  \\
Number of exocometary orbits 
	& 351 & 598 & 243 &
	& 683 & 1408 & 496 \\
Mass released in one orbit (kg) 
	& $2.5\times 10^{13}$ & $1.2\times 10^{13}$ & $5.4\times 10^{13}$ &
	& $2.6\times 10^{13}$ & $5.5\times 10^{12}$ & $5.1\times 10^{13}$ \\
Flux ratio at 2\,$\mu$m (first orbit) 
	& $2.0\times 10^{-6}$ & $1.9\times 10^{-6}$ & $2.1\times 10^{-10}$ & 
	& $5.4\times 10^{-6}$ & $3.4\times 10^{-6}$ & $2.8\times 10^{-10}$ \\
Total flux ratio at 2\,$\mu$m 
	& $1.8\times 10^{-4}$ & $3.8\times 10^{-4}$ & $9.2\times 10^{-9}$ &
	& $4.8\times 10^{-4}$ & $1.5\times 10^{-3}$ & $1.3\times 10^{-8}$ \\
Total flux ratio at 10\,$\mu$m 
	& $3.1\times 10^{-3}$ & $3.7\times 10^{-2}$ & $4.1\times 10^{-7}$ &
	& $7.2\times 10^{-3}$ & $9.7\times 10^{-2}$ & $2.5\times 10^{-7}$ \\
\hline
\end{tabular}
\end{center}
\end{table*}

\end{appendix}

\end{document}